\documentclass{aa}
\usepackage{graphicx}
\usepackage{natbib}

\def\arxivprefixe{arXiv}
\def\arxivprefixesep{:}

\usepackage[separate-uncertainty]{siunitx}
\graphicspath{{.}}

\usepackage{tikz}
\usetikzlibrary{positioning}

\usepackage{txfonts}
\usepackage{hyperref}
\hypersetup{pdfauthor=Remo Burn, pdftitle=New generation planetary population synthesis. VI., breaklinks=true, colorlinks=true, urlcolor=blue, linkcolor=blue, citecolor=blue}

\renewcommand*\P[1]{Paper \uppercase\expandafter{\romannumeral #1\relax}}\defcitealias{Emsenhuber2020a}{Paper I}
\def\paperone{\citetalias{Emsenhuber2020a}\relax}
\defcitealias{Emsenhuber2020b}{Paper II}
\def\papertwo{\citetalias{Emsenhuber2020b}\relax}
\newcommand*\Tra{\mbox{TRAPPIST-1} }

\newcommand*\frev[1]{#1}
\newcommand*\frevi[1]{#1}
\newcommand*\flrev[1]{#1}

\begin{document}

   \title{The New Generation Planetary Population Synthesis (NGPPS)}
   \titlerunning{The New Generation Planetary Population Synthesis (NGPPS). IV.}

   \subtitle{IV. Planetary systems around low-mass stars\thanks{The data supporting these findings are available online at \mbox{\url{http://dace.unige.ch}} under section ``Formation \& Evolution''.}}

   \author{R. Burn \inst{\ref{unibe},\ref{mpia}}, M. Schlecker \inst{\ref{mpia}}, C. Mordasini \inst{\ref{unibe}}, A. Emsenhuber \inst{\ref{unibe},\ref{lpl}}, Y. Alibert \inst{\ref{unibe}}, T. Henning \inst{\ref{mpia}}, H. Klahr \inst{\ref{mpia}} \and W. Benz \inst{\ref{unibe}}}
   \authorrunning{R.~Burn et al.}

   \institute{Physikalisches Institut  \& Center for Space and Habitability, Universit\"at Bern, CH-3012 Bern, Switzerland \label{unibe} 
         \and
             Max-Planck-Institut f\"ur Astronomie, K\"onigstuhl 17, 69117 Heidelberg, Germany,\label{mpia}
             \email{burn@mpia.de}
         \and
                 Lunar and Planetary Laboratory, University of Arizona, 1629 E. University Blvd., 85721 Tucson, AZ, USA\label{lpl}
             }

   \date{\today}
 
  \abstract
   {Previous theoretical works on planet formation around low-mass stars have often been limited to large planets and individual systems.
   As current surveys routinely detect planets down to terrestrial size in these systems, \flrev{models have shifted} toward a more holistic approach that reflects their diverse architectures.
}
   {
        Here, we investigate planet formation around low-mass stars and identify differences in the statistical distribution of modeled planets. We compare the synthetic planet populations to observed exoplanets and we discuss the identified trends.}
   {We used the Generation III Bern global model of planet formation and evolution to calculate synthetic populations, while varying the central star from Solar-like stars to ultra-late M dwarfs. This model includes planetary migration, N-body interactions between embryos, accretion of planetesimals and gas, and the long-term contraction and loss of the gaseous atmospheres.   }
   {We find that temperate, Earth-sized planets are most frequent around early M dwarfs (\SIrange{0.3}{0.5}{M_{\odot}}) and that they are more rare for Solar-type stars and late M dwarfs. 
   The planetary mass distribution does not linearly scale with the disk mass. 
   The reason behind this is attributed to the emergence of giant planets for $M_\star \ge \SI{0.5}{M_{\odot}}$, which leads to the ejection of smaller planets.
        Given a linear scaling of the disk mass with stellar mass, the formation of Earth-like planets is limited by the available amount of solids for ultra-late M dwarfs.
        For $M_\star \ge \SI{0.3}{M_{\odot}}$, however, there is sufficient mass in the majority of systems, leading to a similar amount of Exo-Earths going from M to G dwarfs. In contrast, the number of super-Earths and larger planets increases monotonically with stellar mass.
   We further identify a regime of disk parameters that reproduces observed M-dwarf systems such as TRAPPIST-1. However, giant planets around late M dwarfs, such as GJ 3512b, only form when type I migration is substantially reduced.}       
   {
   \frev{We} are able to quantify the stellar mass dependence of multi-planet systems using global simulations of planet formation and evolution.
   The results fare well in comparison to current observational data and predict trends that can be tested with future observations.
   }

   
   \maketitle
%
\section{Introduction}
\label{sec:introduction}
M dwarf stars are the most abundant stars in the Milky Way \citep{Winters2014} and represent a unique laboratory for testing current planet-formation theories. Following the discovery of the first planet around such a star \citep{Marcy1998}, they are now known to be the most frequent hosts of exoplanets \citep[e.g.,][]{Gaidos2016}. Currently, the observational sample of planets around M dwarfs is rapidly increasing, as a number of new detection surveys are being conducted or planned. Programs using the transit detection method include space-based TESS, which is more sensitive to longer wavelengths than its predecessor, \textit{Kepler} \citep{Ricker2014}, or surveys on the ground, such as MEarth \citep{Nutzman2008}, TRAPPIST \citep{Gillon2016,Gillon2017}, NGTS \citep{Wheatley2018}, \flrev{EDEN \citep{Gibbs2020} and SPECULOOS \citep{Burdanov2018,Delrez2018a}. The latter makes use of the purpose-built Saint-Ex telescope \citep{Demory2020}}. In addition, the radial velocity program CARMENES \citep{Quirrenbach2020} or the MARROON-X \citep{Seifahrt2016} and NIRPS instruments \citep{Bouchy2017} have already led to many planet detections around low-mass stars. Together, they are capable of exploring a parameter space of the planetary population that has thus far been fairly inaccessible  -- rocky exoplanets around low-mass stars. The fundamental benefit in the search for potentially habitable planets around low-mass stars for both transit and radial velocity surveys lies in the fact that the planet to star radius ratio, or respectively the mass ratio, becomes larger for lower stellar masses. This results in a higher measured signal-to-noise ratio. Additionally, the temperate zone \citep{Kasting1993,Tasker2017} is located at shorter orbital periods around M dwarfs due to the lower temperature of the central star. Therefore, less observation time is needed for the discovery of planets receiving stellar irradiation fluxes comparable to Earth and they are thus, in the current stage, the best candidates for the search for  extraterrestrial life with the potential to emerge under these different conditions \citep[e.g.,][]{Shields2016}.

Theoretical works have previously addressed planet formation around different stellar masses: for instance, \citet{Laughlin2004} found that giant planet formation is reduced around low-mass stars. Early works on population synthesis were carried out by \citet{Ida2005} and \citet{Alibert2011}, also mainly focusing on the most heavy planets in a system. \frev{Similarly, \citet{Payne2007} developed a semi-analytic model for planet growth akin to that of \citet{Ida2004,Ida2005} and applied it to stellar masses extending to brown dwarfs.}

In contrast to these works, \citet{Raymond2007} discussed the formation of terrestrial planets around low-mass stars following the dissipation of the disk (i.e., without migration) by injecting about 100 N-body particles -- called planetary embryos -- into regions chosen to cover the temperate zone and the water iceline. They found fewer planets in the temperate zone with decreasing stellar mass and go on to note in their conclusions that disk migration could help to explain the observed system around the M3 dwarf Gliese 581 \citep{Bonfils2005,Udry2007}. This is further supported in the light of the discoveries of systems around even lower-mass stars, such as \Tra \citep{Gillon2017}, or by observational trends derived from the \textit{Kepler} sample \citep{Mulders2015a}. Overall, that highlights the need to include a migration mechanism in modern planet formation models to match the observed systems. However, this is still not undisputed as \cite{Hansen2015} reported in situ planet formation models around a \SI{0.5}{M_{\odot}} star matching the observed planetary periods.

More recent theoretical planet formation works have discussed the solid delivery process. While \frev{\citet{Ormel2017},} \citet{Schoonenberg2019}, and \citet{Coleman2019} focus on the \Tra system in the framework of (hybrid) planetesimal and pebble accretion, \frev{\citet{Liu2019,Liu2020a}} consider distributions of initial stellar and disk properties in the pebble accretion scenario. \frev{\citet{Liu2019,Liu2020a}} calculate a population of single planets, that is, neglecting N-body interactions, around stars with masses from \SI{0.1}{M_{\odot}} to \SI{1}{M_{\odot}}. The resulting planetary populations based on pebble accretion could be compared to planetesimal accretion, which is considered to be the dominating solid mass delivery mechanism here. However, we caution that for smaller rocky planets, the scenario of a single planet forming in the disk gives different results compared to simulations taking into account the interaction and growth competition between planets \citep{Alibert2013}. Therefore, it is more insightful to compare pebble and planetesimal accretion in the context of the same model, as recently done by \citet{Coleman2019} and \citet{Brugger2020}.

\citet{Miguel2020} focus on forming rocky planets around stars of masses up to \SI{0.25}{M_{\odot}} by accreting planetesimals \citep[following][]{Ida2004} and explicitly exclude the accretion of gaseous envelopes. In contrast, we include the growth of atmospheres around small planets\frev{ which allows for} the formation of giants that are observed at stellar masses larger than \SI{0.3}{M_{\odot}}. This extends the parameter space \frev{for which our model can be applied} to the range from \SIrange{0.1}{1}{M_{\odot}}.

A preceding synthetic population of planets around a \SI{0.1}{M_{\odot}} star was presented in the letter from \citet{Alibert2017}, who found more water-rich compositions of planets forming around very low-mass stars compared to those around solar-type stars. Here, we use an updated version of our population synthesis model compared to \citet{Alibert2017} and \citet{Alibert2011}, which affects the conclusions to a certain degree, as discussed in Sect. \ref{sec:discussion}. These updates are presented as part of a series of papers: In \citet[][hereafter referred to as \paperone{}]{Emsenhuber2020a}, the updated model is described in detail and \citet[][hereafter \papertwo{}]{Emsenhuber2020b}, the statistical properties of the populations around solar-type stars are discussed.

This work begins with a brief description of the model and the adopted stellar mass dependencies of initial conditions in Sect. \ref{sec:model}, followed by the presentation of general results in Sect. \ref{sec:results}. In Sect. \ref{sec:discussion}, we compare selected aspects of the resulting planetary population to observations and previous theoretical works. In particular, we discuss the synthetic systems with respect to the TRAPPIST-1 system in Sect. \ref{sec:trappist}. Finally, we summarize our findings and conclusions in Sect. \ref{sec:conclusion}.
%
\section{Formation models}
\label{sec:model}

We use the Generation III \textit{Bern} model of planet formation and evolution. It originates from the model of \citet{Alibert2004a,Alibert2005}, which was extended to incorporate the long-term evolution of planets \citep{Mordasini2012c,Mordasini2012b} and simultaneously developed to include N-body interactions \citep{Alibert2013,Fortier2013}. The third generation employed here combines the two branches, including a multitude of additional and updated physical mechanisms, as  described in detail in \paperone{}. We computed a population of planets given a variety of initial disk conditions that are based on the observed population of disks. This approach is referred to as a planetary population synthesis \citep{Ida2004,Ida2004a,Ida2005,Mordasini2009,Mordasini2009b,Benz2014,Mordasini2015,Mordasini2018}.

Here, we briefly summarize the relevant physical processes included in the Bern model. For a detailed, complete description, we refer to \paperone{}. The protoplanetary disk is modeled following the viscous $\alpha$-disk model \citep{Shakura1973,Pringle1981}. In addition to energy dissipation due to viscosity and shear, the disk is also heated by stellar irradiation. To offer a description of this, we follow \citet{Hueso2005} and \citet{Nakamoto1994}, who give analytic expressions for the disk midplane temperature, assuming an isothermal disk in the vertical direction. In contrast to those two works, we consider the starting time of our disk to come after a significant infall of gas onto the disk has stopped.

The final phase of the disk's life is dominated by disk photo-evaporation. We modeled internal photo-evaporation by the star following \cite{Clarke2001} and we use a simple prescription to include external photo-evaporation due to the radiation by nearby stars \citep{Matsuyama2003}.

To model planetary growth by planetesimal accretion, protoplanetary embryos with an initial mass of \SI{0.01}{M_\oplus} are injected at random locations in the disk (see Sect. \ref{ssec:initial_embryo_placement}). The embryos are gravitating bodies tracked by the MERCURY N-body code \citep{Chambers1999}. They can accrete planetesimals inside their feeding zone from a planetesimal disk that is described statistically as a surface density with an evolving dynamic state, that is, with eccentricity and inclination \citep{Fortier2013}. The gravitational forces of the planetesimal disk onto the embryos is not taken into account. In addition to the gravitational forces of the central star and other embryos, the planets are subject to the torque of the gaseous disk (see Sect. \ref{ssec:disk_migration}).

Concurrently, the Bern model solves the one-dimensional internal-structure equations \citep{Bodenheimer1986} for each embryo at every timestep for the solid core and the gaseous envelope assuming hydrostatic equilibrium. The energy input of the accreted planetesimals is assumed to be deposited at the core-envelope boundary of the embryo. To accrete gas, it has to cool and contract by radiating away the potential energy of accreted planetesimals and gas. This then determines the envelope mass. The cooling becomes efficient at $\sim$\SI{10}{M_\oplus}, which can then lead to a planet undergoing "runaway" gas accretion \citep{Mizuno1980,Pollack1996}. As a consequence, the planet contracts and is thus considered to be detached from the disk. In this stage, gas accretion is limited by what the disk can provide \mbox{\citep{Machida2010}}. This transition roughly coincides with the time the planet changes the migration regime (type I to type II) \citep{Alibert2005}.

\frev{Despite the importance of the envelope structure for gas accretion, the radii of planets in the terrestrial mass regime are dominated by the core structure. To calculate realistic radii in this case, we follow the model of \citet{Mordasini2012b}, which uses the equation of state from \citet{Seager2007} for the considered iron, silicate, and water phases. They are layered in this order and we assume no mixing and no thermal expansion. If an envelope is present, we set the outer pressure for the core structure to the pressure obtained from the envelope model, otherwise a value of \SI{e-7}{bar} is used.}

\frev{The required water and iron mass fractions are tracked over the course of the formation phase of the model and depend on where the planet has accreted planetesimals. Therefore, two planetary cores of equal mass can have different densities at the same location if they followed a different migration path.}
        
\frev{Although we consider this approach sufficient for our purposes and required precision, we stress that in recent years, significant advances have been made in the field of interior structure modeling \citep[see][for recent reviews]{Hirose2013,VanHoolst2019,Taubner2020}: these include calculations of the pressure and temperature dependent mineralogy \citep{Dorn2015} or hydration of the core and the mantle \citep{Shah2021}. Furthermore, recent updates \citep{Bouchet2013,Hakim2018,Mazevet2019} and collections \citep{Zeng2019,Haldemann2020} of equations of state were presented.}

After the gaseous disk is gone, we continue running the N-body integration up to \SI{20}{Myr} to track dynamical instabilities occurring after the dissipation of the disk. After that, only the evolutionary calculations are performed, which include the continuation of solving the internal structure equations -- thus tracking the long-term cooling and contraction of planets -- as well as the evaporation of atmospheres by X-ray and extreme-ultraviolet driven photo-evaporation along with tidal migration. At all times, although this is of greater significance in the evolution phase, the star is evolving in radius and luminosity following the stellar evolution tracks of \citet{Baraffe2015}. This leads to an evolving radiative energy input to the planetary structure.

Overall, we chose the nominal physical parameters and processes of \paperone{} and \papertwo,{} but extended the explored parameter space to lower stellar masses. Apart from the different initial disk conditions described in Sect. \ref{sec:scaling}, the stellar mass directly enters in different important processes:
First, the luminosity and radius of the star as well as its evolution is altered, following the model of \cite{Baraffe2015}. Second, the evolution of the protoplanetary disk is modified for lower-mass stars, as the viscous heating (which depends on the Keplerian frequency, \citealt{Pringle1981}) and the irradiation depend on the mass, radius, and effective temperature of the central object \citep{Hueso2005,Nakamoto1994}. Next, the radius of disk-embedded planets is equal (for masses that are not overly small) to a fraction of the Hill radius \citep{Mordasini}, which depends on the mass of the central body as $R_\mathrm{H} = a \left(\frac{M}{3 M_\star}\right)^{1/3}$. In addition, the evolution of the planetesimal disk (in terms of $e$ and $i$) is a function of the planetesimals' Hill radii  \citep{Adachi1976,Inaba2001,Rafikov2004}. Similarly, the accretion rate of planetesimals scales linearly with the planet's and the planetesimals' mutual Hill radius \citep{Fortier2013}. Furthermore, the fraction of ejected planetesimals is a function of the escape speed from the primary ($v_{\mathrm{esc,}\star} = \sqrt{2 G M_\star/a}$) \citep{Ida2004}. Also, the computation of the planetary orbital evolution due to disk-planet interactions is modified for low mass stars (see Sect. \ref{ssec:disk_migration}). Last, the tidal interaction with the star changes depending on the radius and mass of the star \citep{Ogilvie2014}, leading in our simplified model \citep{Benitez-Llambay2011} to slower tidal migration for lower stellar masses.

\subsection{Disk migration}
\label{ssec:disk_migration}
Planets embedded in a disk will be subject to the torques of the disk. Depending on the mass of the planet, the migration regime is classified as type I and type II for no gap-opening or gap-opening, respectively. As described in detail in \paperone{}, we follow \citet{Paardekooper2011} with regard to the eccentricity and inclination damping from \citet{Coleman2014} for the type I regime and \citet{Dittkrist2014} for type II.

For type I, the formulas for planets on circular orbits follow \citet{Paardekooper2011}. Therefore, an overall factor of:\ 
\begin{equation}
\label{eq:gamma_0}
\Gamma_0 = \left(\frac{M a}{M_{\star} H} \right)^2 \Sigma_\mathrm{g} a^4 \Omega_\mathrm{K}^2\,,
\end{equation}
where $H$ is the disk scale height and $\Omega_\mathrm{K}$ is the Keplerian orbital velocity at the planet location, can be identified to get a grasp of the general scaling of the total torque $\Gamma$ with the stellar mass ($\propto M_\star^{-1}$) at fixed semi-major axis and gas surface density.

\frev{However, the migration rates in the type I regime follow a more complex pattern due to the presence of the corotation torque. This torque -- typically leading to outward migration -- originates from gas parcels in the horseshoe region of the planet describing U-turns. In the presence of entropy or vortensity gradients, the angular momentum exchange on an outward U-turn and an inward one is not balanced and leads to a net change in angular momentum. As soon as the gradients vanish, for example, if the horseshoe motion takes longer than the local viscous diffusion timescale of the gas, the corotation torque saturates and goes to zero.}

\begin{figure}
        \centering
        \includegraphics[width=\linewidth]{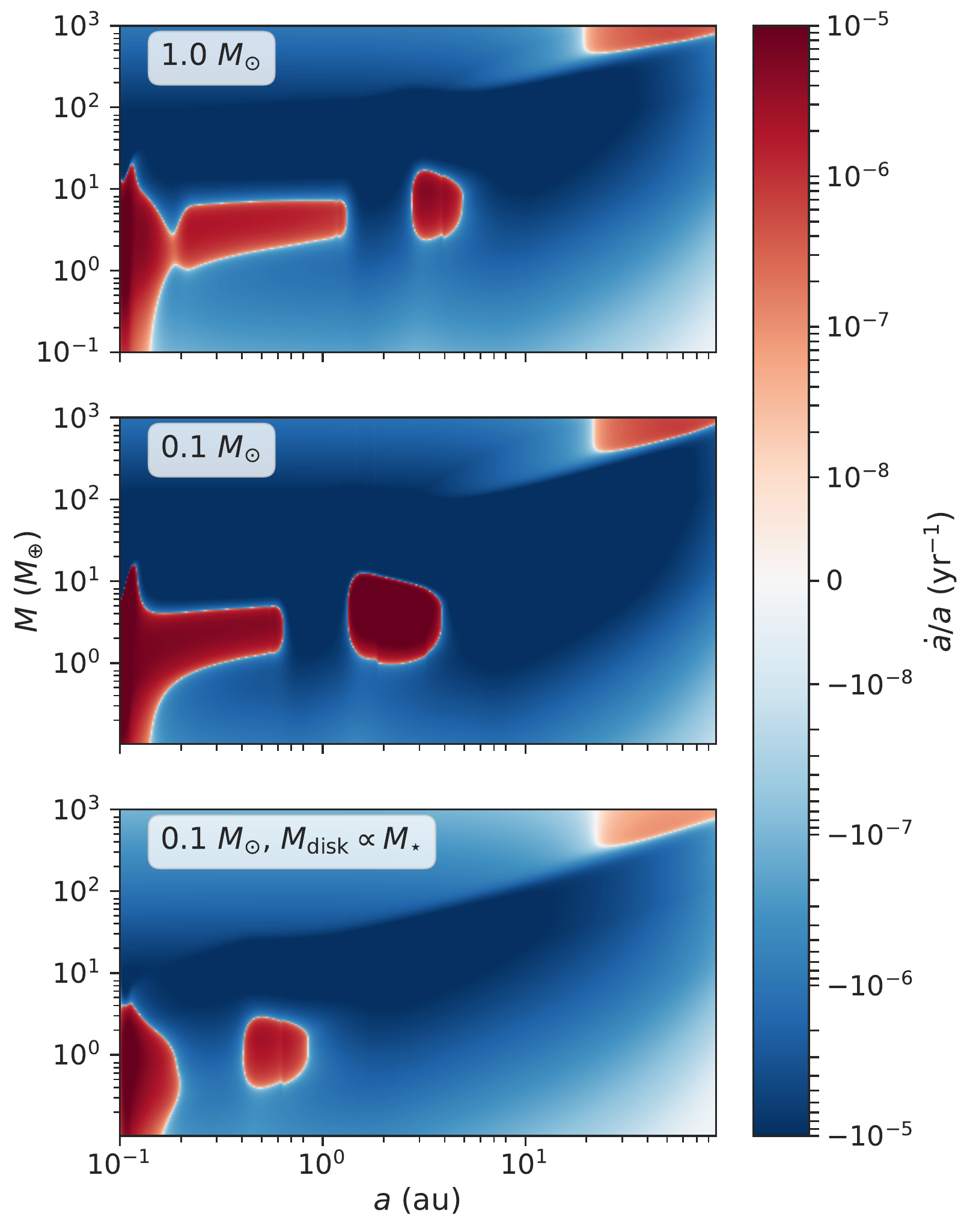}
        \caption{Normalized \frev{type I and type II} migration rate for three different stellar masses and disks. In the top panel, a disk with a total mass of \SI{0.02}{M_{\odot}} around a solar mass star after an evolution of \SI{100}{kyr} is displayed; the central panel shows the same disk at the same time, but the stellar mass is reduced to \SI{0.1}{M_{\odot}}; and the normalized migration rates for a disk with ten times less mass around a \SI{0.1}{M_{\odot}} star can be seen in the bottom panel. Regions with most strongest colors (blue or red) are regions where migration is fastest.
                With scaled disk mass, the outward migration zones (red) move to lower planet masses and closer orbits. This generally causes an earlier inward migration.
        }
        \label{fig:migration_map}
\end{figure}

Figure \ref{fig:migration_map} shows the resulting migration map \frev{of a disk with a power law slope of \SI{0.9}{}, a gas surface density of \SI{200}{\gram\per\centi\meter\squared} at \SI{5.2}{au} and an exponential cut-off radius $R_\mathrm{disk}$ at \SI{30}{au} (see Eq. \ref{eq:initial_surf_dens}). This translates to a total disk mass of \SI{0.02}{M_{\odot}}. The} normalized migration rate $\dot{a}/a\propto \Gamma/\sqrt{M_\star} \propto M_\star^{-1.5}$  is encoded in color. For the second rough proportionality relation, we inserted $\Gamma_0$ (Eq. \ref{eq:gamma_0}) for the total torque $\Gamma$. Thus, the proportionality only holds for the Lindblad torque.

\frev{The red-colored regions corresponding to outward migration in the low-mass regime stem from dominating corotation torques. The regions are limited in vertical direction by viscous saturation. The particular pattern in horizontal direction emerges from the temperature structure of the disk. The inner outward migration zone stops due to the increase in the dust opacity -- leading to additional heating  -- toward the iceline at \SIrange{200}{160}{\kelvin} \citep{Bell1994}. The outer zone ends at the location where the temperature profile becomes dominated by irradiation instead of viscous heating \citep[see also][]{Dittkrist2014}.}

The impact of the stellar mass on migration is apparent in Fig. \ref{fig:migration_map}: A decrease of the stellar mass leads to an increase in the migration rate for the same disk (top and central panel). We can observe that the outward migration region is shifted toward the star by about \SI{1}{au}, which we attribute to the cooler temperature structure due to decreased viscous heating ($\dot{E}_\mathrm{visc}\propto\Sigma_\mathrm{g}\nu\Omega_K^2\propto M_\star^{3/2}$) at fixed semi-major axis and $\Sigma_\mathrm{g}$ for lower stellar mass.

In this work, we assume that the disk mass scales linearly with the stellar mass (see Sect. \ref{ssec:scaling_disk_mass}). Therefore, we include the case of a disk with a mass that is reduced to \SI{10}{\percent} compared to the other two depicted disks in the bottom panel of Fig. \ref{fig:migration_map}. According to the scaling with $\Gamma_0$, we expect migration rates that are more similar to the top panel than the middle panel ($\dot{a}/a \propto \Sigma_\mathrm{g}/M_\star^{1.5}$), which roughly holds in the type I regime wherever corotation torques are weak. The outward migration regions are further shifted toward the star compared to the case of more massive disks, which impacts the resulting population of planets by a large degree since individual planets tend to pile up at the outer edge of outward migration zones, which only change on typical timescales of the disk evolution ($\sim$\SI{}{Myr}). Furthermore, the mass at which the corotation torque saturates, which marks the upper limit of the outward migration zones, is lower for our lower-mass disks. This moves the typical mass of fast-migrating outer planets (the "horizontal branch", \citealp{Mordasini2009}) from typically $\sim$\SI{10}{M_\oplus} to $\sim$\SI{3}{M_\oplus}.

For type II migration, we follow \citet{Dittkrist2014}. Thus, the overall torque is proportional to $\Omega_\mathrm{K} \nu$. The alpha-viscosity $\nu=\alpha c_s^2/\Omega_\mathrm{K}$ is sensitive to the temperature structure of the disk via the isothermal sound speed $c_s$. The temperature differs for varied stellar masses due to the change of direct illumination by the star -- caused by the different stellar radius and temperature -- as well as due to a lower viscous dissipation rate $\dot{E}_\mathrm{visc}$. \frev{To transition from type I to type II, we use the gap opening criterion of \citet{Crida2006}. Planets with masses below this critical mass are not able to carve a deep gap in the gaseous disk and are therefore moving according to type I migration.}
\frev{In Fig. \ref{fig:migration_map}, the transition can be seen at masses of \SIrange{e2}{e3}{M_\oplus} for the top panel. Moving from lower to higher planetary masses, the color code gets a visibly lighter tone corresponding to a significant migration rate drop. In the outermost disk, gas moves radially outwards which leads to an outward migration in type II (top right corner in Fig. \ref{fig:migration_map}) but not in type I.} We observe a more pronounced transition in the bottom panel of Fig. \ref{fig:migration_map}, which can be explained by the much cooler disk and the different scaling of the two regimes. \frev{It} thus becomes more relevant in the low-mass star disks. . Furthermore, the transition location is moved to lower masses $\sim$\SI{30}{M_\oplus} for the bottom panel.

Overall, disk migration can take place more quickly around low-mass stars if the disk mass is the same as it is for solar-type stars. However, for typical disk masses, we expect slightly slower migration in type I and type II regimes, with outward migration occurring for smaller planets, and the relevant zones lying closer to the star.

\subsection{Transit radii}
\label{ssec:transit_radii}
To better compare observed radii measured by transit surveys with radii of synthetic planets, we calculate the "transit radii" of the modeled planets following \citet{Guillot2010}. For this purpose, we can use the internal and atmospheric structure of each planet. Simply taking the arbitrary numerical outer boundary of the structure is not appropriate. Instead, an estimate for the radius of the shadow that a planet casts when passing in front of its host star needs to be employed.
 
\citet{Hansen2008} found that the optical depth along a chord is enlarged by a factor of $\gamma \sqrt{2\pi R/H_0}$ compared to the optical depth integrated radially outwards from a radius $R$. Here, $\gamma$ is a factor relating the opacity in the visual to the opacity in the thermal wavelength range \citep[][Table 2]{Jin2014} and $H_0$ is the local scale height in the envelope at a given radial location.

We chose an optical depth $\tau = 2/3$ in our gray atmosphere as the outer Eddington boundary condition and we note that for a percent-level comparison to a particular observation, a non-gray atmosphere and the instrument specific wavelength band would have to be included. For planets without a H/He envelope, the transit radius is equal to their composition-dependent core radius calculated using a three layer model (considering iron, silicates, and whether a planet incorporates \flrev{ice} from outside of the iceline(s)) \citep{Mordasini2012c}.

\subsection{Initial conditions and Monte Carlo variables}
\label{sec:scaling}
Owing to the statistical approach of population synthesis, we treat the initial conditions as random variables that we draw from probability distributions for each simulation. The randomized variables, called Monte Carlo variables, are the gas disk mass $M_\mathrm{gas}$ (sometimes expressed as initial surface density $\Sigma_\mathrm{g}$ at \SI{5.2}{au}), the metallicity [Fe/H], the inner edge $r_\mathrm{in}$ of the disk, a parameter scaling the strength of photo-evaporation $\dot{M}_\mathrm{wind}$, and the starting locations of the planets $a_\mathrm{start}$. For consistency within the paper series, we employ the values from \papertwo{} for the case of solar-type stars and scale them with stellar mass. 
\begin{figure*}
    \centering
    \includegraphics[width=\linewidth]{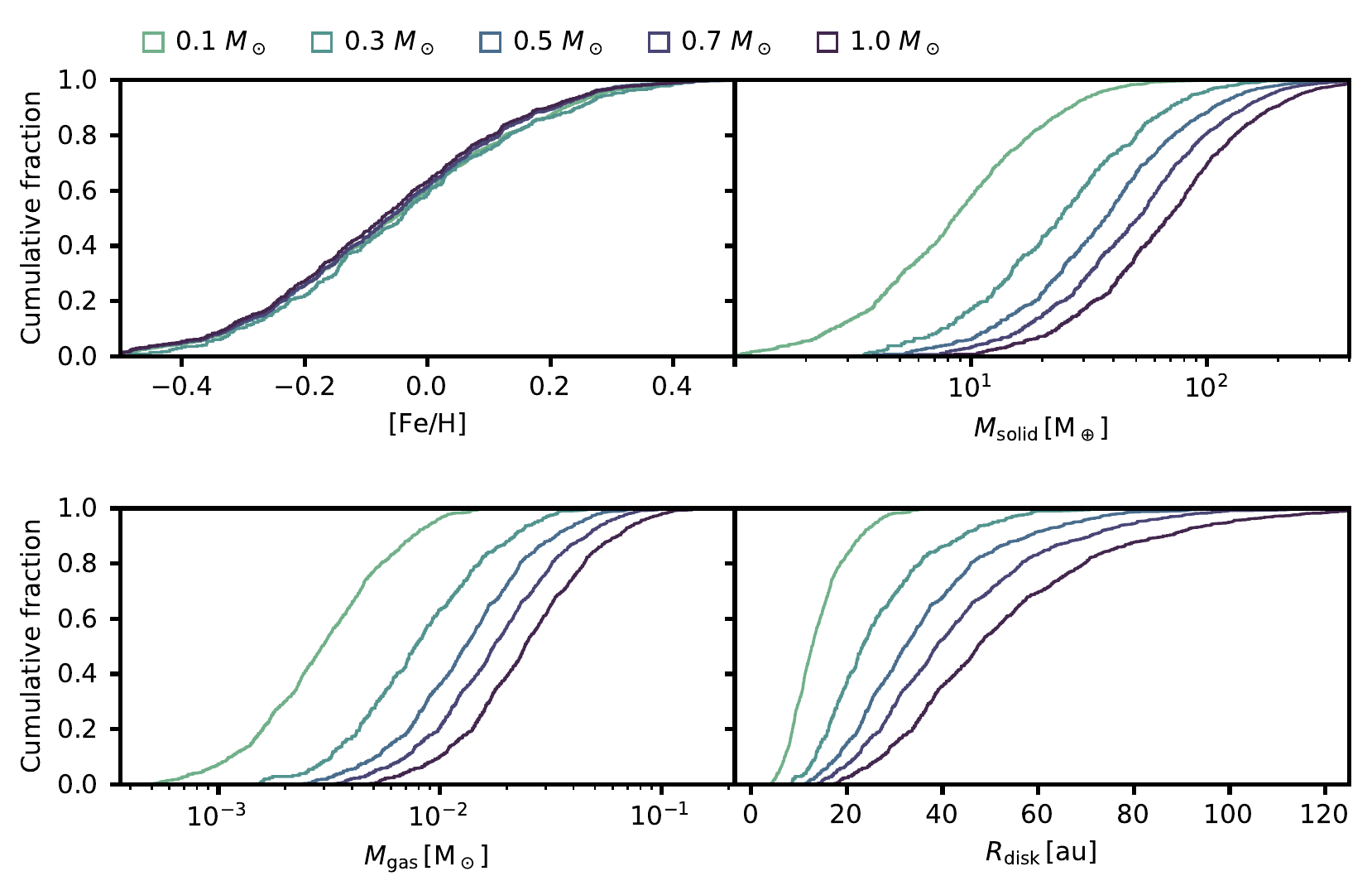}
    \caption{Cumulative distributions of initial conditions and resulting quantities for different $M_\star$. The Monte Carlo parameters [Fe/H] and $M_\mathrm{gas}$ can be converted to initial solid disk masses ($M_\mathrm{solid} = 10^{[\mathrm{Fe}/\mathrm{H}]} f_{dg,\odot} M_\mathrm{gas}$, where $f_{dg,\odot} = 0.0149$ following \citealp{Lodders2003}) and disk truncation radii \citep[$R_\mathrm{disk}\propto M_\mathrm{gas}^{-1.6}$, following][]{Andrews2010}. We chose to distribute [Fe/H] equally for all host star masses.}
    \label{fig:initialConditions}
\end{figure*}

Figure \ref{fig:initialConditions} shows for each host star mass bin, the probability distributions of the randomized variables used in this study as well as closely related, resulting quantities. While we assume that the distribution of the metallicity \citep{Santos2003} remains the same, all other parameters scale with stellar mass. We note that M dwarfs are generally older than solar-type stars (in terms of the mean), therefore the metallicity could change with the evolution of stellar clusters. Nevertheless, we assume that this effect is small compared to other unknown scalings of the protoplanetary disks discussed below.

Apart from randomized variables, we also fix a set of parameters, which are summarized in Table \ref{tab:modelParams}. Importantly, we follow \papertwo{} in setting the radius of planetesimals to \SI{300}{\meter} and initializing the gas surface density profile as \citep{Veras2004}:
\begin{equation}
\label{eq:initial_surf_dens}
\Sigma_\mathrm{g}(t=0) = \Sigma_{\mathrm{g},0} \left(\frac{r}{r_{0}}\right)^{-\beta_\mathrm{g}}\exp{\left(-\left(\frac{r}{R_\mathrm{disk}}\right)^{2-\beta_\mathrm{g}}\right)}\left(1-\sqrt{\frac{r_\mathrm{in}}{r}}\right)
,\end{equation}
where $r_0=\SI{5.2}{au}$ is the reference distance, $\beta_\mathrm{g}=0.9$ the power-law index \citep{Andrews2010}, $R_\mathrm{disk}$ the cutoff radius for the exponential decay and $r_\mathrm{in}$ is the inner edge of the disc.

Before discussing the individual choice of distributions for the observation-informed parameters, we note that the solar-type initial conditions of \papertwo{} include a steeper slope for the planetesimal disk $\beta_\mathrm{pls}$ than the gas disk ($\beta_\mathrm{pls}=-1.5$ instead of -0.9).
This is motivated by results from planetesimal formation models \citep{Drazkowska2016,Drazkowska2017,Schoonenberg2017,Lenz2019}. We assume that this steepening is valid for all stellar masses and adopt a $\beta_\mathrm{pls}=-1.5$ for all simulations (except in the scenario described in \frev{Sect. \ref{ssec:giants_extreme}}). 

\newcommand\T{\rule{0pt}{2.6ex}}       
\newcommand\B{\rule[-1.2ex]{0pt}{0pt}} 
\begin{table}
        \caption{Model Parameters}
        \label{tab:modelParams}
        \centering
        \begin{tabular}{r c c}
                \hline\noalign{\smallskip} 
                Parameter & Symbol &  Value\\
                \hline\hline\noalign{\smallskip}  
                Disk Viscosity & $\alpha$ & $\SI{2e-3}{}$ \\
                Power Law Index (Gas) & $\beta_\mathrm{g}$ & $0.9$$^{(a)}$\\
                Outer edge of planetesimal disk & & $R_\mathrm{disk}/2$$^{(b)}$\\
                Power Law Index (Solids) & $\beta_\mathrm{pls}$ & $1.5$$^{(c)}$\\
                
                Radius of Planetesimals & &\SI{300}{\meter} \\
                Number of Planet Seeds (Embryos) & & $50$ \\
                Mass of Embryos & & \SI{0.01}{M_\oplus} \\
                Embryo placement time & & \SI{0}{yr} \\
                Envelope opacity reduction & & 0.003$^{(d)}$ \\
                N-body integration time & & \SI{20}{Myr} \\
                \hline
                \multicolumn{3}{l}{ $^{(a)}$ \citet{Andrews2010}; } \\
                \multicolumn{3}{l}{ $^{(b)}$ \citet{Ansdell2018}; } \\
                \multicolumn{3}{l}{ $^{(c)}$ \citet{Drazkowska2017,Lenz2019};}\\
                \multicolumn{3}{l}{ $^{(d)}$ \citet{Mordasini2014} } \B\\
                \hline
        \end{tabular}
  \end{table}

\subsubsection{Disk mass}
\label{ssec:scaling_disk_mass}
The initial gas disk mass for solar type stars is based on the Class I disk observations \citet{Tychoniec2018} (see \mbox{\papertwo{}}). \mbox{\citet{Tychoniec2018}} do not split their sample of Class I disks into different stellar masses, but they do mention that they observe a weak correlation with the bolometric luminosity which can be seen as a proxy for the stellar mass. However, the non-triviality of the mass-luminosity relation for very young protostars makes inferring the slope of this weak correlation a task outside the scope of this work. \frev{While sensitive to dust and not  gas\textit{}}, the scaling of disk mass with stellar mass can be inferred from recent ALMA measurements. For more evolved disks, \mbox{\citet{Pascucci2016}}, \citet{Barenfeld2016}, as well as \citet{Ansdell2017}, found a dust mass dependency on stellar mass, which is steeper than linear. \frev{These measurements cover stellar masses down to \SI{0.1}{M_{\odot}}. Therefore, they probe deep into the M dwarf class. Within the scatter of the observations, the linear slope fits well the different spectral types without any visible transition.} \citet{Testi2016} and \citet{Sanchis2020} extended the observations to the brown dwarf regime and found statistically consistent results.

While \citet{Barenfeld2016} had not yet found a clear difference in slopes of the dust mass to stellar mass relation when comparing their results for Upper Sco with disk masses for the younger Lupus cluster, \citet{Ansdell2017} and \citet{Pascucci2016} reported a time dependency of the slope by enlarging the sample to more stellar clusters. These latter findings point toward a stellar mass dependent time evolution of the dust and, therefore, do not well constrain the initial dust mass. Tentatively interpolating back this steepening of the disk mass to stellar mass relation to time zero, we adopted a linear dependency of the disk mass on the stellar mass $M_\text{gas} \propto M_\star$ consistent with \citep{Andrews2013}. In addition, this is in line with previous theoretical works \citep[slopes of 0.5 to 2 in][]{Raymond2007}.

\begin{figure}
        \centering
        \includegraphics[width=\linewidth]{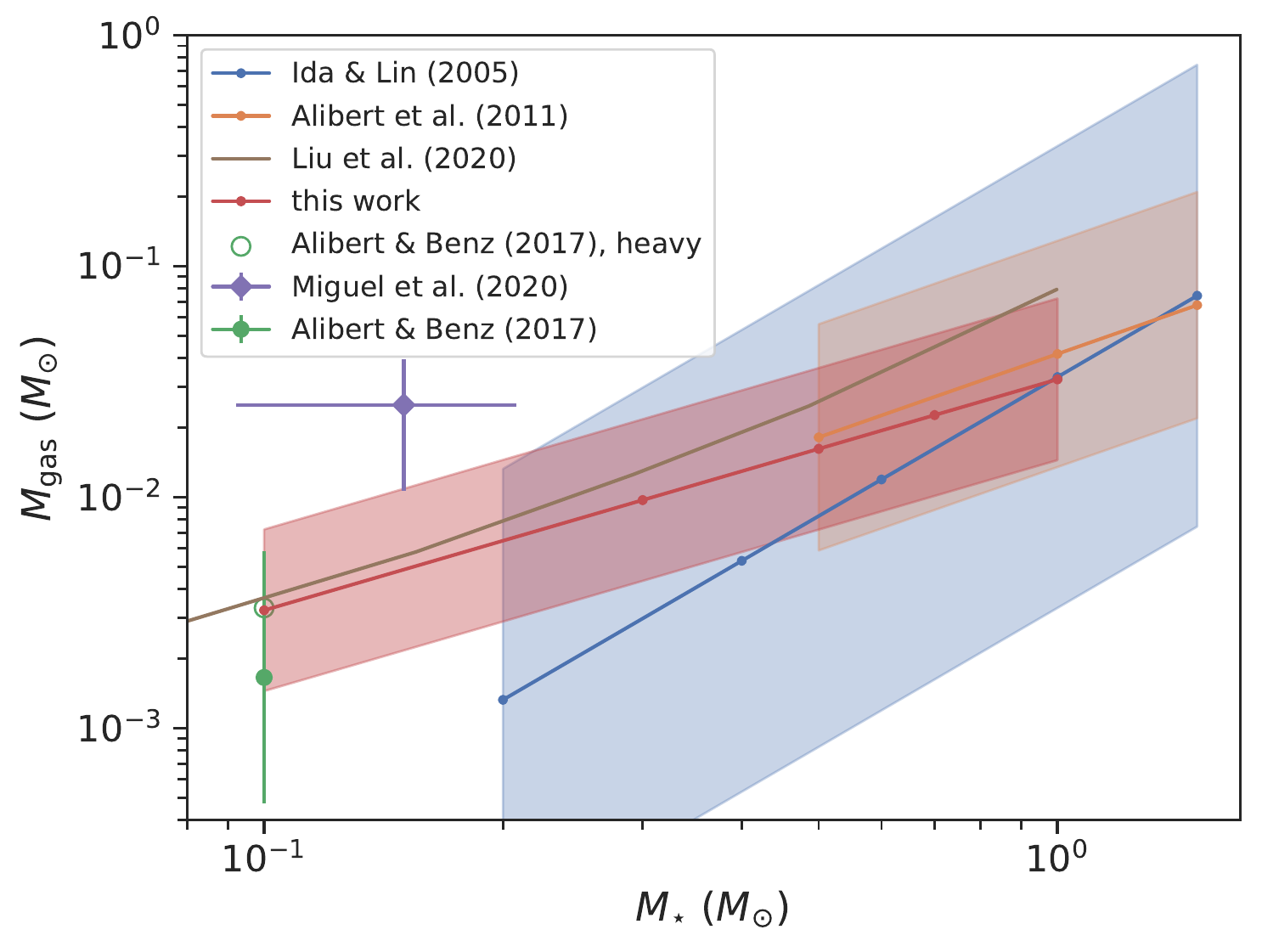}
        \caption{\frev{Disk masses used in different theoretical planetary population synthesis works. Shaded regions or error bars show the 1$\sigma$ scatter in the distributions. Log-uniform distributions are used to sample disk masses in all works except for \citet{Miguel2020} who draw from a linear-uniform distribution (\SIrange{1e-4}{5e-2}{M_{\odot}}). They are also the only work which samples over stellar masses instead of using fixed values. \citet{Liu2020a} prescribe stellar accretion rates as opposed to disk masses; the data shown here corresponds to integrating from time zero to \SI{10}{Myr}. The "heavy disk" case of \citet[][open-circle, scatter-omitted]{Alibert2017} draws disks with twice their nominal mass. Apart from this case, we show the distributions labeled as nominal by the respective authors. The disk masses used in this work are in agreement to previous assumptions.}}
        \label{fig:disk_mass_literature_comparison}
\end{figure}

\frev{Compared to other works, these choices lead to a similar mean disk mass for the \SI{0.1}{M_{\odot}} stars of \SI{0.0032}{M_{\odot}} as in the "heavy disk" case of \citet{Alibert2017}. They use a stellar accretion rate informed scaling with a power of -1.2, but lower values for the solar mass stars. In Fig. \ref{fig:disk_mass_literature_comparison}, we compare the disk gas masses drawn for this work with those used in previous planetesimal-based population synthesis works of \citet{Ida2005},\citet{Alibert2011},\citet{Alibert2017}, and \citet{Miguel2020}. There is also a considerable agreement with the initial conditions used by \citet{Liu2020a} who prescribe an accretion rate following the Orion data set by \citet{Manara2012}. To get an estimate of the disk mass, they integrated the stellar accretion rate over time assuming no photoevaporation or accretion onto the planets.}

As in \papertwo{}, we then multiply the gas disk mass with spectrally measured stellar metallicities \citep{Santos2003} to get the initial dust disk mass, which is available for the formation of planetesimals. We assume no dependency of the metallicity on the stellar mass. Additionally, an efficiency of transforming the dust to planetesimals of \SI{100}{\percent} was chosen.
The planetesimal surface density is reduced at radii closer to the star if for a given element no chemical species containing it can condense out at the local disk temperature \citep[see][]{Thiabaud2014}. This leads to sharp transitions in the surface density profile of solids, most prominently the water iceline, whose location depends on the temperature at time zero of the simulation.

\subsubsection{Disk radius}
Although it is challenging to measure the initial radius of the gaseous disk, trends about the disk size of -- notably evolved -- disks with disk mass were found by \citet{Andrews2010}. The scaling relation of disk mass to disk radius follows $M_\mathrm{gas}\propto R_\mathrm{disk}^{1.6}$, which was recovered later \citep{Andrews2018}. We use this constraint to calculate the gas disk size from the randomized mass without introducing additional scatter.

For the dust disk size, a direct correlation of dust disk radii with stellar mass using ALMA data could not be found by \citet{Ansdell2018} who advocate more high-resolution CO line observations to give clearer constraints. However, it became clear thanks to the ALMA measurements that dust radii are smaller than gas radii by about a factor of 0.5 \citep{Ansdell2018}, which we adopt to truncate the planetesimal disk. \frev{The recent study of \citet{Sanchis2021} expand and confirm these results and further find no dependence of the CO to dust size ratio on the stellar mass.}

\frev{With this approach, the exponential cut-off radii of gaseous disks around stars with lower stellar masses follow the relation $R_\mathrm{disk}\propto M_\star^{0.625}$. The numerical values drop from a mean of \SI{62.1}{au} for \SI{1.0}{M_{\odot}} to \SI{14.8}{au} for \SI{0.1}{M_{\odot}}.}

\subsubsection{Inner edge}
\label{ssec:scaling_inner_edge}
The numerical inner edge $r_\mathrm{in}$ is a free parameter of our model and a key factor in the final orbital positions of many of the innermost planets. This is especially important when comparing synthetic results to transit and radial velocity surveys which are most sensitive to the innermost region.
In this section, we analyze where $r_\mathrm{in}$ should lie as a function of stellar mass.

The physical motivation for an inner edge is a magnetospheric cavity \citep{Bouvier2007}, where ionized material of the disk is lifted by the magnetic field lines from the midplane and accreted onto the star. This typically happens at the corotation radius \citep[e.g.,][]{Gunther2013}, where the magnetic field rotates at the same speed as the gas. It is therefore reasonable to not extend the modeled disk closer to the star than its corotation radius. The order of percent sub-Keplerian speed of the gas disk is negligible for this consideration and thus we take $r_\text{in}$ at the location where the Keplerian orbital period is equal to a stellar rotation period drawn from measured distributions. 
We do not take into account the vaporization of silicates and ionization of the disk gas in the innermost regions of the disk.

The rotation periods of young stars can be derived from periodic variations of objects in young stellar clusters, such as the Orion Nebula Cluster \citep{Herbst2002}, NGC 6530 \citep{Henderson2011}, NGC 2264 (\citealp{Lamm2005};\citealp{Affer2013};\citealp{Venuti2017}), NGC 2362 \citep{Irwin2008}, and NGC 2547 \citep{Irwin2007}. \citet{Irwin2008}, as well as \citet{Henderson2011}, discuss an increasingly steep slope in the rotation period versus stellar mass diagrams with increasing age. However, for the youngest clusters (Orion and NGC 6530) no decrease of the rotation period with decreasing stellar mass was found \citep{Henderson2011}. Therefore, this feature can be attributed to a faster spin-up of low mass stars and is not considered an initial condition for planet formation. As initial condition, we therefore chose the same rotation periods for all stellar masses. Thus, the orbital distance of the inner edge of the disk scales at $\propto M_\star^{1/3}$.

\begin{figure}
    \centering
    \includegraphics[width=\linewidth]{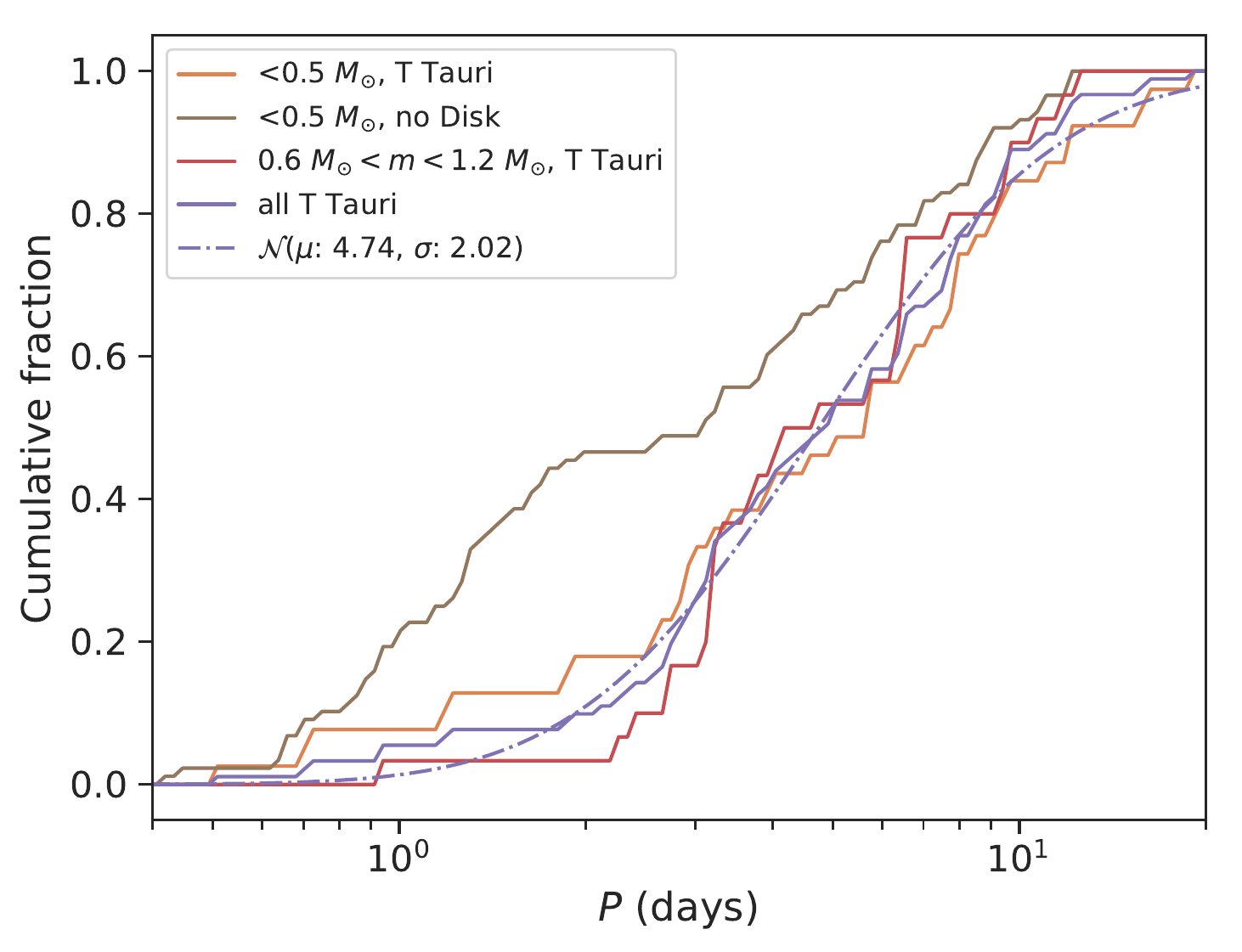}
    \caption{Cumulative distribution of rotation periods of disk-bearing and diskless stars in NGC 2264 (estimated age $\sim$\SI{3}{Myr}) found by \citet{Venuti2017}. An error function corresponding to the normal distribution with the indicated mean and standard deviation was fitted to the logarithm of the rotation periods of all T Tauri stars (dash-dotted line). This distribution defines the corotation radius, which we set as the inner disk edge.}
    \label{fig:rot_periods_venuti}
\end{figure}

Despite the long history of observations in the field, the exact distribution of classical T Tauri rotation periods is still subject to a lot of statistical uncertainty. \citet{Venuti2017} used data from 38 days of CoRoT observations to constrain the rotation periods in NGC 2264, which has an estimated age of $\sim$\SI{3}{Myr}. As is the case with other authors \citep[e.g.][]{Henderson2011}, they found that stars that still show signs of accretion (i.e., classical T Tauri stars) have slower rotation periods than diskless stars. In Fig. \ref{fig:rot_periods_venuti}, we show the data of \citet{Venuti2017} in two mass bins for diskless and disk-bearing stars. No clear difference was found between the different masses in the case of disk-bearing stars. Therefore, we adopted a distribution of rotation periods used to constrain the inner edges following the full T Tauri sample of \citet{Venuti2017}, that is, a log-normal distribution with a mean of \SI{4.74}{days} and a spread $\sigma$ of \SI{0.3}{dex} (\SI{2.02}{days}).

To summarize, in period space, the same distribution as of the inner disk edge is used for all stellar masses in terms of period, implying a dependency in terms of orbital distance as $M_\star^{1/3}$. This also means that for \SI{1}{M_{\odot}}, the distribution used in \papertwo{} is recovered.

\subsubsection{Initial embryos}
\label{ssec:initial_embryo_placement}
The initial seeds of planetary growth, called embryos, are placed randomly into the protoplanetary disk at $t=0$ starting from the inner edge of the disk, $r_\text{in}$, out to an upper limit. In \papertwo{}, the upper limit is chosen to be \SI{40}{au}, which we adopt for the solar-mass populations. This outer edge is then multiplied with $\left(M_\star/M_\odot\right)^{1/3}$. Therefore, it is kept at fixed orbital period. Many timescales relevant to planet formation, such as the growth timescale of the core by planetesimals, scale with the orbital period. This motivates the move to keep it the same for better comparability amongst the populations.

Based on simulations of runaway planetesimal accretion and beginning of oligarchic growth \citep{Kokubo1998,Chambers2006}, the locations of the initial embryos are drawn from a log-uniform distribution between these two boundaries. If a planet would be placed within 10 Hill radii of an already placed embryo, a new location is drawn.

We chose to place 50 embryos into each protoplanetary disk at time zero of our simulations. This is a compromise because placing more embryos leads to longer simulation times. The influence of the number of initial embryos is studied in detail in \papertwo{}. For the simulations shown in \frev{Sect. \ref{ssec:giants_extreme}}, we do single-embryo calculations.

As in \papertwo{}, the initial mass of the embryo is set to be $\SI{e-2}{M_\oplus}$. This mass is not scaled with the stellar mass, which is a choice that changes the initial mutual Hill spacing with varying stellar mass and thus the gravitational interactions between the embryos (see the discussion in Sect. \ref{ssec:dependence_on_placement}).

\subsection{Disk observables}
\label{sec:disk_properties}
\subsubsection{Disk lifetime}\label{ssec:disk_lifetime}
Although the disk lifetime is not a direct Monte Carlo variable, \frev{it depends strongly on the photo-evaporation variable $\dot{M}_\mathrm{wind}$. We follow \citet{Matsuyama2003} and remove mass with an equal rate per area due to FUV radiation from different stars (external) as:
\begin{equation}
\dot{\Sigma}_\mathrm{wind} = \frac{\dot{M}_\mathrm{wind}}{\pi\left( (\SI{1000}{au})^2-\beta^2_\mathrm{I}r^2_\mathrm{g,I}\right)}\,.
\end{equation}
This is applied outside of a modified gravitational radius $r_{\mathrm{g,I}}=0.14 \frac{G M_\star}{c_{s,I}^2}$ \citep[see][for a discussion of the prefactor]{Alexander2014}, where $c_{s,I}$ is the sound speed of dissociated hydrogen at a temperature of \SI{1e3}{\kelvin}. The amount of external photo-evaporation would depend on the proximity to more massive stars. Here, we consider this as a free parameter \citep[see e.g.,][for a more modern approach to constrain it]{Haworth2018}.} 

\begin{figure}
        \centering
        \includegraphics[width=\linewidth]{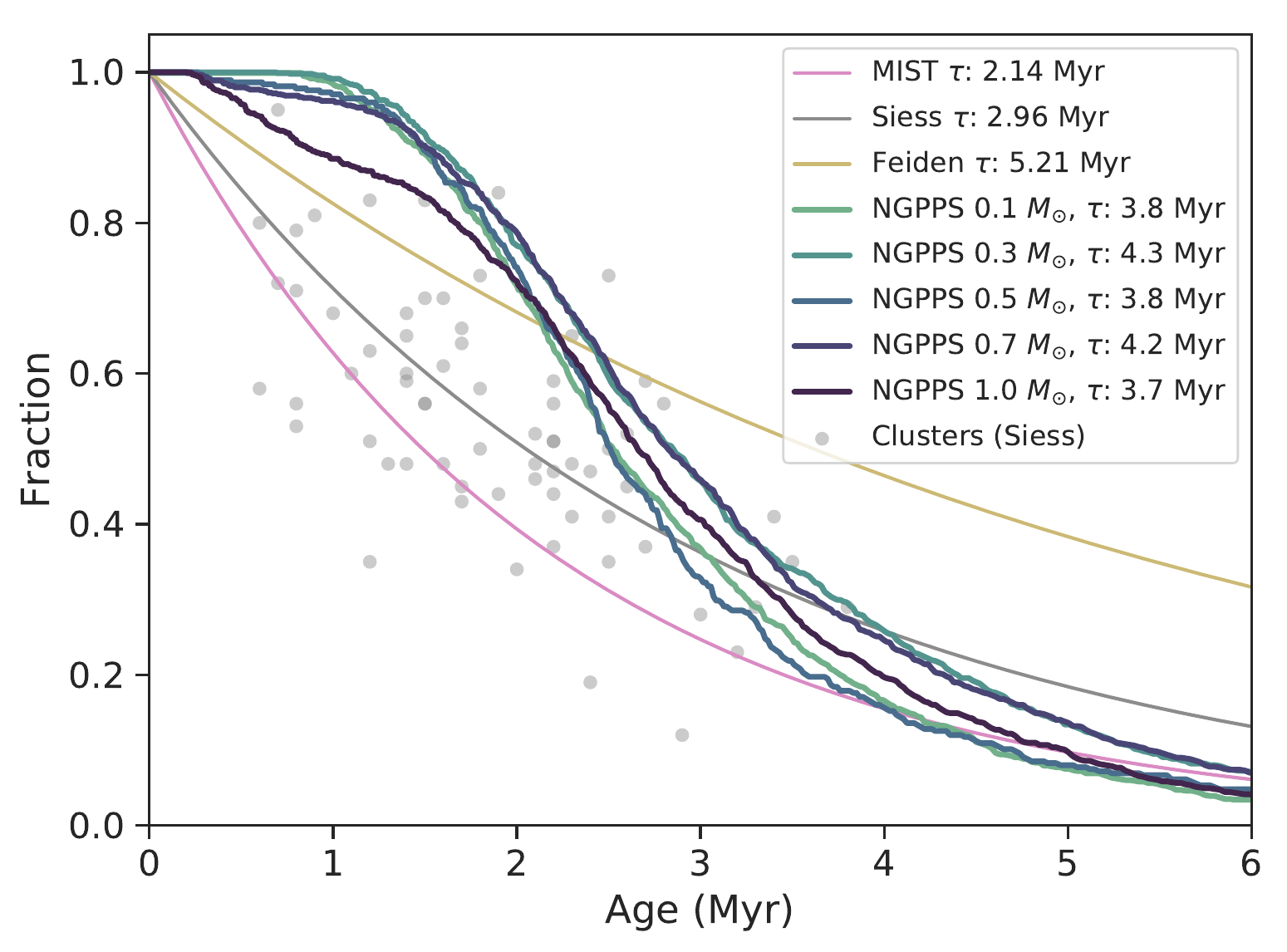}
        \caption{Fraction of disk-bearing stars as a function of time. Both observational data assembled by \citet{Richert2018}, as well as the synthetic lifetimes for different stellar masses, are shown. The age determination of clusters is sensitive to the employed pre-main-sequence model. The observational data and an exponential fit to it is shown using the age scale from \citet{Siess2000} as well as fits to the same cluster data but using the dating of \citet{Feiden2016} and the MIST collaboration \citep{Choi2016}.}
        \label{fig:lifetimes}
\end{figure}

\frev{We choose $\dot{M}_\text{wind}$ such that} the distribution of the lifetimes of the synthetic disks is in agreement with observations, as can be seen in Fig. \ref{fig:lifetimes}. For that, we keep the unknown viscous $\alpha$ parameter at $\alpha = \SI{2e-3}{}$ for all stellar masses. By construction, the lifetimes of the different stellar mass bins are similar. Disk lifetimes obtained from observed fractions of disk-bearing stars in young stellar clusters \citep{Strom1989,Haisch2001,Mamajek2009,Fedele2010,Ribas2014,Richert2018} are sensitive to the pre-main-sequence evolution model of the stars used to determine the cluster age \citep{Richert2018}. Therefore, a larger uncertainty than the empirical scatter results.

In Fig. \ref{fig:lifetimes}, we show the results of \citet{Richert2018}, who used three different pre-main-sequence evolution models to get typical disk lifetimes varying by more than a factor of two. For comparison of the modeled disk lifetimes to measurements, it is necessary to estimate the time during which a simulated disk would be detected by typical survey used to get the observational data. We follow \citet{Kimura2016} to consider a disk as dispersed at the moment the disk becomes transparent (optical depth smaller than unity, \frev{$\tau = \kappa\Sigma_g/2  < 1$, where $\kappa$ is the \citet{Bell1994} opacity}) everywhere in the region \frev{where $T_{\mathrm{mid}}>\SI{300}{\kelvin}$}. This is a broad estimate for near-infrared observations, which are the basis of disk lifetimes studies. For low-mass stars, this observable disk
lifetime substantially differs from the different criterion used in \paperone{} and \papertwo{}, where the disk is considered to be dispersed at the moment the total mass $M_\mathrm{gas} < \SI{1e-6}{M_{\odot}}$ or when the surface density is $\Sigma_\mathrm{g} <\SI{1e-3}{\gram\per\centi\meter\squared}$ everywhere in the disk.

Given the large uncertainty in the age determination of the clusters, our modeled disks reproduce the observations to a satisfying degree. However, there is a lack of short-lived disks for some clusters. Nevertheless, the scatter is very large at early times, indicating that the environments where the stars are born play a large role and should be investigated in the future.

Regarding the dependency with stellar mass, \citet{Richert2018} do not find a significant difference when going from solar mass stars to lower stellar masses. Therefore, we assume that the lifetime of disks does not depend strongly on the stellar mass. We point out that there is a clear dependency of the lifetime of a disk on stellar mass when looking at $M_\star > M_\odot$ \citep{Haisch2001,Ribas2015}, but we are not aware of any reason to extrapolate this dependency to lower stellar masses given the lack of observational evidence.

\subsubsection{Stellar accretion rates}\label{ssec:accretion_rate}

Similarly to the observed lifetimes, stellar accretion rates have to be matched by the simulated disks \citep{Manara2019}. The process driving accretion rates in the simulations is the viscous evolution. Numerically, we take the disk gas mass that flows into the innermost numerical cell as gas accretion onto the star $\dot{M}_{\rm acc}$. In general, this flux is not radially constant in the disk due to it not being in equilibrium at all times \citep[Eq. 25 in][]{Mordasini2012c}.
\begin{figure}[htbp]
        \centering
        \includegraphics[width=\linewidth]{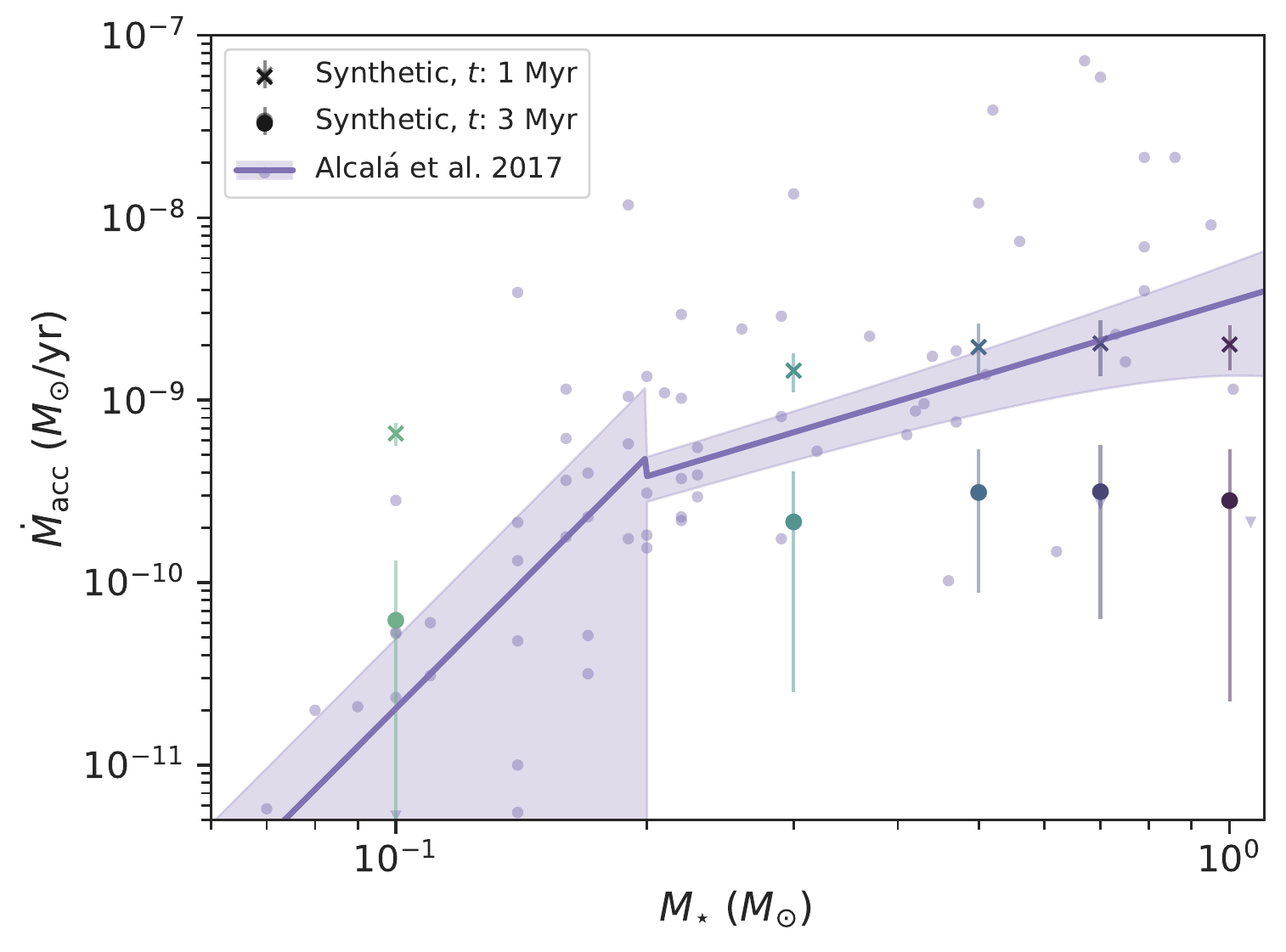}
        \caption{Stellar accretion rates of observations and synthetic disk populations as a function of stellar mass. For the synthetic data points at two different times (colored crosses and large dots), the mean of the distribution is plotted and the standard deviation is indicated with error bars. The observational data (small points and triangles for upper limts) and the broken power-law fit with its estimated errors is taken from \citet{Alcala2017} for the Lupus cluster (estimated age of \SI{1}{} to \SI{3}{Myr}). The break for this particular fit was chosen to lie at \SI{0.2}{M_{\odot}} \citep[motivated by][]{Vorobyov2009}. }
        \label{fig:macc}
\end{figure}

In Fig. \ref{fig:macc}, we show the resulting $\dot{M}_{\rm acc}$ of our synthetic populations for the various stellar masses at two different times. This can be compared to the Lupus data obtained by \citet{Alcala2017}. For planet formation, the most important stages are early in the disk evolution when most of the mass is still present. Therefore, a comparison to clusters older than Lupus (\SI{1}{} to \SI{3}{Myr}) would not be as relevant.

We find mass accretion rates on the same orders of magnitude as were observed in Lupus. The intrinsic scatter of the synthetic populations is lower than in the observed sample. This is a known discrepancy between simple models using a single viscous $\alpha$ value (\SI{2e-3}{} in our case) and observations  (see for example \citealp{Rafikov2017} or \citealp{Manara2020} and references therein). Furthermore, the evolution with time seems to be rather rapid compared to observations, given that at \SI{3}{Myr,} a lot of the more massive stars have already accreted most of the disk mass. 
We note that we consider here our numerical simulation time: realistic cluster ages would include the early star formation stages, which can lead to a shift of a few \SI{100}{kyr}.
The scaling of $\dot{M}_{\rm acc}$ with stellar mass in the synthetic work is more shallow than the fitted observational data.
An in-depth comparison of disk properties to disks resulting from population synthesis work that extends the approach of \citet{Manara2019} will be addressed in a future paper.

%
\section{Results}
\label{sec:results}
We now turn our attention from disk initial conditions and parameters for the given stellar masses to the resulting synthetic planetary population. To help identify the different simulations, we list in Table \ref{tab:sims} the identifiers and stellar masses of the different simulations discussed here. 
We present our results as a function of the stellar mass focusing on the type of planets that form (Sect. \ref{sec:types_planets}), the mass -- semi-major axis plane (Sect. \ref{sec:m-a}), planetary mass functions (Sect. \ref{sec:planetary_mass_distribution}), and planet radii and compositions (Sect. \ref{sec:composition}).

\begin{table}[htbp]
        \caption{Overview of the simulation names, stellar masses and effective temperatures, number of initial embryos $N_\mathrm{emb,ini}$, and number of simulation runs for all simulations discussed in this work}
        \label{tab:sims}
        \centering
        \begin{tabular}{l c c c c}
                \hline\noalign{\smallskip} 
                Name & $M_\star$ & $T_{\mathrm{eff,}\star\mathrm{,5 Gyr}}^{(a)}$  & $N_\mathrm{emb,ini}$ & Simulations\\
                \hline\hline\noalign{\smallskip}  
                NGM10 & \SI{0.1}{M_{\odot}} & \SI{2811}{\kelvin} & 50 & 1000\\
                NGM14 & \SI{0.3}{M_{\odot}} & \SI{3416}{\kelvin} & 50 & 997\\
                NGM11 & \SI{0.5}{M_{\odot}} & \SI{3682}{\kelvin} & 50 & 1000\\
                NGM12 & \SI{0.7}{M_{\odot}} & \SI{4430}{\kelvin} & 50 & 999\\
                NG75$^{(b)}$  & \SI{1.0}{M_{\odot}} & \SI{5731}{\kelvin} & 50 & 1000\\
                \hline
                \T
                NG76$^{(b)}$ & \SI{1.0}{M_{\odot}} & \SI{5731}{\kelvin} & 100 & 1000\\
                \hline
                \T
                NGM17$^{(c)}$ & \SI{0.1}{M_{\odot}} & \SI{2811}{\kelvin} & 1 & 10000\\
                NGM21$^{(c)}$ & \SI{0.1}{M_{\odot}} & \SI{2811}{\kelvin} & 1 & 9867\\
                NGM22$^{(c)}$ & \SI{0.1}{M_{\odot}} & \SI{2811}{\kelvin} & 1 & 9939\\
                \hline
\multicolumn{5}{l}{$^{(a)}$ following \citet{Baraffe2015}} \\
                \multicolumn{5}{l}{ $^{(b)}$ also discussed in \papertwo{} } \\
\multicolumn{5}{l}{ $^{(c)}$ with modified parameters (see \frev{Sect. \ref{ssec:giants_extreme}})}

                \B\\
                \hline
        \end{tabular}
\end{table}

\subsection{Mass-distance diagrams}
\label{sec:m-a}
\begin{figure}
        \centering
        \makebox[\linewidth]{
                \begin{tikzpicture}
                \node (img) {\includegraphics[trim={0cm 0.3cm 0cm 0.26cm},clip,width=\linewidth]{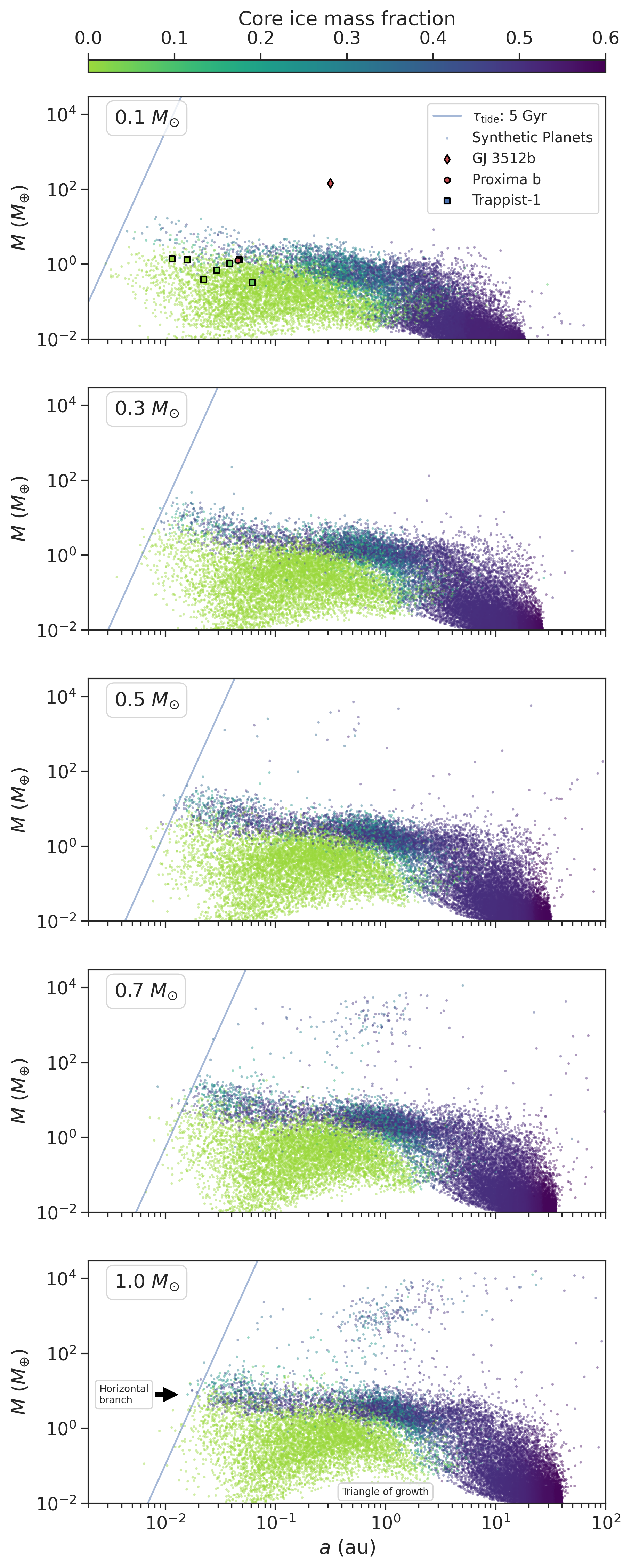}};
                \end{tikzpicture}
        }
        \caption{Synthetic populations of planets as a function of $a$ and $M$ with summed-up mass fraction of all ice species in the planetary core in color. Their NGPPS identifiers are NGM10, NGM14, NGM11, NGM12, and NG75.
                Some observed planets around very low-mass stars are shown \citep{Anglada-Escude2016,Morales2019,Agol2020}.
                Planet masses increase with host star mass, but no giant planets occur for $M_\star$<\SI{0.5}{M_{\odot}}. 
                The radial position where the tidal orbital decay timescale reaches \SI{5}{Gyr} is indicated.
        }
        \label{fig:aM_all}
\end{figure}

The mass-and-semi-major axis diagrams of the populations are shown in Fig. \ref{fig:aM_all}. Each system starts with 50 embryos of \SI{0.01}{M_\oplus} which collide over time, thus the number of points in each of the plots is on the order of \frev{\SI{20000}{}}. The composition measured by the volatile- or ice-mass fraction in the solid core of the planets is color-coded.

General trends for all stellar masses are as follows. First, the ice-rich planets at large semi-major axes. This population is dominated by low-mass planets and the high ice content is an imprint of the lower local temperatures. Second, the effect of migration visibly brings ice-rich planets at Earth to super-Earth masses closer to the stars. Lastly, we identify a distinct population of giant planets that is separated by a runaway gas accretion desert \citep{Ida2004} from the solid-dominated population.

Although these features can be seen for all the different stellar masses, there are clear differences between the populations, which show the influence of the reduced stellar and disk mass. A first feature is that the "horizontal branch" \citep{Mordasini2009}, a population of icy super-Earths that migrated toward the star in the type I regime, is located at different planetary masses. In Fig. \ref{fig:aM_all}, it can be identified by the blue colored points at small semi-major axes and is marked in the bottom panel. The origin of the difference between the stellar masses can be explained as follows: Given a lower disk mass, the corotation torque saturates and planets start migrating at lower planetary masses (see Sect. \ref{ssec:disk_migration}). Thus, the population of close-in, ice-rich planets extends to lower masses for the lower stellar mass populations (down to $\sim$\SI{1}{M_\oplus} for the \SI{0.1}{M_{\odot}} population NGM10 instead of down to only $\sim$\SI{3}{M_\oplus} for the solar mass case, NG75). However, the scatter is quite large and to \flrev{assess} this in a more quantitative way, a larger set of statistics would be needed, especially for the low stellar mass case, where few ice-rich planets migrated to the inner parts of the disk.

A second distinct trend with the stellar mass is a reduction of the number of giant planets with decreasing stellar mass (\citealp{Laughlin2004}, see also Sect. \ref{ssec:types:giant_planets} for a quantitative discussion). Interestingly, the semi-major axis distribution of the giants differs quite a bit when comparing the \SI{0.7}{M_{\odot}} (NGM12) case with the \SI{1.0}{M_{\odot}} (NG75) case. Giant planets are more frequently scattered for the more massive case, since there is more often a second or third giant planet forming, which then leads to more frequent interactions. Therefore, the distribution of giants in NGM12 is more localized at around \SI{1}{au} compared to the solar-mass case. In the lowest stellar-mass population, not a single giant planet was able to form.

Another quite weakly accentuated feature due to little statistics is an under-density due to tidal migration at very close orbits of a few \SI{e-2}{au}, where tides push some planets into the star and leave a void of massive, close-in planets. This can be seen as a fuzzy diagonal cut-off in the mass-and- semi-major axis diagram, which increases to higher planetary masses with increasing semi-major axis \citep{Schlaufman2010,Benitez-Llambay2011}.

Additionally, all populations show a similar "triangle of growth" at very low planetary masses indicated in the bottom panel of Fig. \ref{fig:aM_all}. This absence of planets means that there is a region where embryos grow for all sampled disk conditions. The region spans from \SIrange{0.1}{10}{au}, which are the regions most favorable for planetesimal accretion where growth timescales are short and feeding zones large enough for planetary growth to occur.

Overall, we recover with the use of the mass-distance diagrams, the expected trends of fewer giant planets with decreasing stellar mass and the imprint of stellar mass dependent migration. \frev{The observed masses of the TRAPPIST-1 system \citep{Gillon2017} as well as Proxima b \citep{Anglada-Escude2016} are reproduced. However, this is not the case for the recently discovered giant planet orbiting the late (M5.5) M dwarf GJ 3512 \citep{Morales2019}. This warrants further discussion in Sect. \ref{ssec:giants_extreme}.} After this qualitative first look at the results, we present a quantitative analysis of planetary types and masses in the following sections.

\subsection{Types of planets}
\label{sec:types_planets}
As a second step to explore the synthetic populations of planets around different host star masses, we categorize the planets into the following groups:
\begin{itemize}
\item Planets with masses larger than Earth ($M>\SI{1}{M_\oplus}$)
\item Earth-like planets defined as planets with masses ranging from \SI{0.5}{M_\oplus} to \SI{2}{M_\oplus}
\item Super Earths (\SIrange{2}{10}{M_\oplus})
\item Neptunian planets (\SIrange{10}{30}{M_\oplus})
\item Sub-Giants (\SIrange{30}{100}{M_\oplus})
\item Giant planets (masses larger than \SI{100}{M_\oplus}).
\end{itemize} 
Additionally, we add a category of temperate, Earth-mass planets that are introduced and discussed in Sect. \ref{sec:hab_planets}.

Compared to the analysis performed in \papertwo{}, the categories are identical with the exception of including the few brown dwarf mass planets as giant planets and choosing a different temperate zone. The distinction of giants and core-accretion brown dwarfs is not relevant for the purpose of this paper and the second change is introduced because accounting consistently for the scaling of the temperate zone with stellar mass requires the use of a more complex model.

\begin{figure*}
\centering
\makebox[\linewidth]{
        \begin{tikzpicture}
        \node (img) {\includegraphics[trim={0cm 0.4cm 0cm 0cm},clip,width=\linewidth]{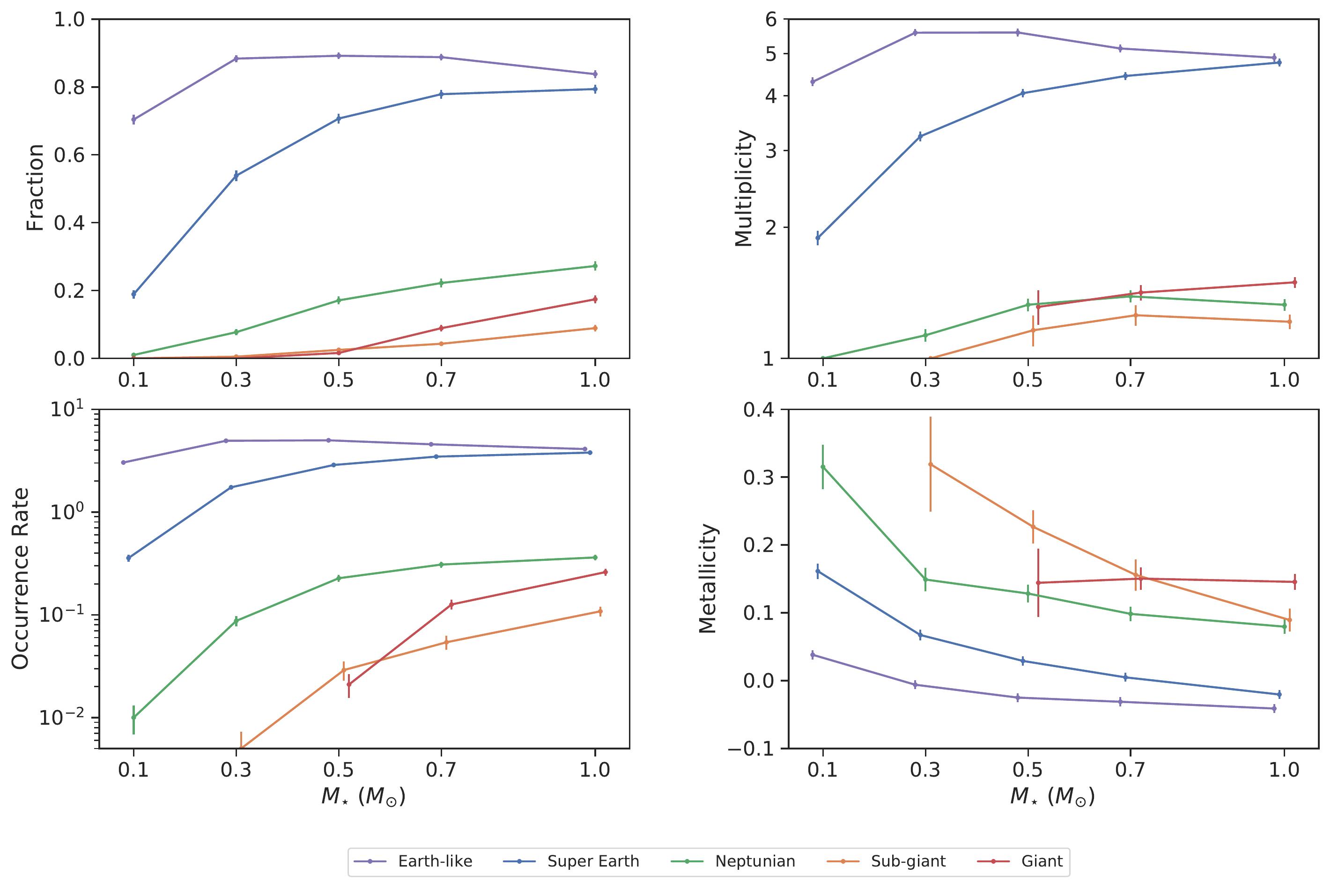}};
        \end{tikzpicture}
}
\caption{Fraction of systems, multiplicity, and occurrence rate for five planet-mass categories as a function of the stellar mass. 
        The standard error of the mean is indicated. Where necessary, the markers are slightly shifted in x-direction to better distinguish them.}
\label{fig:frac_mult_occ_type}
\end{figure*}

For the following analysis, we use the term fraction of systems, $f_s$, as defined by the ratio of systems containing one or more planets of a given planetary type divided by the total number of simulated systems $N_\mathrm{sys,tot}$. The occurrence rate $p_\mathrm{occ}$ of a planetary type is the number of formed planets of this kind in any simulated system divided by $N_\mathrm{sys,tot}$.
The ratio $p_\mathrm{occ}/f_s$ results in the mean multiplicity, that is, the mean number of this type of planet per system. The resulting $f_s$ for the different planetary types are shown in Table \ref{tab:fractions} and their mean multiplicity is shown in Table \ref{tab:multi}. Additionally, a visual representation of the same data also including $p_\mathrm{occ}$ is shown in Fig. \ref{fig:frac_mult_occ_type}. To get an idea of the dynamics, the eccentricities of the different types can be found in Table \ref{tab:eccentricity} and the host star metallicity [Fe/H] in Table \ref{tab:metallicity} and in the bottom right panel of Fig. \ref{fig:frac_mult_occ_type}. The metallicity is calculated based on the drawn dust to gas ratio $f_{dg}$ (see Sect. \ref{ssec:scaling_disk_mass}) which then translates to $[\mathrm{Fe}/\mathrm{H}] = \log_{10} \left(  f_{dg} / f_{dg,\odot} \right)$, where $f_{dg,\odot} = 0.0149$ \citep{Lodders2003}.

\begin{table}[ht]
\caption{Fraction of systems with specific planetary types for the different stellar mass populations with initially 50 lunar-mass embryos}

\begin{tabular}{lrrrrr}
        \hline \noalign{\smallskip}
        &\multicolumn{5}{c}{Stellar mass (\SI{}{M_{\odot}})} \\
        Type & 0.1 & 0.3 & 0.5 & 0.7 & 1.0 \\
        \hline\hline\noalign{\smallskip}
        $M> \SI{1}{M_\oplus}$& 0.44 & 0.77 & 0.88 & 0.91 & 0.95 \\
        Earth-like& 0.70 & 0.88 & 0.89 & 0.89 & 0.84 \\
        Super Earth& 0.19 & 0.54 & 0.71 & 0.78 & 0.79 \\
        Neptunian& 0.01 & 0.08 & 0.17 & 0.22 & 0.27 \\
        Sub-giant& 0.00 & 0.00 & 0.02 & 0.03 & 0.05 \\
        Giant& 0.00 & 0.00 & 0.02 & 0.09 & 0.19 \\
        Temperate zone& 0.35 & 0.66 & 0.70 & 0.66 & 0.57 \\
        \noalign{\smallskip}
        \hline
\end{tabular}

\label{tab:fractions}
\end{table}

\subsubsection{Giant planets}
\label{ssec:types:giant_planets}
As can be seen in Table \ref{tab:fractions} and Fig. \ref{fig:frac_mult_occ_type}, \frev{the giant planet occurrence drops with increasing stellar mass} in our simulations\frev{. No} planets with masses above \SI{100}{M_\oplus} form around stars with masses below \SI{0.5}{M_{\odot}}. At this transition stellar mass, the synthetic population with nominal parameters contains only \frev{21 giant planets in 16 systems}.

At \SI{1}{M_{\odot}}, \SI{17}{\percent} of the stars have one or more giant planets (see also \papertwo{}). This is in agreement with observational estimates \citep[\SI{14}{\percent} were found by][]{Mayor2011}. This planet type shows a clear preference for high \frev{stellar} metallicities, with a mean of about 0.15 for all stellar masses where giants can form. We can thus recover the well-known observational result \citep[e.g.][]{Santos2003}. \frev{Additionally, giant planets are} characterized by rather significant eccentricities ($\sim$0.15) compared to lower-mass planets, and a rather low multiplicity ($\sim$1.4), again compared to super-Earth and Earth-like planets.

\frev{In Sect. \ref{ssec:giant_planet_discussion}, we discuss the trends with respect to previous literature. Furthermore, in light of recent discoveries of massive planets around low-mass stars, we explore possible ways to increase the giant planet occurrence rate.}
\begin{center}
        \begin{table}[ht]
                \caption{Multiplicity of specific planetary types for all populations}
                
                \begin{tabular}{lrrrrr}
                        \hline \noalign{\smallskip}
                        &\multicolumn{5}{c}{Stellar mass (\SI{}{M_{\odot}})} \\
                        Type & 0.1 & 0.3 & 0.5 & 0.7 & 1.0 \\
                        \hline\hline\noalign{\smallskip}
                        $M> \SI{1}{M_\oplus}$& 3.04 & 5.47 & 6.51 & 6.89 & 7.01 \\
                        Earth-like& 4.31 & 5.58 & 5.59 & 5.14 & 4.89 \\
                        Super Earth& 1.89 & 3.23 & 4.06 & 4.44 & 4.77 \\
                        Neptunian& 1.00 & 1.13 & 1.33 & 1.39 & 1.33 \\
                        Sub-giant& -- & 1.00 & 1.14 & 1.06 & 1.17 \\
                        Giant& -- & 1.00 & 1.30 & 1.58 & 1.63 \\
                        Temperate zone& 1.23 & 1.53 & 1.74 & 1.81 & 1.98 \\
                        \noalign{\smallskip}
                        \hline
                \end{tabular}

                \label{tab:multi}
        \end{table}
\end{center}

\subsubsection{Neptunian planets and sub-giants}
The frequency of simulated Neptunian planets and Sub-giants declines with decreasing stellar mass. The slope is steeper for smaller masses where these kind of planets become very rare. In strong contrast to the Earth-like and super Earth planets, \frev{Neptunian} planets and sub-giants are both most commonly the only ones of their kind in a system and their orbits are more eccentric than those of other types, which holds for all stellar mass bins.

The orbits of sub-giants are more eccentric than those of Neptunian planets. This does not seem to be the case for the lowest stellar mass where sub-giants emerge. However, more statistics would be required to make a strong point because only $\sim$5 simulations exist where sub-giants formed around $\SI{0.3}{M_{\odot}}$ stars. For stellar masses larger than \SI{0.3}{M_{\odot}}, sub-giants are present in \SIrange{5}{10}{\percent} of the systems.

For both sub-giants and Neptunian planets, the mean metallicity of their host stars decreases with increasing stellar mass.

\begin{table}[ht]
        \caption{Mean eccentricities of different planetary types for all populations}
        
        \begin{tabular}{lrrrrr}
                \hline \noalign{\smallskip}
                &\multicolumn{5}{c}{Stellar mass (\SI{}{M_{\odot}})} \\
                Type & 0.1 & 0.3 & 0.5 & 0.7 & 1.0 \\
                \hline\hline\noalign{\smallskip}
                $M> \SI{1}{M_\oplus}$& 0.07 & 0.05 & 0.04 & 0.04 & 0.04 \\
                Earth-like& 0.07 & 0.05 & 0.04 & 0.04 & 0.04 \\
                Super Earth& 0.08 & 0.05 & 0.04 & 0.04 & 0.03 \\
                Neptunian& 0.08 & 0.10 & 0.10 & 0.10 & 0.09 \\
                Sub-giant& -- & 0.05 & 0.12 & 0.16 & 0.13 \\
                Giant& -- & 0.01 & 0.19 & 0.14 & 0.17 \\
                Temperate zone& 0.11 & 0.06 & 0.04 & 0.03 & 0.02 \\
                \noalign{\smallskip}
                \hline
        \end{tabular}

        \label{tab:eccentricity}
\end{table}

\subsubsection{Earths and super-Earths}
\label{ssec:types:earths}
We find that in our synthetic populations, Earth-like planets are most common around stars with a mass ranging from \SIrange{0.3}{0.7}{M_{\odot}} and -- conversely to the initial solid mass trend -- become less frequent around stars with masses above \SI{0.7}{M_{\odot}}. In contrast, the frequency-peak for Super Earths lies at the highest stellar mass bin (\SI{1.0}{M_{\odot}}). However, the frequencies of super-Earths are very similar for the two highest stellar mass bins, pointing to a flattening of the curve, much like in the Earth-like planet case.

Similarly, the multiplicity keeps increasing for super-Earths but peaks for Earth-like planets. The highest multiplicities of Earth-like planets are present for \SIrange{0.3}{0.5}{M_{\odot}}, which is lower compared to the stellar mass where this planet type is most common. 

\frev{The culprits associated with the decreasing trend are the giant planets ofsuch systems. As discussed in \citet{Schlecker2020}, the presence of giant planets often leads to dynamical instabilities that end up removing smaller planets from the system. This is reflected in the combined occurrence rates of Earths and super-Earths: for stars with $M_\star>\SI{0.5}{M_{\odot}}$, they range from \SIrange{7.9}{9.1}{} in systems without giant planets but are lowered to \SIrange{2.4}{2.7}{} if a giant planet is present.}

Most of the Earth-like planets and super-Earths are on relatively circular orbits. However, the eccentricity scatter is of the same magnitude as the mean. A trend toward lower eccentricities for higher stellar masses can be seen.

In terms of host star metallicities, high metallicity is required to form Earth-like planets or super-Earths around the very-low-mass stars, whereas for stellar masses larger than \SI{0.5}{M_{\odot}}, the mean metallicity of Earth-like planet and super-Earth hosts is close to the mean of the whole population. This outcome indicates that growth to these masses is not limited by the available amount of solids at the higher stellar masses. For all the terrestrial planets, the mean metallicity decreases monotonically over the stellar mass range. This decrease can be attributed to either or both of: (a) disks containing enough solid mass to form the planetary types at lower metallicities; or (b) the destruction by scattering or ejection of planets due to the presence of more massive planets around high-metallicity stars. Given the absence of giant planets around low-mass stars, the former is sure to be the driver for the metallicity dependency there. On the other hand, the presence of giants plays an important role in disks where they can form ($M_\star \geq \SI{0.5}{M_{\odot}}$).

\begin{table}[ht]
\caption{Mean metallicity [Fe/H] of stars hosting specific categories of planets}

        \begin{tabular}{lrrrrr}
                \hline \noalign{\smallskip}
                &\multicolumn{5}{c}{Stellar mass (\SI{}{M_{\odot}})} \\
                Type & 0.1 & 0.3 & 0.5 & 0.7 & 1.0 \\
                \hline\hline\noalign{\smallskip}
                $M> \SI{1}{M_\oplus}$& 0.10 & 0.02 & 0.00 & -0.01 & -0.01 \\
                Earth-like& 0.04 & -0.01 & -0.03 & -0.03 & -0.04 \\
                Super Earth& 0.16 & 0.07 & 0.03 & 0.00 & -0.02 \\
                Neptunian& 0.32 & 0.15 & 0.13 & 0.10 & 0.08 \\
                Sub-giant& -- & 0.24 & 0.23 & 0.16 & 0.11 \\
                Giant& -- & 0.44 & 0.17 & 0.15 & 0.14 \\
                Temperate zone& 0.07 & 0.00 & -0.03 & -0.05 & -0.09 \\
                \noalign{\smallskip}
                \hline
        \end{tabular}

\label{tab:metallicity}
\end{table}

\subsubsection{Earth-mass planets in the temperate zone}
\label{sec:hab_planets}
\frevi{There is a wealth of processes that have to be assessed in order to constrain the habitability of planets around low-mass stars \citep{Kaltenegger2017}. In particular, the activity levels of M dwarfs are elevated, which might inhibit the emergence of life \citep[see the review of][]{Shields2016}. Nevertheless, constraints on the frequency of planets on which life similar to Earth has a chance to emerge is of interest.} \frev{We can assess whether liquid water may exist on the surface of the synthetic planets using either mass or radius limits and select orbital separations according to the maximum and runaway greenhouse limits by \citet{Kopparapu2014}.}
For simplicity, the reported dependence of the zone on the planetary mass is not taken into account. Instead, the limits used are those calculated for masses of \SI{1}{M_\oplus}. We note that the parameters in \citet{Kopparapu2014} differ on a $\sim$\SI{5}{\percent} level compared to the parameters in \citet{Kopparapu2013,Kopparapu2013a}. Furthermore, the authors suggest avoiding the use of the moist greenhouse limit presented in \citet{Kopparapu2013} due to large differences to other works. For this reason, we chose the runaway greenhouse limit as the inner boundary.

The temperate zone \frev{defined in this way} moves with time based on the luminosity and the radius evolution of the star, for which we use the evolutionary tracks of \citet{Baraffe2015}. In this module of the code, we use identical luminosities and radii \frev{for each stellar mass} at all times. The resulting temperate zone limits after \SI{5}{Gyr} of evolution are:
\begin{itemize}
        \item \SIrange{0.03}{0.06}{au} for \SI{0.1}{M_{\odot}}
        \item \SIrange{0.11}{0.21}{au} for \SI{0.3}{M_{\odot}}
        \item \SIrange{0.20}{0.38}{au} for \SI{0.5}{M_{\odot}}
        \item \SIrange{0.39}{0.72}{au} for \SI{0.7}{M_{\odot}}
        \item \SIrange{0.96}{1.70}{au} for \SI{1.0}{M_{\odot}}.
\end{itemize}
Thus, the temperate zone moves with increasing stellar mass from orbital periods on the order of days for \SI{0.1}{M_{\odot}} stars to orbital periods on the order of years for Solar-type stars. In our simulations, this implies that the zone is displaced from close to the former inner edge of the disk to the proximity of the former disk snowline.

\begin{figure}
        \centering
        \includegraphics[width=.99\linewidth]{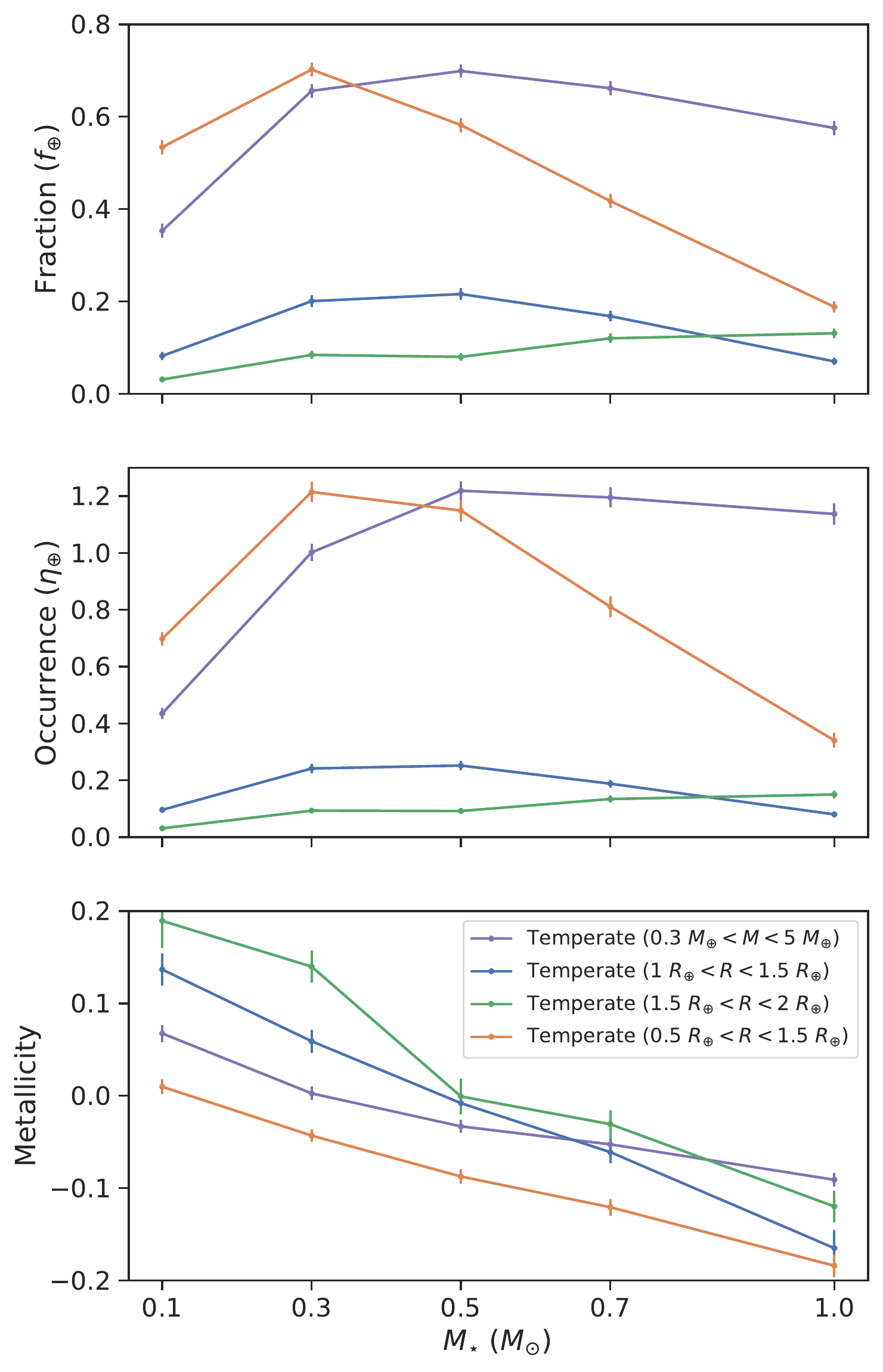}
        \caption{Fraction (upper), occurrence rate (middle), and mean host-star metallicity of synthetic planets with masses or radii similar to Earth in the temperate zone \citep{Kopparapu2014}. The error-bars depict the standard error of the mean. \frev{The cuts based on radii are the same as in \citet{Dressing2015} and \citet{Bryson2020}. The resulting $f_{\oplus}$ and $\eta_{\oplus}$ sensitively depend on this choice.}
        }
        \label{fig:frac_temp}
\end{figure}

An interesting pattern can be seen for the fraction of systems with Earth-sized planets in the temperate zone (see Fig. \ref{fig:frac_temp}): The stellar mass with the highest fraction of systems with temperate, low-mass (potentially habitable) planets is \SI{0.5}{M_{\odot}} when using the nominal mass cuts.  \frev{This peak can be partially attributed to the general trend of higher occurrence of Earth-like planets around these stars (Sect. \ref{ssec:types:earths}) as well as to the shift of the temperate zone toward less massive hosts}.

However, the multiplicity keeps increasing with stellar mass (Table \ref{tab:multi}). \frev{This can be expected as the width of the temperate zone increases monotonically from \SI{25}{} to \SI{44}{} Hill radii (Eq. \ref{eq:rhill}) of Earth-mass planets from \SI{0.1}{M_{\odot}} to \SI{1.0}{M_{\odot}}. Therefore, there is more dynamical space available to accommodate objects.} Despite that, the occurrence rate retains a peak at \SI{0.5}{M_{\odot}}; although, the data is consistent with a constant occurrence for $M_\star > \SI{0.5}{M_{\odot}}$ within confidence intervals.

In addition to the data based on cuts in planetary mass, Fig. \ref{fig:frac_temp} also employs cuts based on the planetary radii used in the works of \citet{Dressing2015} and \citet{Bryson2020}. The difference in the occurrence rates for the varied ranges in radii and masses can be explained by the fact that in the simulations \frev{around solar mass stars}, many planets in the \frev{temperate zone} contain ices or are enveloped in gas. Therefore, their radius is increased and they no longer fall in the radius selection despite having a mass below \SI{5}{M_\oplus}. The number of hydrogen-bearing planets is much lower for lower stellar masses due to lower mean masses and the proximity to the star. This shifts the most common temperate planet host to lower stellar masses (\SI{0.3}{M_{\odot}}).

Due to the dynamically distinct locations of the temperate zone, we do not expect the respective planets' eccentricity to be similar. Indeed, a trend toward lower eccentricities with increasing stellar mass is recovered (see Table \ref{tab:eccentricity}). In terms of mean host-star metallicity (bottom panel of Fig. \ref{fig:frac_temp}), a similar picture as for the Earth-like planets emerges with increased mean metallicities for low-mass stars and reduced metallicities for solar-type stars. 
This has strong implications for planetary searches aimed at maximizing the yield of potentially habitable planets.

\frev{Here, we put these results into context with existing observational constraints.} \frevi{For Earth-like planets with radii ranging from \SIrange{1}{1.5}{R_\oplus}, which corresponds to the blue line in Fig. \ref{fig:frac_temp}, \citet{Dressing2015} derive an occurrence rate of $0.16^{+0.17}_{-0.07}$ around cool stars ($T_\mathrm{eff} < \SI{4000}{\kelvin}$). For their larger, super-Earth size bin ranging from \SIrange{1.5}{2}{R_\oplus} (green line in Fig. \ref{fig:frac_temp}), they report $0.12^{+0.10}_{-0.05}$. Our synthetic occurrence rates with \SI{0.24}{} and \SI{0.09}{} for Earth-analogues and super-Earths respectively around \SI{0.3}{M_{\odot}} and \SI{0.25}{} and \SI{0.09}{} for \SI{0.5}{M_{\odot}} lie within error bars. This very precise match is quite contrasting to the overestimation of general occurrence rates reported by \citet{Mulders2019}.}

\frevi{Recently, \citet{Bryson2020} find consistently increasing habitable zone occurrence rates with stellar effective temperature in the range of \SIrange{3900}{6300}{\kelvin}. The size limits they chose correspond to the orange line in Fig. \ref{fig:frac_temp} (\SIrange{0.5}{1.5}{R_\oplus}). Although the uncertainties derived by \citet{Bryson2020} are quite large, an increasing trend of temperate planet occurrence with stellar effective temperature is apparent. The authors note, however, that it is weaker than expected from pure geometrical enlargement of the habitable zone with stellar temperature. Furthermore, they chose the optimistic habitable zone limits \citep{Kopparapu2014} to discuss the stellar temperature dependency, which increases this enlargement of the habitable zone with stellar mass. In contrast to their findings, in our simulations, there is an opposite clear decreasing trend of occurrence rates with stellar mass in the range considered by \citet{Bryson2020}. This trend is dominated by the population of small planets ranging from \SIrange{0.5}{1.0}{R_\oplus}. The skewness of the distribution of planetary radii especially around lower-mass stars makes the comparison to observations that favor the detection of larger planets statistically more challenging. Altogether, if future studies }\frev{confirm the increasing occurrence trend of temperate planets with stellar mass}\frevi{, the existence of such a numerous sub-population of smaller planets needs to be reevaluated.}

\begin{figure}[htbp]
        \centering
        \includegraphics[width=\linewidth]{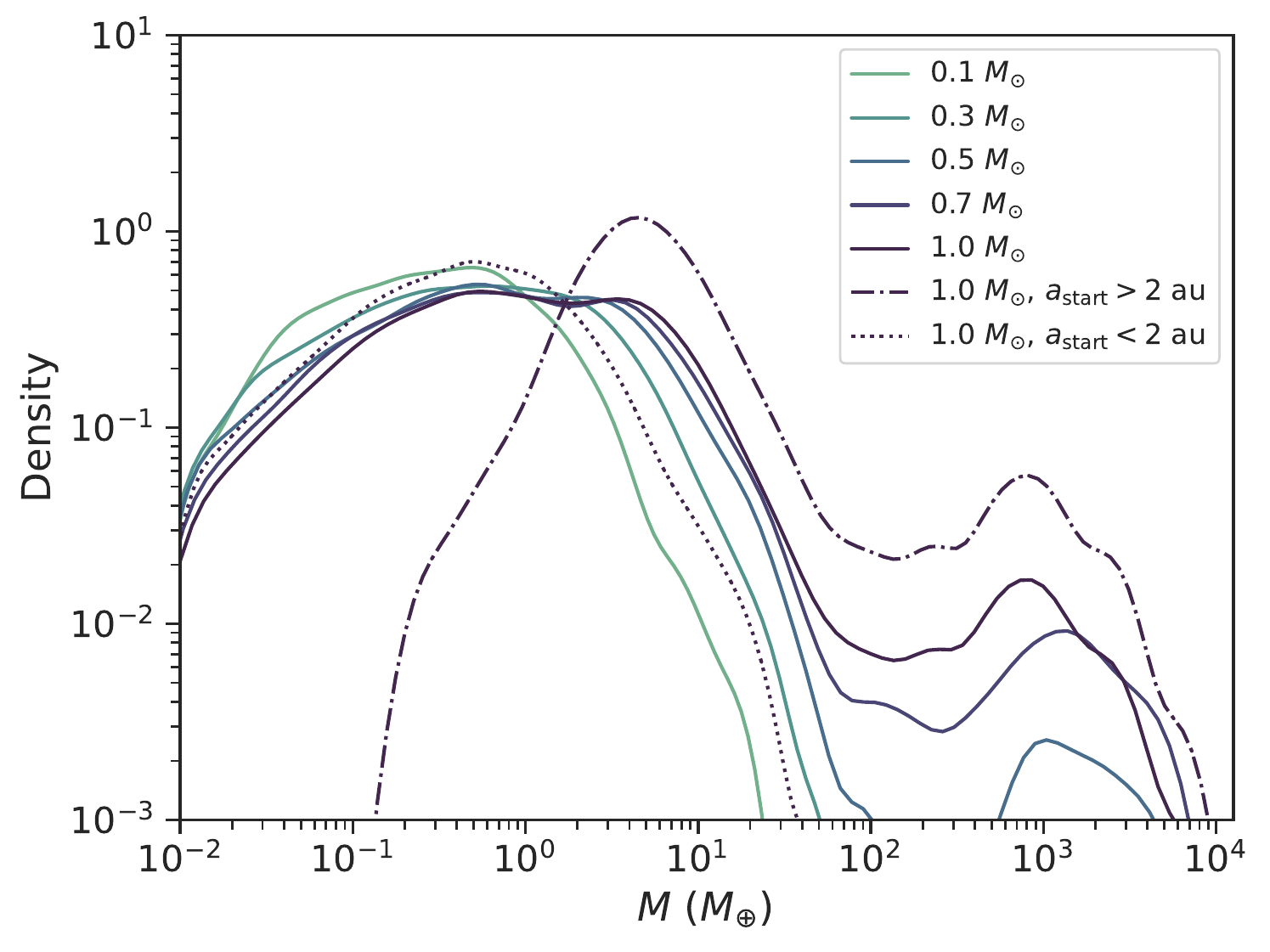}
        \includegraphics[width=\linewidth]{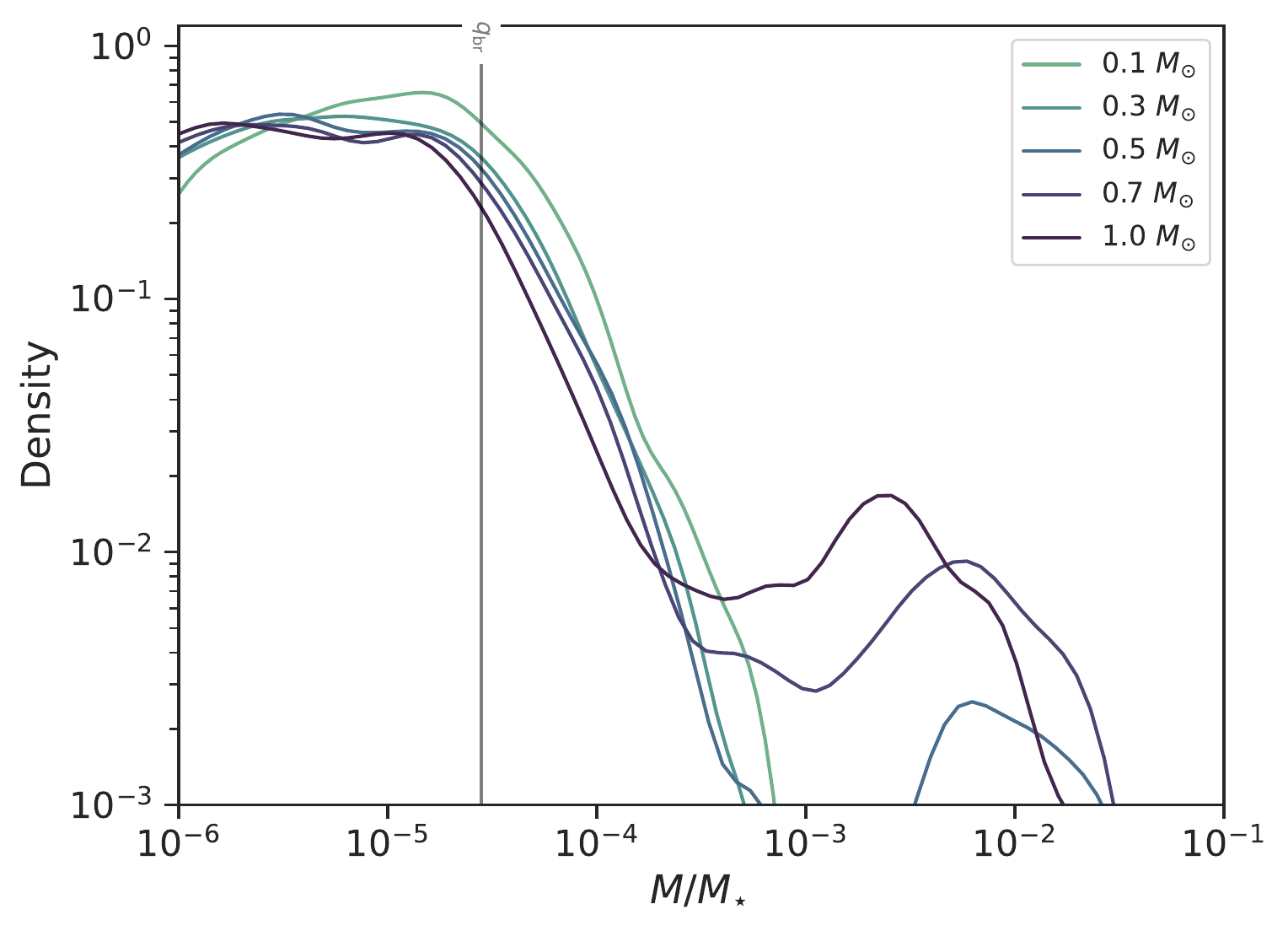}
        \caption{Kernel-density estimates of the close-in ($P<\SI{300}{\day}$) planetary mass distribution (top) and the planetary to stellar mass ratio (bottom) for populations of planets around different stellar masses. The bandwidth of the Gaussian kernel was chosen following the Normal reference rule \citep{Scott1992}. In the top panel, the population around solar-type stars is shown three times: once in full, once only including planets that started \frev{within (i.e., $a(t=0)<\SI{2}{au}$),} respectively, \frev{outside} \SI{2}{au}. In the bottom panel, the break in the distribution at $q_\mathrm{br}=\SI{2.8e-5}{}$ that was found by \citet{Pascucci2018} is depicted as a gray, vertical line.}
        \label{fig:kde_mass_astart}
\end{figure}
\subsection{Planetary mass distribution}
\label{sec:planetary_mass_distribution}

Because some regions at lower planetary masses in Fig. \ref{fig:aM_all} are saturated with points, we derive kernel-density estimates of the probability distribution of the synthetic planets' masses for better visibility and comparability. The resulting distribution of planetary masses and planet to star mass ratios for different stellar masses are shown in Fig. \ref{fig:kde_mass_astart}. For better comparability with results from the \textit{Kepler} mission, we only include planets with periods $P<\SI{300}{\day}$.

Starting at low planetary masses, the density estimates increase up to a plateau spanning from \SIrange{0.1}{10}{M_\oplus}. The plateau-like shape originates from two more narrow distributions of planets originating from further out and those growing at small orbital distances. This is revealed by splitting the population for \SI{1.0}{M_{\odot}} at initial locations of \SI{2}{au}. For lower stellar masses, the same pattern can be found, but the relative contributions of the two subsets varies.

Moving on to larger masses, there is a gap in the distribution at \SI{100}{M_\oplus} before giant planets become more frequent at larger masses. The exact shapes of the distributions of giant planets are influenced by low-number statistics and should not be interpreted quantitatively. Qualitatively, most giants originated from outside \SI{2}{au} and we find that the number of giant planets increases with stellar mass.

Similar patterns can be seen in the planet to star mass ratio $q$ distribution. It becomes clear that the distributions are not exactly overlapping. Instead, the population of small mass planets is shifted to  slightly larger $q$ around lower-mass stars. Such a trend is not in agreement with the studies of \citet{Wu2019} and \citet{Pascucci2018}, who found no dependency on the stellar mass in the \textit{Kepler} database We investigate in greater detail the origins of the mass distribution in Sect. \ref{sec:growth_regimes} and make a comparison with the observations in Sect. \ref{sec:frequency_of_earths}.

\subsection{Planetary composition and radius}
\label{sec:composition}
Even though the mass of a planet is the more fundamental parameter in terms of formation, transit measurements are sensitive to the radius of the planets. Therefore, we use the structure of the modeled planetary hydrogen-helium envelopes to calculate planetary transit radii following the prescription outlined in Sect. \ref{ssec:transit_radii}. 

\begin{figure}
        \centering
    \includegraphics[width=\linewidth]{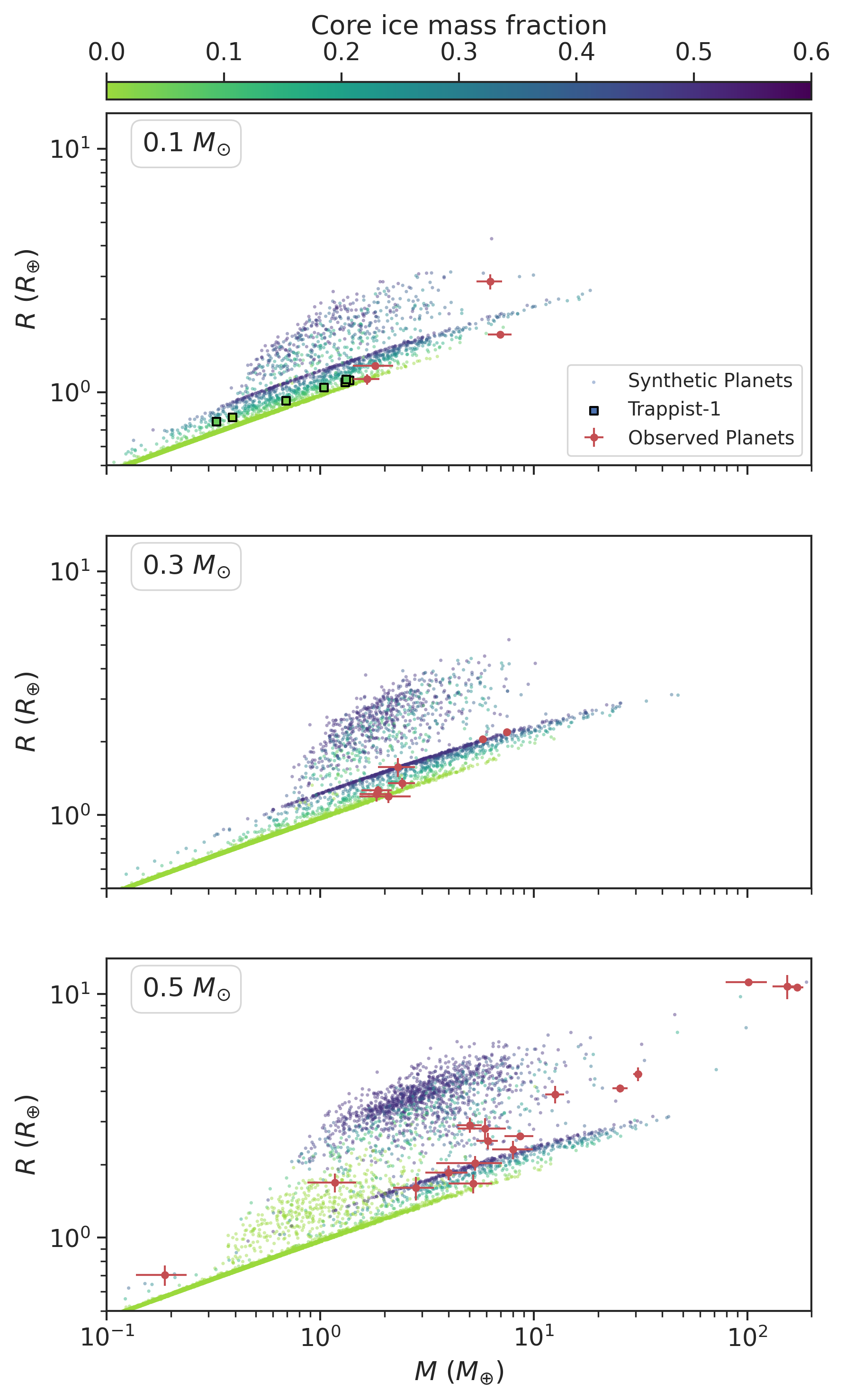}
    \caption{Populations of synthetic planets as a function of planet mass and transit radius. Only planets with orbital periods $P<\SI{300}{\day}$ are included. The core ice mass fraction is shown in color and observational data from the NASA Exoplanet Archive (accessed \frev{8.3.2021}) is over-plotted in red for comparison.}
    \label{fig:M_R}
\end{figure}

Figure \ref{fig:M_R} shows the masses and derived transit radii of the synthetic planets with $P<\SI{300}{\day}$ for the populations \frev{NGM10, NGM14, and NGM11} with respective stellar masses of \frev{\SI{0.1}{M_{\odot}}, \SI{0.3}{M_{\odot}}, and \SI{0.5}{M_{\odot}}.} The red markers show the observational data from the NASA Exoplanet Archive, where only planets with relative errors in radius, mass and stellar mass \frev{lower than \SI{50}{\percent}} are included. \frev{For stellar masses below \SI{0.2}{M_{\odot}}, shown in the top panel, the planets that are shown are GJ 1132 b \citep{Bonfils2018}, GJ 1214 b \citep{Harpsoe2012}, and LHS 1140 b and c \citep{Ment2019}. Outside the shown parameter space, the object 2MASS J02192210-3925225 b found by direct imaging close to the brown dwarf limit \citep{Artigau2015} is listed in the archive. Objects like this, with large planet-to-star mass ratios, probably formed in a process more similar to binary formation than planet formation \citep [as in][]{Bate2012}.} Additionally, the \Tra system is included with masses based on \citet{Agol2020} and with the color-coded ice mass fraction derived assuming an iron core mass fraction of \SI{32.5}{\percent}.

\frev{For the \SI{0.3}{M_{\odot}} case, we selected observed planets around stars ranging from \SIrange{0.2}{0.4}{M_{\odot}}. The two datapoints with remarkably small errors at dynamically estimated masses of \SI{5.77\pm0.18}{M_\oplus} and \SI{7.50\pm0.23}{M_\oplus} are the two planets orbiting K2-146 \citep{Hamann2019}. Given incident fluxes above \SI{10}{S_\oplus}, their precise alignment with the ice-rich, envelope-free, synthetic planets might be a coincidence.}

\frev{Concerning the synthetic data, t}he two straight lines at the low-mass end in Fig. \ref{fig:M_R} correspond to the compositions of pure rocky or ice-rock mixture with $\sim$\SI{50}{\percent} ice, which is the typical ice fraction of planetesimals outside the water iceline \citep{Thiabaud2014,Marboeuf2014b}. Only a small sub-population has ice fractions in between the two limiting cases. This group of planets accreted a significant amount of mass originating from outside and inside the water iceline. We see, however, that there are more of these planets for the low-stellar-mass cases due to fast type I migration at lower planetary masses. Migration of icy, far-out planets naturally leads to mixed compositions if the embryos reach the inner regions of the disk. \frev{For more massive stars,} most planets only migrate at $\sim$\SI{10}{M_\oplus}, where they can already accrete a light envelope, which leads to a significant increase in radius. For lower-mass disks, migration starts at lower planetary masses, where no significant envelope can be kept (see Sect. \ref{ssec:disk_migration}). Thus the \SI{0.1}{M_{\odot}} population NGM10 includes more intermediate-composition planets without hydrogen-helium envelopes.

Overall, the resulting planets follow roughly speaking the observational data -- where available -- in terms of their mass-radius distribution. \frev{An exception are the outliers 2MASS J02192210-3925225 b \citep{Artigau2015} and -- although of unknown radius -- also GJ 3512 b \citep{Morales2019}. Furthermore, the population of ice- and hydrogen-rich planets is not observed (see also the discussion in the next section).} In the future, objects discovered by TESS and characterized using follow-up programs will populate the mass-radius diagram for low-mass stars and will help to better validate the internal structure and envelope models.

\begin{figure}
\centering
\includegraphics[width=\linewidth]{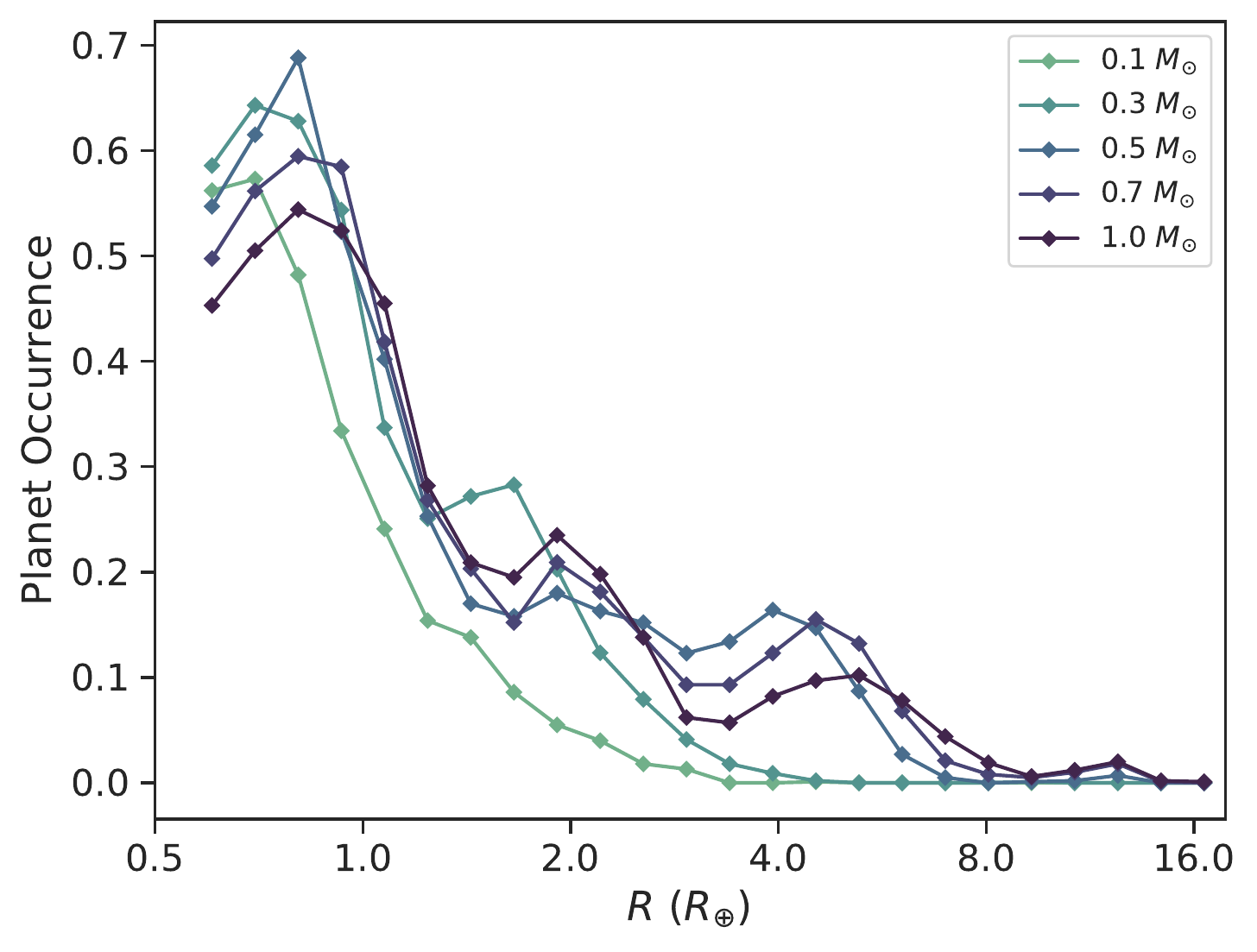}
\caption{Occurrence rates ($p_\mathrm{occ} = N_{\rm planets}/N_{\rm systems}$) of synthetic planets on orbits with periods $P < \SI{100}{days}$ as a function of their transit radii. 
\frev{The data is binned and bin centers are connected and marked with squares for better visibility.}}
\label{fig:R_hist}
\end{figure}

In general, the statistical distribution of planetary radii in Fig. \ref{fig:R_hist} shows features that are very similar to the features seen in mass-space distribution (Sect. \ref{sec:planetary_mass_distribution}). The main difference is that the planets do not span over many orders of magnitude in radii due to the degeneracy in the radii of giant planets (maximum at $R\sim$\SI{13}{R_\oplus}) and the dependency of radii on masses ($\sim M^{1/3}$).

The synthetic radius distribution in Fig. \ref{fig:R_hist} shows hints of two radius valleys for the higher stellar masses (\SIrange{0.5}{1.0}{M_{\odot}}). \frev{In contrast, only a single valley} is found in the observed population of planets \citep{Fulton2018}.
\flrev{A single valley was also predicted by} photo-evaporation \citep{Owen2013,Lopez2013,Jin2014}. Alternative explanations are core-driven mass loss \citep{Ginzburg2016}, impacts \citep{Wyatt2020} or the formation itself \citep{Venturini2020}.

\frev{It is striking that the two synthetic valleys in our simulations are located to either side of \SI{2}{R_\oplus}, where the observed gap is found. This is related to the presence of several synthetic, ice-rich planets close to the inner edge of the disk. As shown by \citet{Jin2018}, hydrogen-free, icy cores populate the radius distribution at exactly the location where the observed radius valley is located. However, the radii would change if a vapor phase in hot, water-rich atmospheres was included (\citealp{Turbet2020}, see also \citealp{Venturini2020} for a work including this effect). Furthermore, too efficient envelope stripping for colliding embryos could re-populate the gaps.} 
We will address this in more detail in a future work.

The observed gap shows an interesting stellar mass dependency \citep{Fulton2018,Wu2019}. \frev{Due to the general mismatch, }we cannot \frev{directly} compare it to \frev{our simulations}. Additionally, the sizes of the super-Earths (i.e., the planets below the radius gap) and the sub-Neptunes (i.e., above the gap) were individually analyzed in \citet{Fulton2018} and both show a trend of increasing mean size with increasing stellar mass. We note that their sample extends to only \SI{0.8}{M_{\odot}} and not to M dwarfs. This is in agreement with our synthetic data, where there is a clear trend toward larger mean size for the sub-Neptunes around more massive stars.

An interesting feature at lower radii is that the planet occurrence rate $p_\mathrm{occ}$ of planets at a specific radius is higher for lower-mass stars. \cite{Mulders2015a} show the same trend for the observed \textit{Kepler} sample. However, they find higher occurrence rates by factor of \SI{3.5}{}  , whereas we only find differences on the \SI{10}{\percent} level, with\frev{ the trend stopping at \SI{0.5}{M_{\odot}}}. 
 
%
\section{Discussion}
\label{sec:discussion}

\subsection{Giant planet occurrence for different stellar masses}
\label{ssec:giant_planet_discussion}
The best constrained occurrence rates for planets with known masses exist for giant planets, which are most readily detectable. 
In general, the frequency of giant planets increases with stellar mass \citep{Endl2006,Butler2006,Johnson2007,Johnson2010,Gaidos2013,Montet2014} up to an inversion at high stellar masses ($\sim$\SI{2}{M_{\odot}}, \citealp{Reffert2015}).
\subsubsection{\frev{Solid mass limited growth}}
At least for stellar masses below this inversion, the increasing trend is well explained by works on giant planet formation based on the core accretion paradigm. \citet{Adams2004} report fast external evaporation of disks around M dwarfs, effectively reducing the available gas to form giant planets. Even without this effect, \citet{Laughlin2004} found much slower growth timescales at a fixed semi-major axis around a \SI{0.4}{M_{\odot}} star, mainly due to the reduced solid surface density and the longer orbital timescale, which is also one of the conclusions of the population synthesis work by \citet{Ida2005} and \citet{Alibert2011}. The latter study stresses the importance of the disk mass on the resulting population of planets. 
As in this study, these works nominally assumed more massive protoplanetary disks around more massive stars motivated by measured stellar accretion rates. In general, we recover the same trends of low giant planet frequencies around low-mass stars (see Fig. \ref{fig:frac_mult_occ_type}). The reasons for this are attributed to the growth being limited by fast type I migration and long solid accretion timescales.

In a quantitative scope, \citet{Alibert2011} were able to approximate the synthetic giant planets resulting from their single-embryo populations around different stellar masses by scaling the distribution resulting from the \SI{1}{M_{\odot}} case. They found planetary masses $M \propto M_{\star}^\gamma$ with $\gamma = 0.9$ in their nominal case (disk mass $\propto M_\star^{1.2}$). The same cannot be recovered in our simulations, where the giant planet distribution seems to peak at the same planetary mass for all stellar masses. The overall frequency of giants is reduced but not their mean mass. However, we stress that our data set is much more sparse since only 1000 stars (each starting with 50 embryos) and about 100 resulting giant planets were simulated compared to the \SI{30000}{} stars each with a single embryo in \citet{Alibert2011}. Therefore, it is possible that some trends are hidden in statistical noise. If there is indeed no dependency of the mean giant planet mass on the stellar mass, N-body effects -- such as the ejection of planets -- might have played an important role in influencing the mass function in the newer simulations. \frev{A dedicated future paper in this series will focus on this point.}

\frev{The trend toward fewer giant planets with decreasing stellar mass is also found in the study using pebble accretion instead of planetesimals by \citet{Liu2019}. Although the growth in that case is no longer limited by the accretion timescale, which is shorter in the pebble accretion scenario, the pebble isolation mass decreases with stellar mass. Therefore, planets do not grow above $\sim$\SI{5}{M_\oplus} around stars with masses $<\SI{0.3}{M_{\odot}}$. At such low planetary masses, gas accretion is still inefficient. This transition stellar mass is similar to our value of $\SI{0.5}{M_{\odot}}$ considering that only few $\sim$\SI{100}{M_\oplus} objects formed at this mass in the simulations from \citet{Liu2019}.}

\frev{Overall, we highlight the fact that despite the range of different models and assumptions about the initial conditions, a number of works recovered the trend of a decreasing number of giant planets for decreasing stellar masses: an insufficiently small solid mass reservoir leads to starved growth which, in turn, inhibits subsequent gas accretion. This indicates that this is a very robust prediction of the core accretion scenario of planet formation.}

\subsubsection{\frev{Massive planets around low-mass stars}}
\label{ssec:giants_extreme}
Even though giant planet occurrence rates do increase with stellar mass, quite a number of systems with giant planets in orbit around M dwarfs of \SIrange{0.3}{0.4}{M_{\odot}} exist. 
The first M dwarf known to host planets is GJ~876 \citep{Marcy1998,Delfosse1998,Marcy2001,Rivera2005,Rivera2010,Millholland2018}, where now four companions have been confirmed. Two of them are giant planets (with masses of \SI{1.95}{M_{Jup}} and \SI{0.6}{M_{Jup}}), orbiting a \SI{0.37}{M_{\odot}} star \citep{vonBraun2013}. Other examples are GJ 849 \citep[\SI{0.65}{M_{\odot}}][]{Stassun2016} with two giants \citep{Butler2006} of masses \SI{0.77}{M_{Jup}} and \SI{0.9}{M_{Jup}} \citep{Montet2014}; GJ 179 a \SI{0.36}{M_{\odot}} star hosting a giant planet with a mass of \SI{0.8}{M_{Jup}} \citep{Howard2010}; and GJ 317 \citep[\SI{0.42}{M_{\odot}}][]{Anglada-Escude2012} hosting a potentially up to \SI{2.5}{M_{Jup}} massive planet and a second giant with $M \sin i = \SI{1.6}{M_{Jup}}$ \citep{Johnson2007,Anglada-Escude2012}.

Below stellar masses of \SI{0.3}{M_{\odot}}, no stars hosting giants were found\footnote{A lower stellar mass of GJ 317 was reported in \citet{Johnson2007}, but was corrected to higher masses by \citet{Anglada-Escude2012}.} until the recent discovery of GJ 3512b, a planet with a minimum mass of \SI{0.463}{M_{Jup}} around a \SI{0.123}{M_{\odot}} star \citep{Morales2019}, which poses the biggest challenge to all current planet formation scenarios. Additionally, it is quite likely that a Saturnian mass companion leads to an inner cavity in the transition disk CIDA 1 \citep{Pinilla2018}.

In our nominal populations presented in Sects. \ref{sec:m-a} and \ref{sec:types_planets}, we do not find giant planets around stars with masses below \SI{0.5}{M_{\odot}}. As a consequence, in order to form a planet like GJ 3512b, either the physical parameters are to be revised in our models or different physical processes are at work, such as a different formation channel (gravitational instability).
\frev{Therefore, we explore variations of parameters in the Bern Model to increase the efficiency of core-accretion planet formation around low-mass stars.}
\frevi{To this end, we chose a stellar mass of \SI{0.1}{M_{\odot}} to model conditions similar to GJ 3512b ($M_\star = \SI{0.123 \pm 0.009}{M_{\odot}}$).}

\frevi{
One trivial pathway to form larger planets would be to increase the disk masses. However, this cannot be considered a free parameter since observational data is available. We use the disk masses derived for the young class I objects from \citet{Tychoniec2018}, which are already larger than those from \citet{Williams2019}. Therefore, a further increase would stand in contradiction to the observations. Furthermore, we are not aware of indications for a shallower scaling than the nominal linear scaling of the disk mass with the stellar mass.
}

\frevi{
Instead, we tested the influence of the placement of planetesimals and embryos and the impact of reducing the type I migration speed. Rapid type I migration is well known to reduce the efficiency of forming giant planets \citep{Alibert2005,Mordasini2009b}. Therefore, we can explore the impact of a reduction factor $f_I=0.1$ for type I migration rates similar to \citet{Mordasini2009b}. Such parameter searches as these, addressing giant planet formation, can be done in the single embryo mode, employing the one-embryo-per-disk approximation (see \papertwo{}).
}

\begin{figure}
        \centering
        \includegraphics[width=\linewidth]{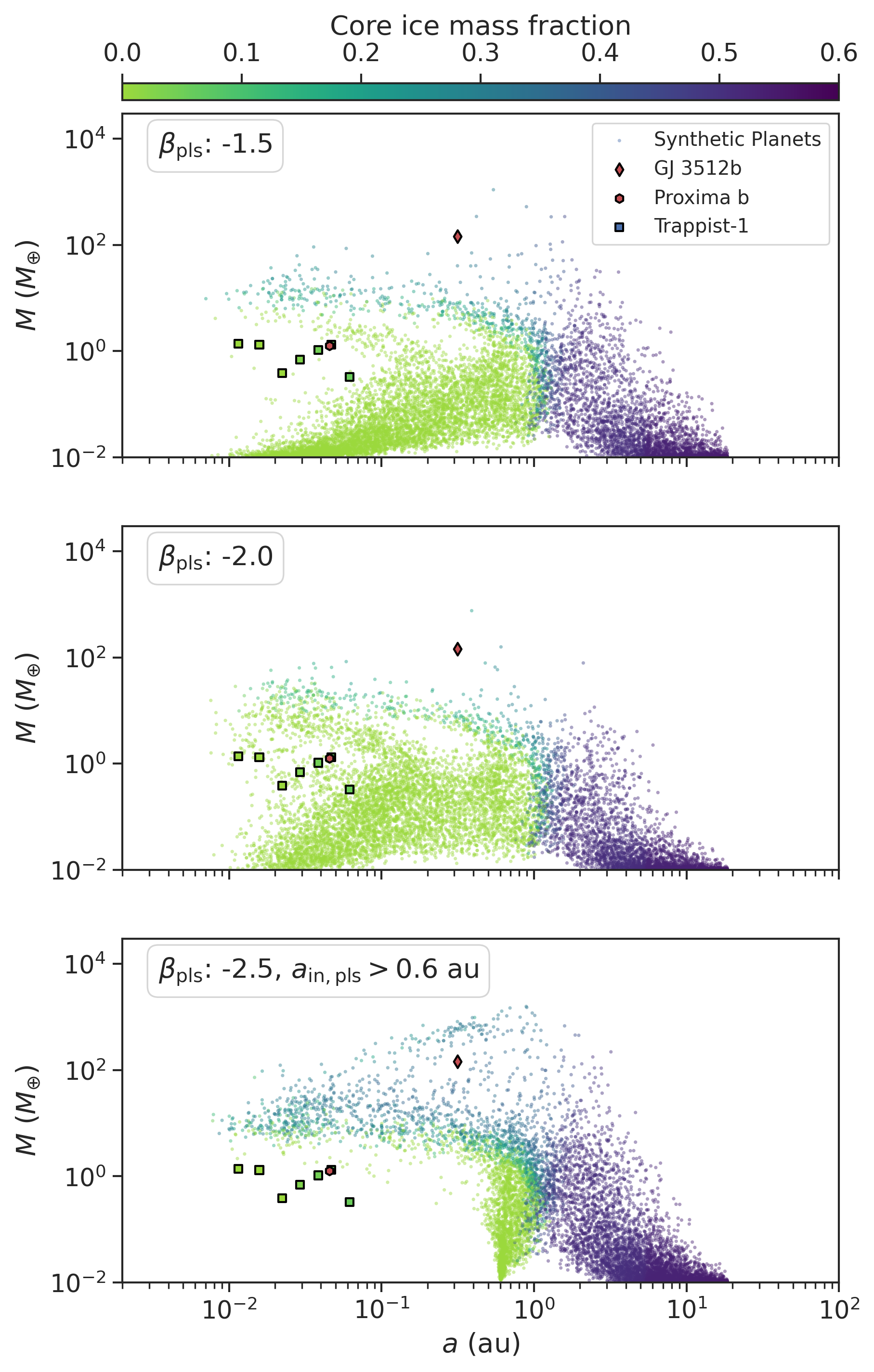}
        \caption{\frevi{Synthetic populations of planets calculated in single-embryo mode with reduced type I migration ($f_I$: 0.1) as a function of $a$ and $M$. The mass fraction of the summed-up ices is shown in color. The top panel shows the population of planets with a nominal planetesimal surface density slope $\beta_{\rm pls}$, whereas the other two panels have steeper planetesimal surface density slopes as indicated in the top left of the panel. The bottom panel displays the population of planets, where embryos and planetesimals are only placed outside of \SI{0.6}{au}.}}
        \label{fig:a_M_single_0p1_compare}
\end{figure}
\frevi{
Figure \ref{fig:a_M_single_0p1_compare} shows three synthetic populations of planets where $f_I=0.1$ for all of them. The number of simulations -- thus also of synthetic planets -- is 10000 for each of them. They differ by the slope $\beta_{\rm pls}$ of the initial radial planetesimal surface density profile and the placement of the planetesimals and embryos: The top panel shows the nominal slope of $\beta_{\rm pls}=-1.5$, the central panel shows a population of planets where the planetesimals were placed with a slope of $\beta_{\rm pls}=-2.0$ and the bottom one with $\beta_{\rm pls}=-2.5$. In the last case, growth would mainly occur close to the star due to the mass concentration there. Therefore, the initial conditions for the population, shown in the bottom panel, were further modified to be optimal for giant planet formation and thus include an inner edge of the planetesimal disk at \SI{0.6}{au} and the same inner boundary for the injection of planetary embryos. The other two simulations do not differ from the nominal simulations in terms of the inner edge of the planetesimal disk or embryo placement (i.e., log-uniform from gas-disk inner edge to $\sim$\SI{20}{au}).}

\frevi{
Compared to the top panel of Fig. \ref{fig:aM_all} with $\beta_{\rm pls}=-1.5$, where type I migration is not reduced, much more massive planets can form with reduced migration (top panel of Fig. \ref{fig:a_M_single_0p1_compare}). Here, \SI{1.87}{\percent} of planets have reached $M>\SI{10}{M_\oplus}$. In the nominal population, this number is as low as \SI{0.02}{\percent} or 11 out of \SI{50000}{}. In very rare cases, the mass of GJ 3512b was already reached. A similar level of efficiency in reproducing GJ 3512b is reached by the simulation shown in the central panel, with $\beta_{\rm pls}=-2.0.$  Also demonstrating an efficiency in producing giant planets is the third population shown in the bottom panel. Overall, giant planets are frequently formed  in the heavier disks.}

\frevi{
In a fourth population where only the planetesimal slope was increased to -2.5 and the placement of planetesimals and embryos constrained to the region outside \SI{0.6}{au} but the type I migration speed was not reduced, more massive planets than in the nominal case can form, but no giant planets formed. Instead, the frequency of icy super-Earths increased drastically.}

\frev{We find that}\frevi{ even for ultra-late M dwarfs, the formation of gas-rich giant planets is possible if two conditions are met: a high disk mass from the upper end of the distribution (for $\beta_{\rm pls}=1.5$ $M_{\mathrm{gas}}>\SI{0.007}{M_{\odot}}$ and $M_{\mathrm{solid}}>\SI{66}{M_\oplus}$ were required) and reduced type I migration. The latter could be due to, for example, trapping in ringed disk structures or different gas surface density profiles, for example, due to disk winds \citep{Suzuki2016Winds,Ogihara2018}.}
Therefore, as long as lower migration speeds or migration traps cannot be excluded, the formation of giant planets by core accretion around very low-mass stars should not be discarded.

However, this pathway does require some "tuning" of parameters (i.e., small planetesimal sizes and low type I migration speeds). It is not yet clear if the gravitational instability pathway \citep{Cameron1978,Boss1997} could more naturally produce giant planets around low-mass stars. For the case of GJ 3512b, the mass-ratio is lower and the planet orbits closer to the star than typical disk instability outcomes \citep[briefly discussed in][]{Morales2019}. For core accretion models, it remains to be checked whether planets could be trapped at a fixed separation from the star in a sufficiently frequent and efficient way to explain the aforementioned large planetary to stellar mass ratio examples. Observations do reveal common ringed structures in protoplanetary disks \citep{Andrews2018}. Those frequent dust rings might trace inverted gas pressure gradients which lead to migration traps \citep[see also][who explore this planet formation pathway]{Hasegawa2011,Coleman2016a,Alessi2020}.

\subsection{Growth regimes}
\label{sec:growth_regimes}
As shown in Sects. \ref{sec:m-a} and \ref{sec:planetary_mass_distribution}, the components that make up the planetary mass distribution (Fig. \ref{fig:kde_mass_astart}) are a population of very low-mass ($\sim$\SI{0.01}{M_\oplus}) "failed cores" at high semi-major axis and low-mass ($\sim$\SI{1}{M_\oplus}) planets mainly growing inside of the water iceline ("terrestrials"). In addition to these, the more massive planets ("horizontal branch" planets, $M\sim\SI{10}{M_\oplus}$, \citealp{Mordasini2009b}), initially growing at larger separations and a few giant planets ($M>\SI{100}{M_\oplus}$), populate the distribution.

In the following, the different origins of the sub-populations are addressed. In particular, we discuss the shape of the planetary-mass and the planet to star mass ratio distribution shown in Fig. \ref{fig:kde_mass_astart}.

\begin{figure}
        \centering
        \includegraphics[width=\linewidth]{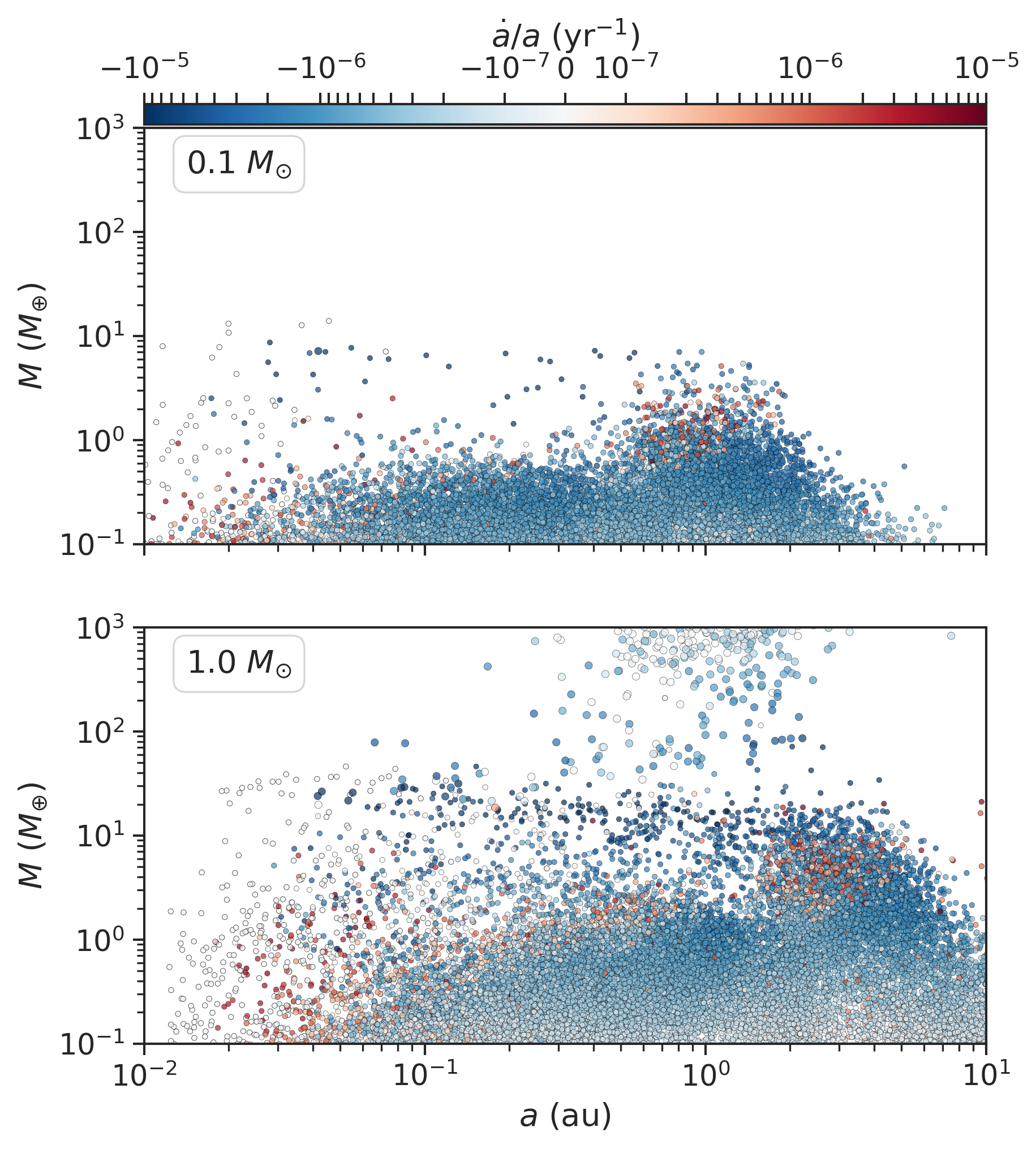}
        \caption{Mass and semi-major axes of the populations around \SI{0.1}{M_{\odot}} and \SI{1.0}{M_{\odot}} after \SI{1}{Myr} of evolution. The normalized migration rate is colored and larger dots denote gap-opening planets that migrate in the type II regime.}
        \label{fig:a_M_mig}
\end{figure}

The initial embryo mass of \SI{0.01}{M_\oplus} is large enough to lie in the oligarchic growth regime \citep{Rafikov2003,Ormel2010b}. At large separations, the oligarchic growth timescale becomes long. This leads to a population of planets that are not able to grow significantly and remain at large distances. These are excluded from our analysis using the cuts in period space. Despite the existence of failed cores \flrev{at almost all of the} sampled starting points and disk conditions, planets are able to grow and favorable conditions with short growth timescales are frequent outside the iceline. This allows for growth up to masses where type I migration becomes important. Therefore, migration could shape the mass distribution. To explore this scenario, we show in Fig. \ref{fig:a_M_mig} the population of planets as it is still growing with color-coded migration rates.

The distribution peaks at the planetary to stellar mass ratio of $q\sim\SI{1e5}{}$. In this sub-population, planets from the "horizontal branch" as well as the "terrestrials" are the primary contributors and, apart from the giant planets, they stand as the most massive planets.

In principle, there could be three reasons for the mass distribution to drop at higher mass ratios than $q\sim\SI{1e5}{}$, leading to a peaked shape: (1) the onset of rapid gas accretion could move planets quickly to larger masses, leading to the typical "desert" in the mass distribution \citep{Ida2004}; (2) the disk conditions and small isolation masses \citep{Kokubo2000} might not allow for further growth for planets growing close to the star, that is, for the "terrestrials;" or (3) fast type I migration could move the planets out of favorable growth regions, thus halting solid accretion.

The first point cannot be the main driver in our simulations because the number of planets that actually grow to become giant planets is too few to influence the shape of the mass distribution at lower masses. From Fig. \ref{fig:a_M_mig}, we can see mechanisms (2) and (3) at work: There is a clear decrease of planets in the inner systems at \SI{1}{Myr} when going to larger masses. Although some more growth and collisions will occur at later times, the disk conditions are clearly limiting growth in the inner system. Additionally, we can identify planets colored in dark blue indicating fast type I migration that originate from the outer system that typically migrate inwards at $\sim$\SI{5}{M_\oplus} ($q\sim\SI{1.5e-4}{}$) in the upper panel (\SI{0.1}{M_{\odot}}), respectively, at $\sim$\SI{20}{M_\oplus} ($q\sim\SI{6e-5}{}$) in the lower panel. The typical masses differ because, as discussed in Sect. \ref{ssec:disk_migration}, migration is a function of the stellar mass, disk mass, and planetary mass. Of particular importance is the mass at which the corotation torque saturates \frev{and planets rapidly drift inwards without time for further growth.} This explains the different locations of the drop in the mass distribution of resulting planets. We note that, as the disk evolves, cools, and thins out, the typical onset of rapid type I migration is shifted to lower values. Therefore, the early snapshot of the first "fast migrators" at \SI{1}{Myr} should give upper limits to the processes.

To summarize, we qualitatively identify that the saturation of corotation torques, as well as the growth-limits in the inner system, is shaping the planetary mass distribution below \SI{50}{M_\oplus}. The non-linearity, despite a linear scaling of the disk mass, is likely due to the non-linear behavior of type I migration with disk mass via the temperature structure and the resulting scale heights. 

At the upper end of the distribution lies a small fraction of planets which can overcome this type I migration barrier. Then, they undergo runaway gas accretion \citep{Mizuno1978}, reach the slower type II migration regime and become giants. A lot of solid mass is required to grow to the type II migration regime more quickly than type I migration timescales ($\sim \num{e4}$ to $\SI{e5}{yr}$).

\subsection{Frequency of Earths and super-Earths}
\label{sec:frequency_of_earths}
After focusing on the origins of low-mass planets in our simulations, we discuss the same population of planets with regards to observations in this section. For planets with masses below $\sim$\SI{100}{M_\oplus}, the best estimates on their frequency can be gained from transit surveys because of their large sample. Using the results from \textit{Kepler}, \citet{Dressing2013} derive an occurrence of about one planet with orbital period shorter than 50 days and radii from \SIrange{0.5}{4}{R_\oplus} per cool star ($T<\SI{4000}{\kelvin}$), while \cite{Gaidos2016} find around two planets per M dwarf with similar radii and orbital periods up to \SI{180}{days}. From Fig. \ref{fig:R_hist}, it is apparent that more planets are formed in the synthetic simulations. This has already been reported in \citet{Mulders2019}, who find a fraction of synthetic stars with planets that is five times greater compared to the \textit{Kepler} sample. This finding still holds for lower-mass stars. The synthetic data on which their conclusion is based on, is very similar to the \SI{1.0}{M_{\odot}} data set shown here.

Planet formation models do more readily yield planetary masses than radii, as gravity is the dominating force that is acting. Consequently, our results can more readily be compared to \citet{Pascucci2018} who use the \textit{Kepler} sample to derive planetary and stellar masses. In contrast to this simple analysis, a future paper in this series will apply the technique described in the companion paper of \citet{Mishra2021} to use our computed planetary radii and derive a biased sample of planets to compare directly to \textit{Kepler}.

\citet{Pascucci2018} find broken scaling relations in planet to star mass ratios obtained from \textit{Kepler} and microlensing surveys, which can be discussed here without the compositional imprints on the radii. In the \textit{Kepler}-based data of \citet{Pascucci2018}, the position of the universal peak lies at  $M/M_\star =\SI{2.8e-5}{}$ indicated in the lower panel of Fig. \ref{fig:kde_mass_astart}. Our synthetic mass function does not show the same prominent peak, but a wider distribution of ratios (see Fig. \ref{fig:kde_mass_astart}). However, if more embryos would have started outside the iceline, a peak similar to the dash-dotted line in the upper panel of Fig. \ref{fig:kde_mass_astart} should have been recovered at a location closer to the \citet{Pascucci2018} peak. In Fig. \ref{fig:kde_mass_astart}, we see that the distribution is not shifted exactly linearly with the stellar mass. Instead, a slight trend toward larger $M/M_\star$ for lower stellar masses can be seen. In \citet{Pascucci2018}, only the most massive F-stars do show a slightly lower mass ratio peak, while the peaks around G, K and M dwarfs lie at the same $M/M_\star$. This could partially be due to a more narrow range of stellar masses that was available for their analysis.

Furthermore, the location of the peak at $M/M_\star = \SI{2.8e-5}{}$ from \citet{Pascucci2018} is not perfectly matched by our models. Instead, the distributions drop at a value of $M/M_\star$ which is a factor $\sim$1.5 lower. This can either point toward migration in the models becoming too efficient at too low planetary masses, too low solid disk masses, or a preferred placement of the embryos could be influencing the results.

Based on these findings, we can conclude that if the mass distribution of planets indeed has a distinct peak and is not a remnant decrease of an inaccessible, broader distribution, then it is most likely due to type I migration halting growth. This pushes many planets of similar mass to the observable regime, which is expected to make up the bulk of the observed data. The planetary mass would then correspond to the mass where the positive corotation torques saturate. As can be seen in Fig. \ref{fig:migration_map}, it is lower for lower-mass stars, leading to a similar value in $q$. 
This would point toward more efficient embryo formation at larger distances and fewer "terrestrial-like" planet growth, which would be in line with works that propose that planetesimal formation preferably occurs outside the water iceline \citep{Drazkowska2017,Schoonenberg2017}. Furthermore, it would strongly favor the migration mechanism in general compared to an in situ growth of planets. This pathway is an alternative to the scenario of pebble-based accretion models, where a peak in the mass distribution can be directly attributed to the pebble isolation mass \citep{Liu2019}.

\subsection{Dependence of dynamical results on initial placement}
\label{ssec:dependence_on_placement}
Planetary orbital parameters, such as the eccentricity or the period ratio of neighboring planets are influenced by close encounters between them. The frequency of close encounters in turn depends sensitively on the initial placement of the synthetic embryos, since placing two embryos in close proximity to each other might lead to interactions already during the early growth stage when migration is still negligible.

One measure that can be used to quantify the probability of dynamical instabilities (i.e., gravitational interactions) is the mutual Hill radius of a pair of planets \citep{Chambers1996}
\begin{equation}\label{eq:rhill}
R_\mathrm{H,mut} = \left(\frac{M_1 + M_2}{3 M_\star}\right)^{1/3} \left( \frac{a_1 + a_2}{2} \right)\,,
\end{equation}
where $M_1$, $M_2$, $a_1$, $a_2$ are the masses and semi-major axes of two planets in a system. Typically, instabilities can occur if two planets are separated by less than $\sim$3.5 mutual Hill radii \citep{Chambers1996}.

We chose the same initial embryo mass for all stellar masses. Thus, the distance, measured in mutual Hill radii, between the initially placed embryos in the simulations increases with increasing stellar mass. This means that in terms of dynamical interactions, the populations are not starting with the same initial conditions and thus, no strong conclusions should be drawn from the nominal populations with 50 embryos each.

\begin{figure}
        \centering
        \includegraphics[width=\linewidth]{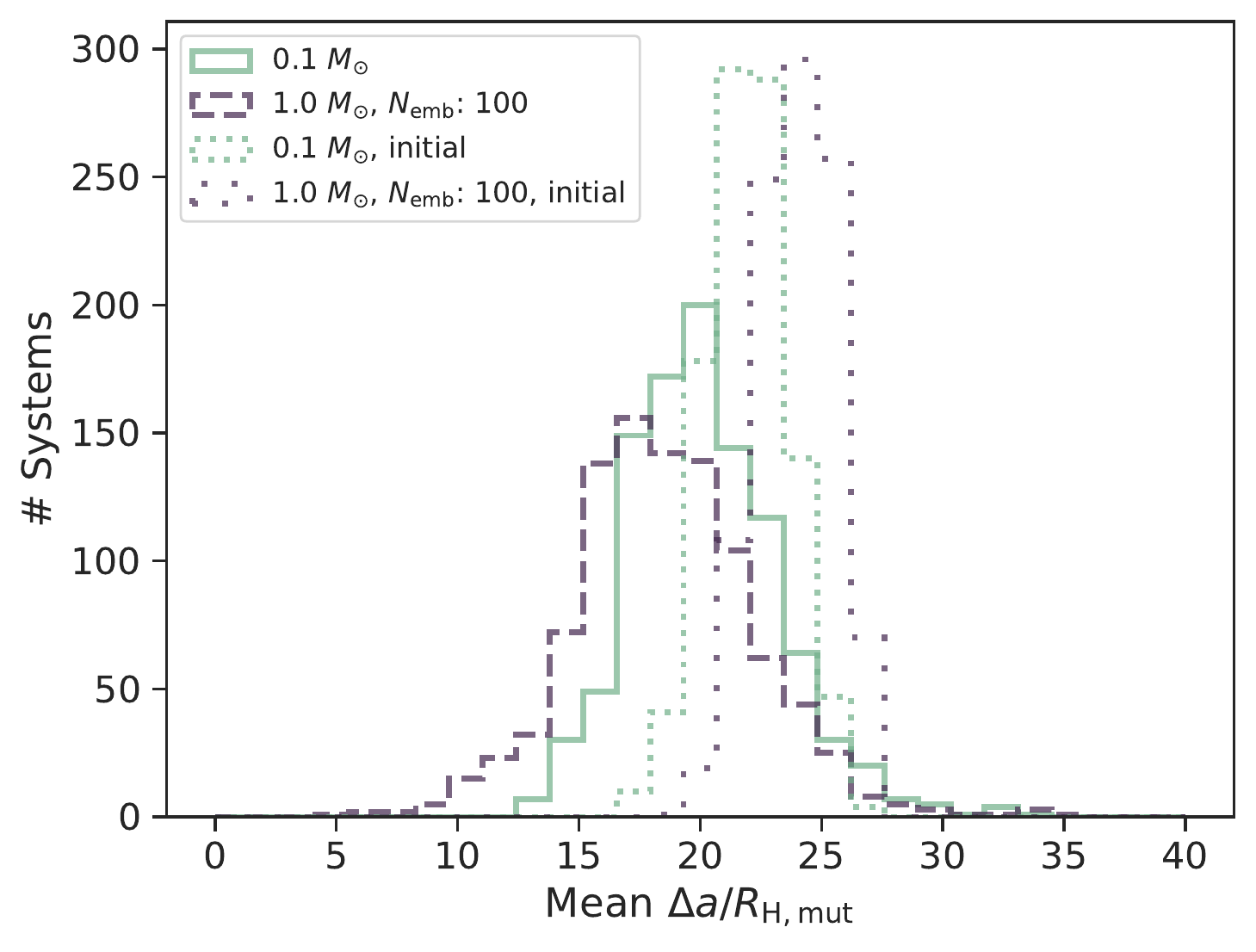}
        \caption{Mean distance between all neighboring planets in a system measured in mutual Hill radii \citep{Chambers1996}. The initial placement corresponds to the dotted lines and the resulting values are shown using a solid line. The number of injected embryos for the population of planets around \SI{0.1}{M_{\odot}} stars is 50, whereas we insert 100 embryos for this population of planets around \SI{1.0}{M_{\odot}} stars.}
        \label{fig:hist_da_mutual_rhills_100emb}
\end{figure}

This discussion touches the topic of the influence of the initial number of embryos, which is discussed in \papertwo{}. There, it is shown that eccentricity and period ratio are quantities sensitive to the number of initial embryos that are placed in the disk. The scaling of the Hill radius $\propto M_\star^{1/3}$ leads to a factor  $\sim$2 in the distance between initial embryos measured in mutual Hill radii between the \SI{0.1}{M_{\odot}} population and the \SI{1.0}{M_{\odot}} population. Coincidentally, a planetary population was calculated and presented in \papertwo{} using 100 embryos around a \SI{1.0}{M_{\odot}} star (Population NG76). We are therefore able to compare here the dynamical outcome between this population and the here presented \SI{0.1}{M_{\odot}} case (NGM10), which have more similar initial dynamical conditions (see Fig. \ref{fig:hist_da_mutual_rhills_100emb}, dotted lines).

Due to growth and migration, the systems get more compact over time leading to lower mean distances measured in mutual Hill radii ($\bar{\Delta a}/R_\mathrm{H,mut}$) but, sporadically, the measure can increase if collisions or scattering of planets occur. The final states of the solar-type star population with 100 initial embryos and the \SI{0.1}{M_{\odot}} population with 50 embryos are still quite close to each other. For \frev{comparison}, the solar-type star population with 50 embryos has a mean over all systems around \SI{27}{} $\bar{\Delta a}/R_\mathrm{H,mut}$ after starting at $\sim$40. This is significantly different from the outcome with 100 embryos, which results in more closely packed systems with a mean over all systems of $\bar{\Delta a}/R_\mathrm{H,mut}$ at $\sim$20.

In Fig. \ref{fig:hist_da_mutual_rhills_100emb}, we can also see a slight trend to more packed systems for the population of planets around solar-type stars. Indeed, even though the initial $\bar{\Delta a}/R_\mathrm{H,mut}$ of the population of planets around solar-type stars is slightly larger than for the population around \SI{0.1}{M_{\odot}}, the resulting mean $\bar{\Delta a}/R_\mathrm{H,mut}$ is lower. We argue that this is due to very little growth in some systems around \SI{0.1}{M_{\odot}}, which then leads to systems with mean $\bar{\Delta a}/R_\mathrm{H,mut}$ closer to the higher initial value. Although planets with arbitrary mass are included in Fig. \ref{fig:hist_da_mutual_rhills_100emb}, the same trends are recovered if small mass planets are removed. Thus, it is relevant for observable planets.

\begin{figure}
        \centering
        \includegraphics[width=\linewidth]{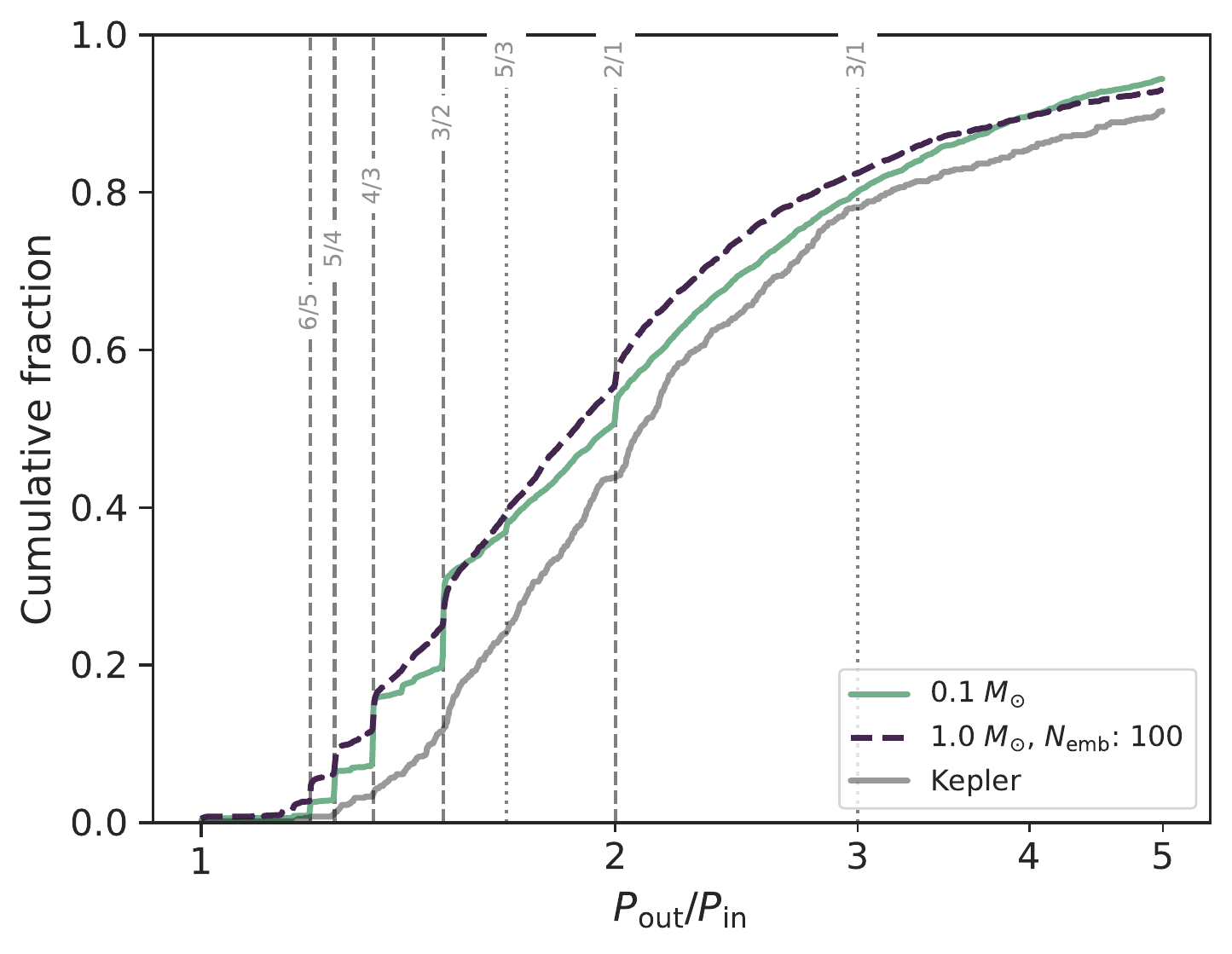}
        \caption{Orbital period ratio of all neighboring planets including planets with $M>\SI{0.1}{M_{\odot}}$ and periods $P<\SI{300}{days}$. Two synthetic populations of planets are shown: one around a solar-type star with initially 100 embryos (NG76) and one around stars with a mass of \SI{0.1}{M_{\odot}} (NGM10). The \textit{Kepler} multi-planetary system period ratios are displayed for reference (NASA Exoplanet Archive, accessed 9.12.2019).}
        \label{fig:period_ratios_100}
\end{figure}

Having established a closer dynamical relationship between the population with 50 embryos around a \SI{0.1}{M_{\odot}} (NGM10) star and the one with 100 embryos around \SI{1.0}{M_{\odot}} (NG76), we can compare the dynamical evolution of the systems of those two populations. Figure \ref{fig:period_ratios_100} shows the period ratio of neighboring planets of any mass and semi-major axis that formed in NG76 (blue line) and NGM10 (brown). An apparent difference is the number of planets in or close to mean-motion resonances seen as vertical jumps of the lines in Fig. \ref{fig:period_ratios_100}. The total number of planet pairs with both $P_1,P_2<\SI{300}{\day}$ and $M_1,M_2>\SI{0.1}{M_\oplus}$ close to integer period ratios (within \SI{2}{\percent}) is \SI{35.5}{\percent} of the planets in the \SI{0.1}{M_{\odot}} population compared to \SI{27.8}{\percent} in NG76. This trend would not have been recovered if we compared the \SI{0.1}{M_{\odot}} population NGM10 to the \SI{1.0}{M_{\odot}} one (NG75) with also 50 initial embryos but an initially larger separation between them measured in mutual Hill radii. There, in NG75, a larger number of planet pairs is close to mean-motion resonances (\SI{47.0}{\percent}) than in NGM10. The trend of fewer planets close to mean-motion resonances with increasing number of embryos was already found by \citet{Alibert2013} by comparing simulations with up to 20 embryos and more recently \citet{Mulders2019} discuss the period ratio distribution of synthetic planets compared to observed systems.

An interesting aspect of the mean-motion resonant chains is that there are a few planet pairs in second-order mean-motion resonance. Second order resonances are less frequently produced in non-eccentric systems and need to be more dynamically excited than what results from low-eccentricity type I migration into resonant chains \citep[see e.g.][]{Coleman2019}. In our simulations, a difference between the stellar masses is already present at the end of the disk lifetime: The \SI{0.1}{M_{\odot}} population includes 133 (\SI{2.8}{\percent}) pairs close to 5:3 period ratios with $P_1,P_2<300$ and $M_1,M_2>\SI{0.1}{M_\oplus}$ whereas there are 149 (\SI{1.6}{\percent}) such pairs in the \SI{1.0}{M_{\odot}} population at this early time. With the disappearance of the damping gas, a reduction in the number of systems in mean-motion resonances occurs \citep{Izidoro2017} and the numbers go down to 78 (\SI{2.1}{\percent}) and 91 (\SI{2.3}{\percent}) for \SI{0.1}{M_{\odot}} and \SI{1.0}{M_{\odot}} respectively. In this context, we recall that the \Tra system planets d and c are close to 5:3 mean-motion resonance, which is rare in the simulations (\SI{3.8}{\percent} of the pairs for planet pairs with masses larger than \SI{0.1}{M_\oplus} and semi-major axes smaller than \SI{0.1}{au}). Because there are only upper limits on the occurrence rates of systems similar to TRAPPIST-1 \flrev{available} \citep{Sestovic2020}, it is not yet possible to conclude if our simulated results concerning mean-motion resonances for very-low-mass stars agree with observations.

Figure \ref{fig:period_ratios_100} additionally shows the observed \textit{Kepler} multi-planetary systems period ratio for reference. We note that the simple cut in masses at \SI{0.1}{M_\oplus} and periods at \SI{300}{days }  that we apply for the synthetic systems is not very well suited for the comparison of the period ratios to the observed systems. Some planetary pairs that are more distant from the star compared to the observed population are included in the synthetic data. Applying the realistic bias from \textit{Kepler} will be addressed in a future study (using the technique of the companion paper \citealp{Mishra2021}).

\begin{figure}
    \centering
    \includegraphics[width=\linewidth]{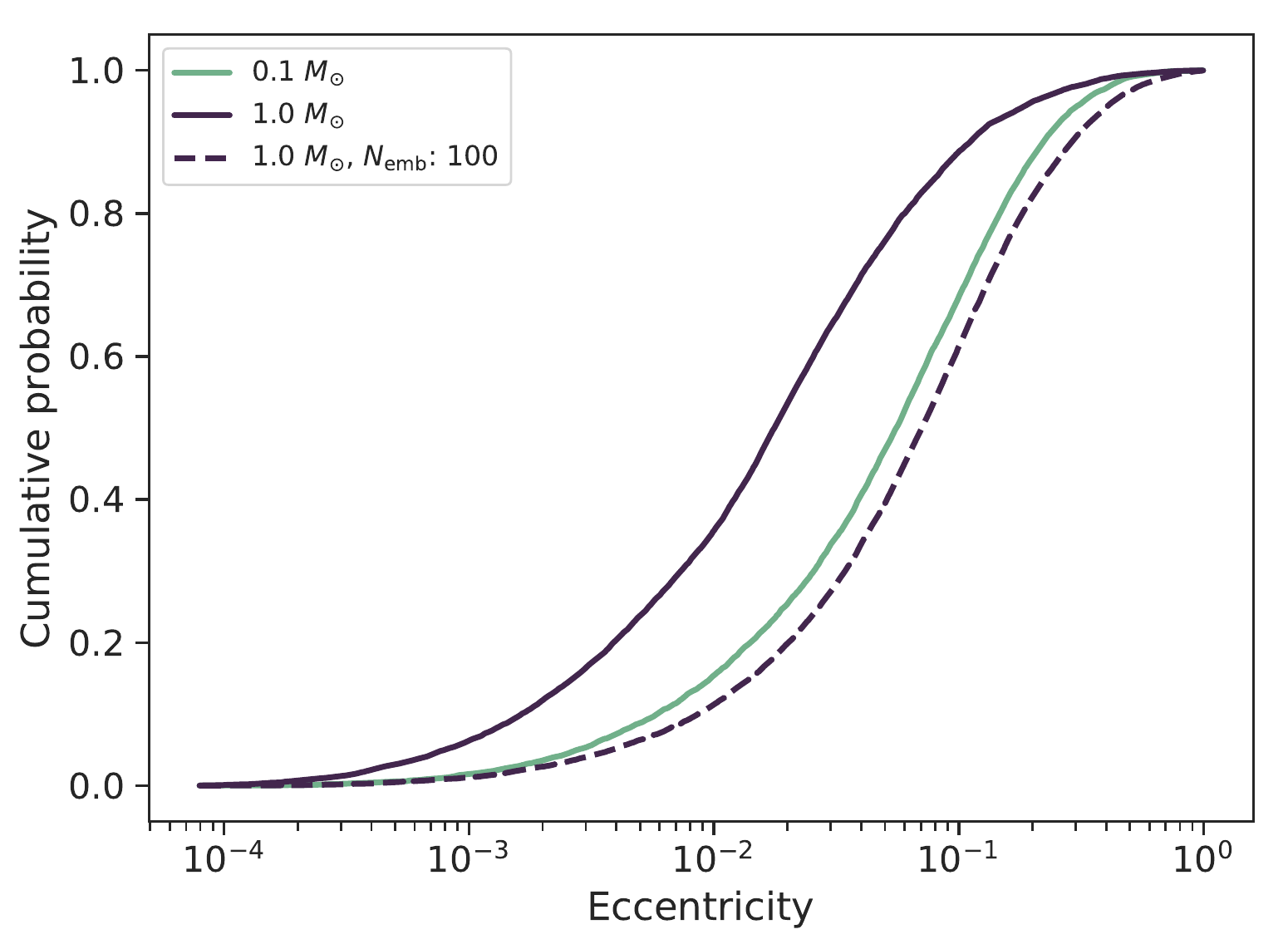}
    \caption{Cumulative distribution function of the planetary eccentricities for synthetic populations NGM10 (\SI{0.1}{M_{\odot}}, 50 embryos, green), NG75 (\SI{1.0}{M_{\odot}}, 50 embryos, dark blue, solid), and NG76 (\SI{1.0}{M_{\odot}}, 100 embryos, dashed, dark blue). Only planets with masses above \SI{0.1}{M_\oplus} are included.}
    \label{fig:cdf_e_50_100}
\end{figure}

The last dynamical property we discuss in this work is the eccentricity of the planetary orbits. Similarly to the discussion above, we can also attribute a large part of a decreasing eccentricity trend with stellar mass in the nominal population to the initial conditions. Indeed, Fig. \ref{fig:cdf_e_50_100} shows the nominal 0.1 and \SI{1.0}{M_{\odot}} populations with 50 embryos each (NGM10, NG75), but additionally the population with 100 embryos (NG76). It is apparent that increasing the number of embryos increases the eccentricity, which is due to an initially closer setting. Reducing the stellar mass while keeping the number of embryos fixed and at the same mass, had a similar effect. Comparing NG76 and NGM10 again, only small differences in eccentricities can be found, with a few highly eccentric planets around solar mass stars that are not present in NGM10. This can be attributed to the systems with giant planets, where the orbits of the smaller planets in the same system can become very eccentric.

Overall, the eccentricity distribution is more sensitive to the initial placement than other parameters. In contrast, it is quite insensitive with regard to the mass of the host star (see also Table \ref{tab:eccentricity}) as long as the initial separation stays similar.

\subsection{Solid mass reservoirs}
Planetesimal accretion is a process regulated by the eccentricities and inclinations of the planetesimals in the proximity of the growing protoplanet \citep{Ida1992a,Ida1992b,Inaba2001,Fortier2013}. For larger planetesimal eccentricities and inclinations, lower accretion rates result and if the planet becomes massive enough, we find that it can even eject a significant amount of planetesimals completely from the system. The region for which one protoplanet perturbs the planetesimal disk expands to a few tens of Hill radii. Therefore, indirect growth reduction for the less massive embryos in this region occurs. This is in line with the findings of \citet{Alibert2013}. One technical difference is that in this third generation of the model, the planetesimal surface density inside a feeding zone shared by several planets is calculated differently. It is not set to the mean value calculated over the whole zone but only to the mean value over the part attributed to one growing embryo (see \paperone{}). This is different from the treatment in \citet{Alibert2013}, where the former mean over the whole merged feeding zone was used. For the models presented here, this leads to a lesser extent of mass transport to the inner regions where planetesimals can be more easily accreted. Nevertheless, the results for both approaches are similar.

\begin{figure}
\centering
\includegraphics[width=\linewidth]{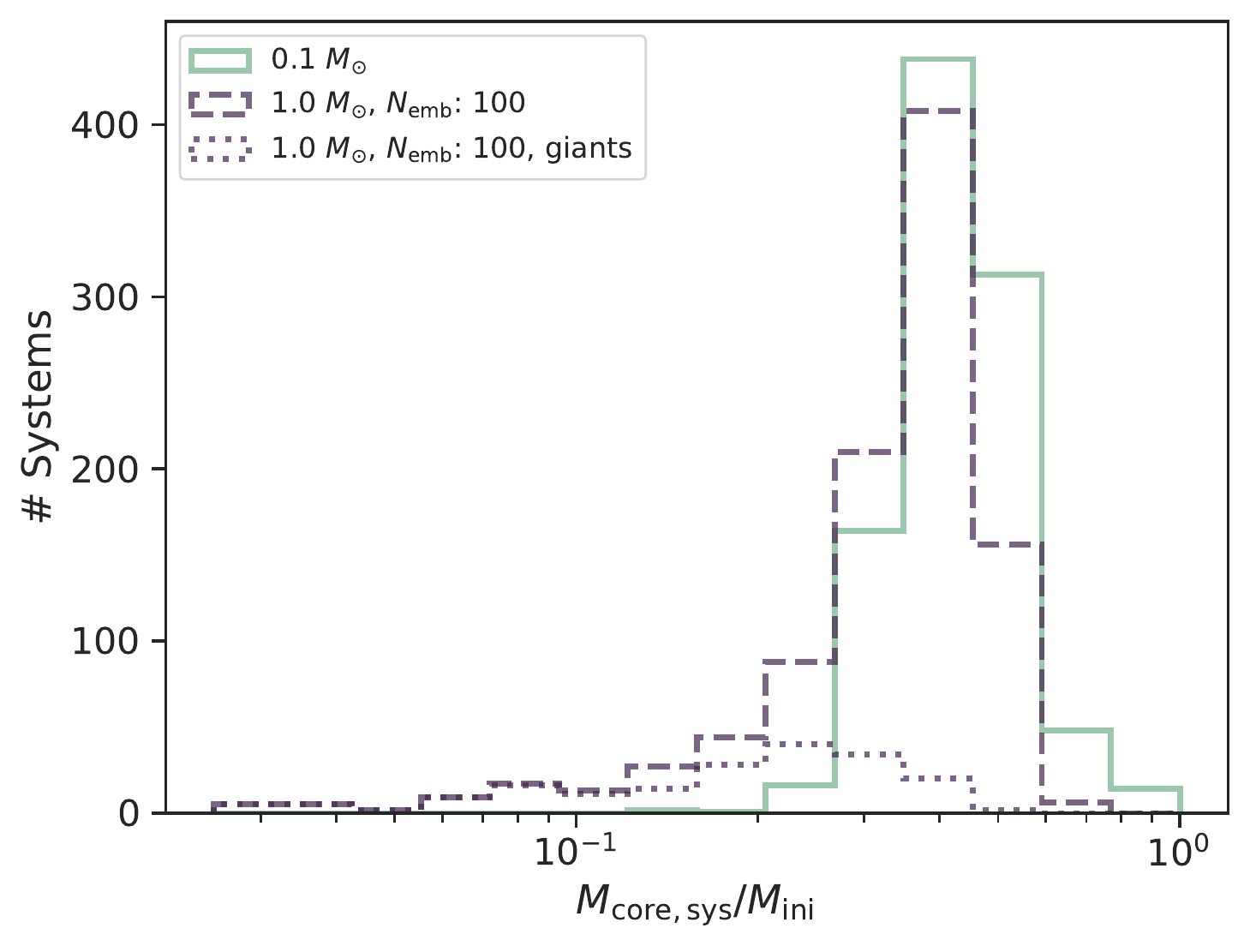}
\caption{Fraction of final solid mass content in planets in a system divided by the initial solid mass in the planetesimal disk. The light green line corresponds to the population around a \SI{0.1}{M_{\odot}} star with 50 embryos (NGM10) and the dashed, dark blue line to the population around a \SI{1.0}{M_{\odot}} star with 100 initial embryos (NG76). The dotted histogram shows the same quantity for the systems with at least one existing giant planet ($M > \SI{100}{M_\oplus}$) in NG76.}
\label{fig:hist_eff_solid}
\end{figure}
For an individual embryo, growth by solid accretion to masses above the classical isolation mass \citep{Lissauer1993} is commonly possible due to migration to non-depleted regions. In addition to the competition for and excitation of planetesimals, this makes an analytic treatment of solid accretion even more difficult. Therefore, we investigate the ratio of the sum of the solids in planetary cores to the solids initially inserted into the planetesimal disk of a system (Fig. \ref{fig:hist_eff_solid}). We term this the efficiency of solid accretion.

In Fig. \ref{fig:hist_eff_solid}, a tail toward very low efficiencies is found in the population around a \SI{1.0}{M_{\odot}} star, which can be attributed to systems with at least one giant planet (dotted line). The overall bulk of the distribution without giant planets peaks at a similar location in the two populations. The reason for the lower apparent efficiency for the systems with a giant planet is mainly ejection of planetary embryos \citep[see also][]{Schlecker2020}. In NG76 ejection of planets occurred in 219 systems. The mean mass of ejected planets was \SI{46.7}{M_\oplus} per system with ejection. In contrast to that, in NGM10 ejection of planets occurred in only 38 systems with a mean mass of ejected planets of \SI{0.27}{M_\oplus}, which is less relative to the initial mass of solids.

From these results, we can qualitatively expect that free-floating planets \citep[e.g.,][]{Sumi2011} predominantly originate  from systems around stars of higher mass with more massive disks if they are mainly produced by planet-planet scattering (see, e.g., \citealp{Veras2012} and \citealp{Stock2020} for discussions on their formation). The dependency on stellar mass is not linear due to gas accretion: Systems with giant planets that underwent rapid gas accretion eject much more planetary mass to interstellar space than systems where no significant gas accretion occurred. Additionally, we find the bulk of the ejected mass to be in the form of embryos and not in the form of ejected planetesimals\footnote{The analytical treatment of planetesimals might influence this result although there are two modes of planetesimal-ejection mechanisms included due to either viscous stirring of eccentricities above unity by planets \citep{Ohtsuki2002} or close encounters \citep{Ida2004}.}. This finding is again a function of the number, spacing, and mass of the initially placed embryos and should be addressed in detail in future works.

\subsection{The case of TRAPPIST-1}
\label{sec:trappist}
\begin{figure}
        \centering
        \includegraphics[width=.98\linewidth]{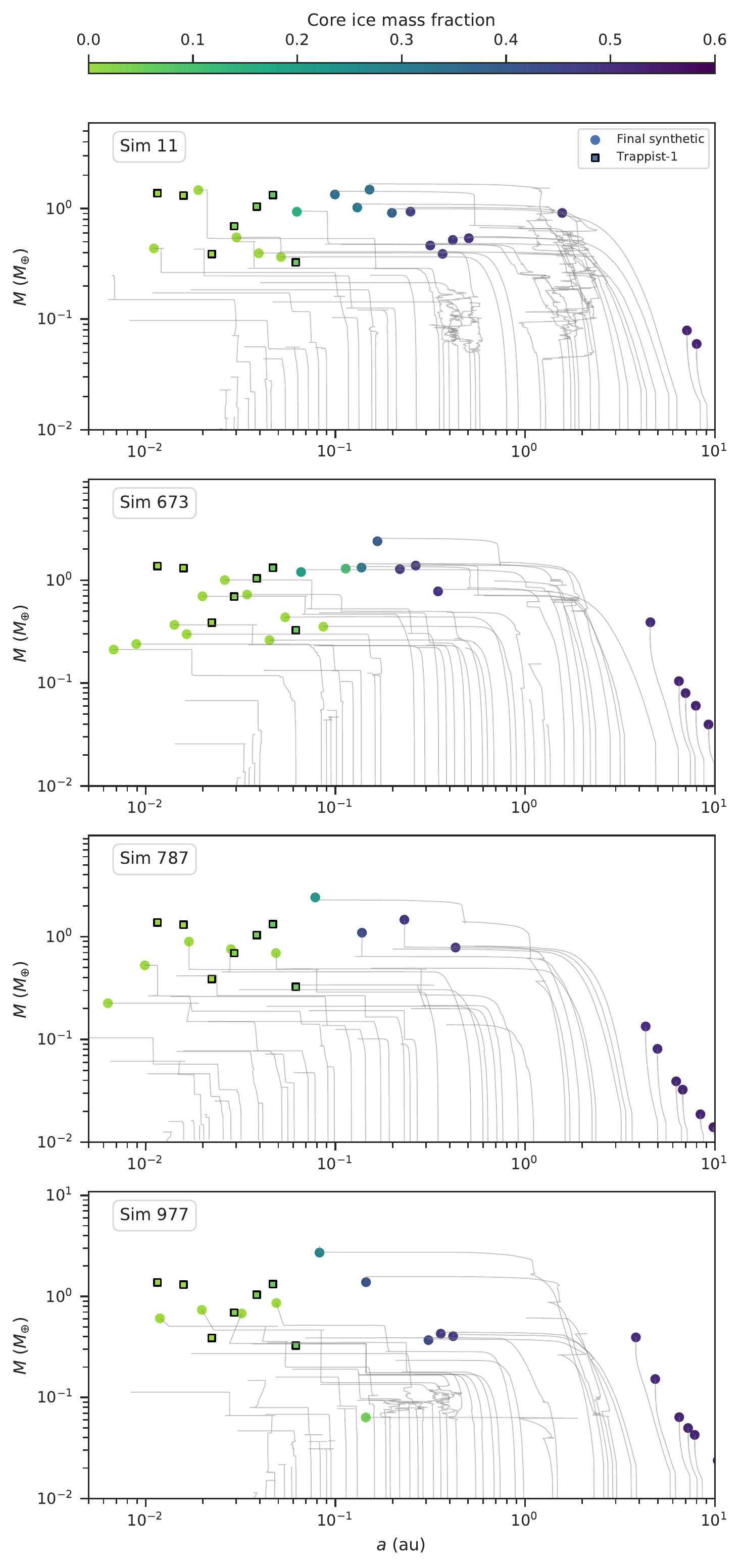}
        \caption{Synthetic systems with final planetary masses most similar to the Trappist-1 system. The mass difference to the observed masses \citep{Agol2020} are \SIlist{-0.17;0.14;-0.15;0.00}{M_\oplus}. Thin lines denote the evolution over time of individual planets. If they stop without ending in a colored dot, the corresponding embryo was ejected or accreted by a more massive planet. Despite closely matching in total mass, significant differences between systems are visible.}
        \label{fig:t1_examples}
\end{figure}

\begin{figure}
        \centering
        \includegraphics[width=\linewidth]{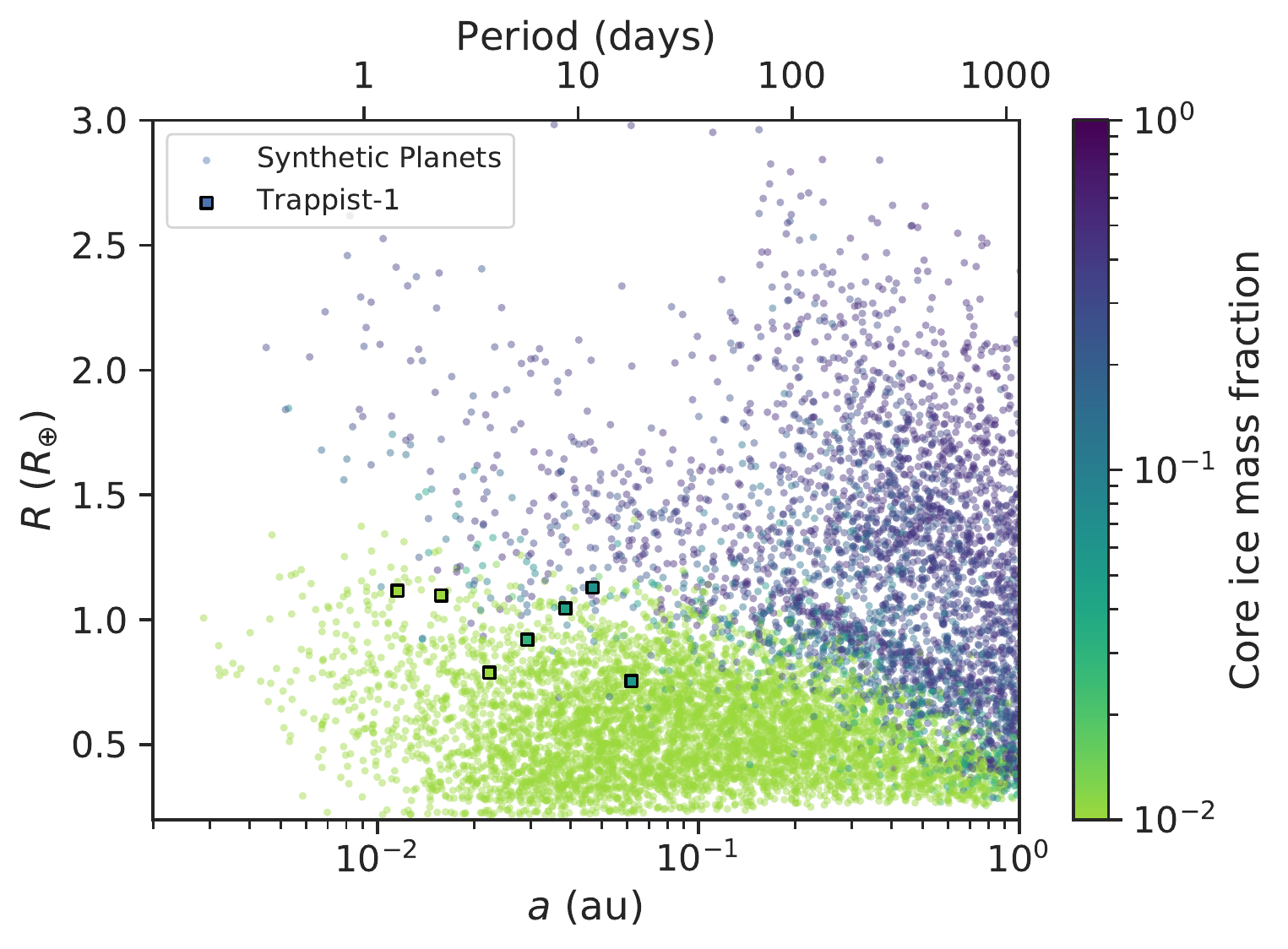}
        \caption{Orbital distance, planetary radius diagram of synthetic planets around a \SI{0.1}{M_{\odot}} star. For comparison, the Trappist-1 system \citep{Agol2020} and the subsample of synthetic planets around them is shown. The core ice mass fraction is here colored using a logarithmic scale for better visibility of the low values derived assuming a core iron mass fraction of \SI{32.5}{\percent} in Trappist-1.}
        \label{fig:a_R_t1}
\end{figure}

The population of synthetic planets around an \SI{0.1}{M_{\odot}} star (NGM10) is well suited for comparison to the \Tra system \citep{Gillon2016, Gillon2017}. \Tra is an ultra-cool dwarf star with an estimated mass of \SI{0.089 \pm 0.006}{M_{\odot}} \citep{Grootel2018}. The system is unique due to the high number of detected transiting planets and their mass constraints \citep{Grimm2018,Agol2020}. Thus, \Tra provides a unique opportunity to benchmark planet formation models against \citep[see also][]{Ormel2017, Alibert2017,Coleman2019,Schoonenberg2019, Miguel2020}.

The TTV fits for the \Tra system exclude additional planets with significant mass for at least the region within \SI{0.1}{au} \citep{Grimm2018}. Therefore, it makes sense to investigate the properties of the synthetic planets with masses above \SI{0.1}{M_\oplus} within \SI{0.1}{au} (here dubbed the \Tra region) in the synthetic \SI{0.1}{M_{\odot}} population NGM10 and compare them with what we know for the \Tra planets. 

\subsubsection{Example systems}
As a first pathway to explore systems similar to TRAPPIST-1, we show a selection in Fig. \ref{fig:t1_examples}. The systems were chosen based on the smallest differences in total mass in the \Tra region (see Sect. \ref{ssec:t1_initial_cond}). Even though the total mass incorporated in the region is similar in all these example systems, very different patterns are visible.

The initial solid mass is not distributed evenly in the initial systems. Our chosen slope of $\beta_{\rm s}=-1.5$ results in isolation masses that increase with semi-major axis with a slope of $0.5$. This could still be visible in the final systems despite the simple arguments that are used to calculate the planetesimal isolation mass \citep{Kokubo2000}. Whereas in the lower three panels of Fig. \ref{fig:t1_examples}, there are hints of an increasing mass gradient with semi-major axis, no such trend can be recognized in simulation 11 and in the \Tra system itself. While not always the case, simulation 11 is an example that collisions and migration can erase the memory of the initial solid mass content at the starting location of the embryos.

The number of planets in the region differs significantly between the systems; simulation 11 and 787 are with seven and six planets close to the seven observed planets. In contrast, simulation 673 produced an astonishing 11 planets in the same region. Due to the short orbital periods, it is unlikely that the \SI{20}{Myr} of numerical N-body integration is not sufficient to reveal if that system is dynamically unstable. Therefore, such systems are at least theoretically possible. It is not clear what distinguishes simulation 673 from the other systems. The inner edge of the disk lies at \SI{0.29}{au} compared to \SI{0.24}{au}, \SI{0.012}{au}, and \SI{0.36}{au} for simulations 11, 787, and 977 respectively. It might just be a coincidence that a last series of collisions due to a system-wide instability did not occur in simulation 673; thus, a large number of planets was retained.

A common occurrence are more ice-rich planets outside the observable region. As embryos are assumed to have the capacity to form everywhere in the disk, this is expected and should be interpreted with this assumption in mind. We also note that all these systems have undergone collisions between embryos in the inner system. In our simulations, these collisions lead to perfect merging, but in reality, fragmentation would occur and typically decrease the resulting planetary mass and enhance the compositional diversity \citep{Emsenhuber2020}.

\subsubsection{Statistics of synthetic planets inside \SI{0.1}{au}}

In our model results, consisting of a total of 1000 systems, typically only a few planets grow to \SI{0.1}{M_\oplus} masses. The most common number of planets in the \Tra region ($M>\SI{0.1}{M_\oplus}$, $a<\SI{0.1}{au}$) is 1, which occurs in 330 systems. In 182 systems, not a single planet is in the \Tra region and only \SI{20}{\percent} of the systems do have more than three planets there. This low number is mainly due to little growth in many systems and further reduced by evolutionary paths that led to a single massive close-in planet that accreted all the other embryos. This second scenario is common in disks with an above-average initial solid mass and supported by fast type I migration. If the migration speed was reduced, for example by a lower viscous $\alpha$ or due to wind-driven accretion \citep{Ogihara2018}, there would be fewer systems where the growing embryos are forced by strong disk torques into each others proximity, which then leads to fewer close encounters and collisions.

Figure \ref{fig:a_R_t1} shows the population of planets in NGM10 in semi-major axis, radius space. The zoom into the region where the \Tra planets lie reveals more clearly that the density of synthetic planets drops toward the star as inner edges of disks become comparable to the location of \Tra b. Additionally, in terms of rocky planets, the \Tra planets populate the parameter space of the largest such bodies. This is consistent with the observational bias as transit surveys are most sensitive to the largest planets.

\begin{figure}
\centering
\includegraphics[width=\linewidth]{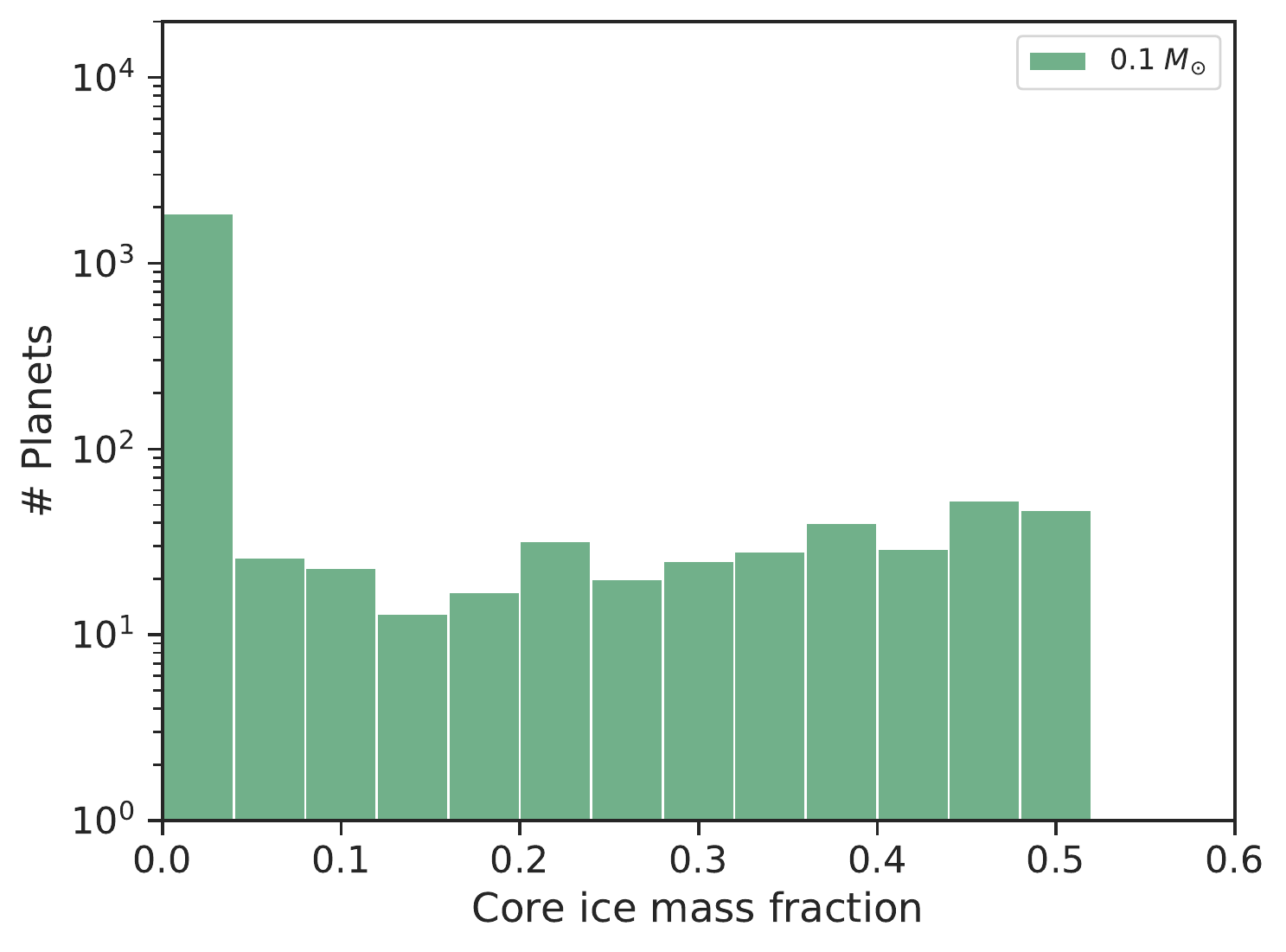}
\caption{Histogram of the core ice mass fractions of the 2222 synthetic planets in the \Tra region ($M>\SI{0.1}{M_\oplus}$, $a<\SI{0.1}{au}$). The shown population of synthetic planets is calculated around a \SI{0.1}{M_{\odot}} star with 50 embryos (NGM10). Most planets are purely rocky, but 314 (\SI{14}{\percent}) show ice fractions larger than \SI{0.1}{}.}
\label{fig:ice_fraction_T1}
\end{figure}

Moving on to the statistics of compositions, the ice mass fraction in the cores of synthetic planets in the \Tra zone are shown in Fig. \ref{fig:ice_fraction_T1}. Currently, the \Tra planets are considered to be quite rocky \citep{Dorn2018,Grimm2018}, even more so in the recent results of \citet{Agol2020}, who find ice mass fractions below \SI{1}{\percent} for low-iron core masses (\SI{18}{\percent}), below \SI{7}{\percent} for Earth-like core mass fractions (\SI{32.5}{\percent}), and still below \SI{14}{\percent} for large iron cores (\SI{50}{\percent}).

Additionally, \citet{Agol2020} find for the updated masses that the ice mass fraction increases with orbital period within error bars throughout the whole \Tra system. This is in much better agreement with theoretical results compared to previous derived compositions \citep{Dorn2018}. Such an outcome is expected from classical planet formation theory and in agreement with the Solar System.

One aspect that is less intuitive is the prevalence of more ice-rich planets with increasing mass due to planetary migration (the "horizontal branch" in Fig. \ref{fig:aM_all}, see also Sect. \ref{sec:m-a} and \citealp{Ida2005,Mordasini2009,Alibert2011,Alibert2017}). From the data in \citet{Agol2020}, it is not yet clear if an increase in the ice mass fraction with planetary mass is also present in the \Tra system. One indication could come from the lower inferred ice mass fraction for the outermost planet h compared to its neighbor g. However, the uncertainties are much too large to confirm the trend of larger ice mass fractions for planets with masses above \SI{1}{M_\oplus}, which we show in Fig. \ref{fig:aM_all}.

\subsubsection{Initial condition regime}
\label{ssec:t1_initial_cond}
\begin{figure}
        \includegraphics[width=\linewidth]{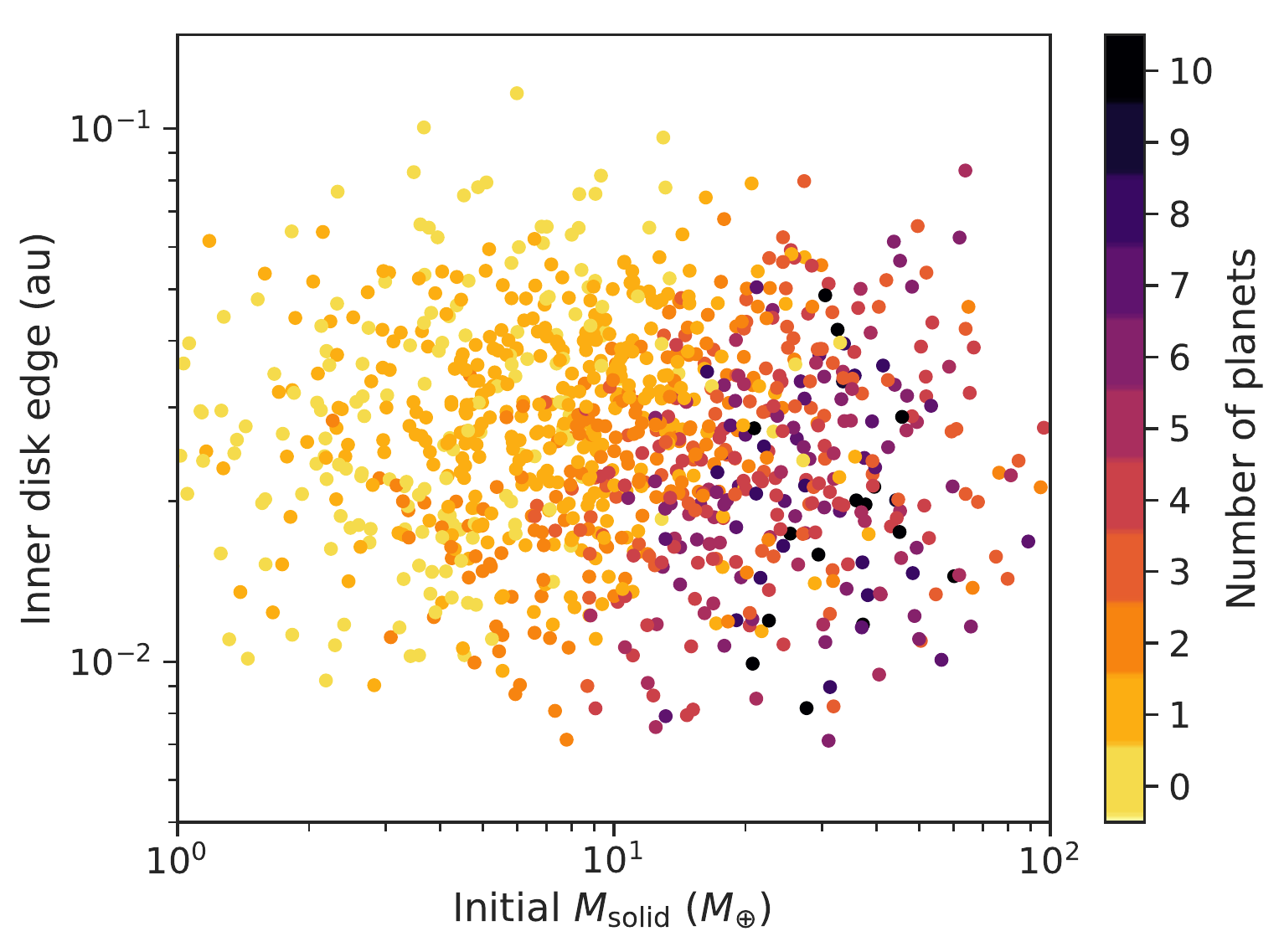}
        \caption{Disk initial conditions color-coded by multiplicity. We plot only the planets in the \Tra region, i.e., with semi-major axis less than \SI{0.1}{au} and mass larger than \SI{0.1}{M_\oplus}. }
        \label{fig:Np_initial_cond_T1}
\end{figure}

\begin{figure}
        \includegraphics[width=\linewidth]{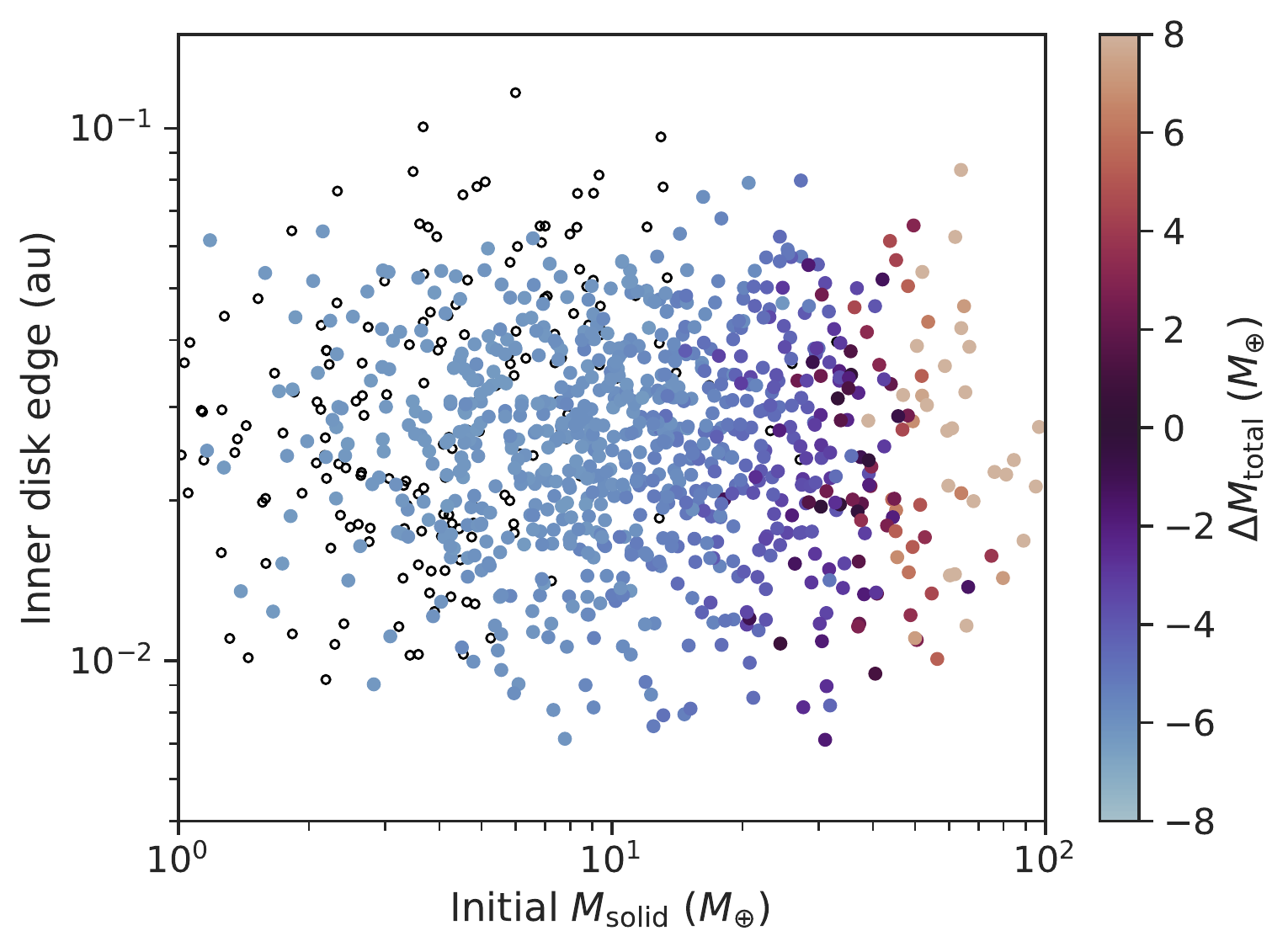}
        \caption{Disk initial conditions color-coded by difference of total mass to the \Tra system. Individual planet masses are only included if their semi-major axis is less than \SI{0.1}{au} and their mass is larger than \SI{0.1}{M_\oplus}. Systems with no planets lying in this regime are drawn as empty black circles. The darkest points show systems most \Tra similar at around \SI{30}{M_\oplus} of initial solid mass.}
        \label{fig:Mtotal_initial_cond_T1}
\end{figure}

To compare the simulated systems with TRAPPIST-1, we chose to take the number $N_\mathrm{p}$ and the difference to \Tra in total mass $\Delta M_{\rm total}$  \citep[where $M_{\rm total}$ of \Tra is 
\SI{6.45\pm0.10}{M_\oplus},][]{Agol2020}
. For both, we only take planets with masses $M>\SI{0.1}{M_\oplus}$ and semi-major axes $a<\SI{0.1}{au}$ into account. This simple approach is a first exploration of the similarity of \Tra and the synthetic data in terms of the most fundamental parameters. A more complex similarity criterion of systems will be applied in future works following \citet{Alibert2019b}. The colors of Figs. \ref{fig:Np_initial_cond_T1} and \ref{fig:Mtotal_initial_cond_T1} display $N_\mathrm{p}$ and $\Delta M_{\rm total}$, where one point corresponds to a synthetic system. It is interesting to show them as a function of the initial solid mass and the inner disk edge to determine the most likely initial parameters of the disk from which the \Tra planets formed. For both, $N_\mathrm{p}$ and $\Delta M_{\rm total}$, there is a clear correlation with the initial solid mass, whereas it is only for the number of planets that there seems to be a correlation with the inner disk edge.

For $\Delta M_{\rm total}$, this can be explained by the mass transport via type I migration. The migration rates and thus the solid mass flux to the inner region is independent of the inner disk edge. Thus, a similar fraction of the total solid mass in the disk is transported inside \SI{0.1}{au} in all disks. The initial mass in planetesimals inside \SI{0.1}{au} only varies negligibly with the inner edge.

However, this mass can be distributed very differently to the planets. In a number of disks at the high-end tail of the solid masses ($M_{\rm solid} > \SI{60}{M_\oplus}$), there are only one to four oligarchs that grew by accreting the rest of the embryos. Due to the higher initial solid mass, larger planets form which correspondingly have bigger Hill spheres and interact more often gravitationally with the other embryos. Consequently, they can be ejected or accreted by other planets, where both outcomes lead to a smaller number of planets in a given zone. Therefore, Fig. \ref{fig:Np_initial_cond_T1} shows a peak in $N_\mathrm{p}$ inside the \Tra zone which is located at moderate, but above-average disk solid masses ($\SI{10}{M_\oplus} < M_{\rm solid} < \SI{60}{M_\oplus}$). At lower than average disk solid masses, the number of planets decreases. This is due to planets not growing to masses above \SI{0.1}{M_\oplus}. Thus, they are not identified to lie in the \Tra zone and additionally, very little transport of solid material due to type I migration takes place.

Overall, one sweet-spot for the formation of a system with a similar amount of planetary mass at comparable semi-major axes as \Tra can be located at $ \SI{30}{M_\oplus} \lesssim M_{\rm solid}  \lesssim \SI{50}{M_\oplus} $, which coincides with the region where seven planets inside \SI{0.1}{au} are frequently present given an inner disk edge below at least \SI{0.06}{au}. These values are larger than the nominal values of \SI{13.3}{M_\oplus} and  \SI{18.3}{M_\oplus}, respectively, applied in the models of \citet{Ormel2017} and \citet{Schoonenberg2019}.

\subsection{\frev{Comparison with previous works}}
\frevi{
We now briefly discuss our results in context of other theoretical works.} \frev{As there are a number of works focusing on ultra-late M dwarfs of \SI{0.01}{M_{\odot}}, this will be the focus of the comparison. In Sect. \ref{ssec:giant_planet_discussion}, we covered the other major branch of works focusing on giant planet formation.}

\frevi{We predominantly find a large rocky population of planets making up about \SI{90}{\percent} of the planets and a smaller, more massive, water-bearing population in the \Tra region (see Fig. \ref{fig:ice_fraction_T1}). This is due to a large amount of accretion inside the water iceline, similarly to what is found in \citet{Schoonenberg2019} but in contrast to the findings of \citet{Alibert2017}, \citet{Miguel2020}, and \citet{Coleman2019} who find mostly water-rich planets. \citet{Coleman2019} assumed an insignificant amount of rocky planetesimals initially, while \citet{Miguel2020} have computed a snowline that lies much closer to the star than in our work. This is due to the differing disk temperature calculations: To estimate disk temperatures, \citet{Miguel2020} use gas accretion rates  $\dot{M}_{\rm acc}$ on the order of \SI{e-10}{M_{\odot}/yr,} which corresponds to a later, cooler stage in the disk evolution compared to our assumed planetesimal formation at the beginning of our simulations (where $\dot{M}_{\rm acc} \simeq \SI{e-9}{M_{\odot}/yr}$). Additionally, the disks presented here are assumed to be twice as turbulent ($\alpha = \SI{2e-3}{}$) compared to the disks in \citet{Miguel2020} ($\alpha = \SI{1e-3}{}$) leading to hotter disks. For these two reasons, \citet{Miguel2020} start their simulation with much fewer rocky planetesimals in comparison to our simulations.}

\frevi{
The best correlation with our model can be found in the work of \citet{Alibert2017}. Our results differ from theirs for three different reasons: (1) \citet{Alibert2017} found that for more massive disks, more rocky planets appear, and they scale the disk mass to the power of 1.2 with the stellar mass. Their "heavy" disk case lies closer to our mass distribution and already produces $\sim$\SI{20}{\percent} rocky planets. (2) We ran simulations with 50 embryos, compared to \citet{Alibert2017}, who ran their simulations with 10 embryos. More embryos lead to more potential accretion close to the star, since the sum of all feeding zone increases, which is the limiting factor for growth in the inner region \frev{in the initial stages}. (3) The slope of the planetesimal surface density differs in the two simulations. While we use a slope of $\beta_\mathrm{pls}=-1.5$, \citet{Alibert2017} used a shallower slope of -0.9. Therefore, by construction, more rocky material is available in our new set of simulations. This third point is an additional relevant difference to the work of \citet{Miguel2020}. Current planetesimal formation models favor steeper slopes, but potentially lower planetesimal formation within the water iceline \citep{Drazkowska2017,Lenz2019,Voelkel2020}.}

\frevi{
As of now, no clear conclusion should be drawn concerning the validity of the pathway of rocky planet formation by almost in situ accretion, which has been seen to be dominant in our work. Due to the moving water iceline, the resulting planetary composition is a strong function of the location and timing of planetesimal formation. This will be incorporated in our model following \citet{Voelkel2020} in the future. For now, we show that if enough planetesimals are assumed to have formed within the water iceline, rocky planets will form directly consistent with the \Tra planets. Otherwise, to reproduce the observed population of rocky planets, the planetesimals or the planets have to desiccate by some additional process (e.g., due to radioactive heating of planetesimals as in \citealp{Lichtenberg2019} or by ablation of pebbles in the planetary envelopes as discussed by \citealp{Coleman2019}).
}

\section{Summary and conclusions}
\label{sec:conclusion}
In this part of the New Generation Planetary Population Synthesis (NGPPS) series, we employed the Generation III Bern model of planet formation and evolution, introduced in \paperone,{} to explore the influence of the stellar mass. We calculated $\sim$1000 synthetic multi-planet systems for a grid of stellar masses (\SIlist{0.1;0.3;0.5;0.7;1.0}{M_{\odot}}) and with an initial 50 planetary embryos per system.
While we linearly scaled the gas and solid disk mass with stellar mass, we assumed physical disk boundaries constant in orbital period and kept the disk lifetime fixed.

This yields a data set for which we found:

\begin{itemize}
\item A larger number of giant planets with larger stellar mass. In particular, no giant planets formed for $M_{\star} < \SI{0.5}{M_{\odot}}$.
\item The most frequent temperate (potentially habitable) planet host to be M dwarfs with masses of \SIrange{0.3}{0.5}{M_{\odot}}. Additionally, we find a strong metallicity dependence of the temperate planet occurrence rate at low stellar masses. Therefore, planet searches aiming for temperate, terrestrial planets around low-mass stars should target metal-rich objects.
\item  The planetary mass function does not shift in a strictly linear fashion with the stellar mass despite the linear scaling of the gas and solid disk mass. This is due to more ejected planets for higher stellar masses because of the growth of giant planets.
\item As consequences of the previous point, there is a reduced apparent efficiency of solid accretion toward higher stellar mass due to ejection of planetary embryos. Directly related to this, we find that planets ejected by planet-planet scattering originate from stars with masses of at least a solar mass.
\item A strong dependency of the dynamical properties such as period ratios and eccentricities on the initial proximity of the embryos measured in mutual Hill radii (initial spacing). Using this measurement, the systems are more compact for higher stellar masses at the end of the simulations.
\item A high occurrence of mean-motion resonances due to migration, which is in contrast to the lower number of observed resonant multi-planetary systems. For a similar initial spacing, $\sim$\SI{10}{\percent} more pairs within \SI{300}{\day} orbital periods are in resonance around ultra-late M-dwarfs compared to solar-like stars.
\item A sweet spot in terms of initial solid mass content (\SI{30}{} to \SI{50}{M_\oplus}) and inner disk edge (closer than \SI{0.06}{au}) to get a system like \Tra, as measured by the observable mass.
\item A clear predominance of rocky compositions for planets in the innermost \SI{0.1}{au} of low-mass stars due to enhanced rocky planetesimal accretion. This is a strong function of how many planetesimals and embryos were initially placed closer to the star than the water iceline.
\item A pathway for the formation of giant planets around very low-mass stars via core accretion: While it is sufficient to reduce type I migration, even more giants form if planetesimals are initially more concentrated toward the star.
\end{itemize}

\frev{Despite the advancements in recent years, there are a number of limitations to the model. Due to computational cost, it was not possible to study the influence of different model parameters (e.g., the viscous $\alpha$ parameter, planetesimal size, or envelope opacity) or initial condition scalings (like the disk mass). Of particular importance is the assumed planetesimal size of \SI{600}{\meter} in diameter, which is at odds with planetesimal formation predictions \citep{Johansen2012,Li2019,Klahr2020}. However, in the Kuiper Belt, there could have existed a numerous population of smaller planetesimals (\citealp{Schlichting2013, Kenyon2020}, see also \citealp{Arimatsu2019,Morbidelli2021}). We expect that for all stellar masses, an increase in the planetesimal size would increase the solid accretion timescale and thus decrease the number and mean mass of the planets \citep{Fortier2013}. Similarly, larger dust opacities in the planetary envelopes slow down cooling and thus gas accretion \citep{Mordasini2014}. We also see considerable potential for improvement in the treatment of the equation of state for water \citep{Haldemann2020} which influences the radius distribution, the formation location of embryos and planetesimals \citep{Voelkel2020,Voelkel2021}, and the accretion of pebbles \citep{Brugger2018,Brugger2020}.}

\frev{Keeping in mind these limitations, our} analysis quantifies, global trends in multiplanet systems as a function of stellar mass, which have previously been studied only in models with a narrower focus.
A major benefit to this work is the availability of enough statistical data and the ability to make use of global simulations of planetary systems to find these aspects in a single, coherent theoretical framework. The results compare reasonably well to de-biased observational data where available. In the future, these comparisons should be examined in greater detail.

\begin{acknowledgements}
Parts of this work have been carried out within the frame of the National Centre for Competence in Research PlanetS funded by the Swiss National Science Foundation (SNSF). R.B. and Y.A. acknowledge the financial support from the SNSF under grant 200020\_172746.
This work was supported by the DFG Research Unit FOR2544 “Blue Planets around Red Stars”, project no. RE 2694/4-1.
A.E. acknowledges the support from The University of Arizona.
A.E. and C.M. acknowledge the support from the SNSF under grant BSSGI0\_155816 ``PlanetsInTime''.
This research has made use of the NASA Exoplanet Archive, which is operated by the California Institute of Technology, under contract with the National Aeronautics and Space Administration under the Exoplanet Exploration Program.
The plots shown in this work were generated using \textit{matplotlib} \citep{Hunter2007} and \textit{seaborn} (\href{https://seaborn.pydata.org/index.html}{https://seaborn.pydata.org/index.html}).
\end{acknowledgements}

\bibliographystyle{aa}
\bibliography{library_update_dec21,library_martin}

\begin{thebibliography}{227}
\expandafter\ifx\csname natexlab\endcsname\relax\def\natexlab#1{#1}\fi

\bibitem[{Adachi {et~al.}(1976)Adachi, Hayashi, \& Nakazawa}]{Adachi1976}
Adachi, I., Hayashi, C., \& Nakazawa, K. 1976, Progress of Theoretical Physics,
  56, 1756

\bibitem[{Adams {et~al.}(2004)Adams, Hollenbach, Laughlin, \&
  Gorti}]{Adams2004}
Adams, F.~C., Hollenbach, D., Laughlin, G., \& Gorti, U. 2004, ApJ, 611, 360

\bibitem[{Affer {et~al.}(2013)Affer, Micela, Favata, Flaccomio, \&
  Bouvier}]{Affer2013}
Affer, L., Micela, G., Favata, F., Flaccomio, E., \& Bouvier, J. 2013, MNRAS,
  430, 1433

\bibitem[{Agol {et~al.}(2021)Agol, Dorn, Grimm, Turbet, Ducrot, Delrez, Gillon,
  Demory, Burdanov, Barkaoui, Benkhaldoun, Bolmont, Burgasser, Carey, de~Wit,
  Fabrycky, Foreman-Mackey, Haldemann, Hernandez, Ingalls, Jehin, Langford,
  Leconte, Lederer, Luger, Malhotra, Meadows, Morris, Pozuelos, Queloz,
  Raymond, Selsis, Sestovic, Triaud, \& Grootel}]{Agol2020}
Agol, E., Dorn, C., Grimm, S.~L., {et~al.} 2021, The Planetary Science Journal,
  2, 1

\bibitem[{Alcal{\'{a}} {et~al.}(2017)Alcal{\'{a}}, Manara, Natta, Frasca,
  Testi, Nisini, Stelzer, Williams, Antoniucci, Biazzo, Covino, Esposito,
  Getman, \& Rigliaco}]{Alcala2017}
Alcal{\'{a}}, J.~M., Manara, C.~F., Natta, A., {et~al.} 2017, A{\&}A, 600, A20

\bibitem[{Alessi {et~al.}(2020)Alessi, Pudritz, \& Cridland}]{Alessi2020}
Alessi, M., Pudritz, R.~E., \& Cridland, A.~J. 2020, MNRAS, 493, 1013

\bibitem[{Alexander {et~al.}(2014)Alexander, Pascucci, Andrews, Armitage, \&
  Cieza}]{Alexander2014}
Alexander, R., Pascucci, I., Andrews, S.~M., Armitage, P., \& Cieza, L. 2014,
  in Protostars Planets VI, ed. H.~Beuther, R.~S. Klessen, C.~P. Dullemond, \&
  T.~Henning (University of Arizona Press), 475

\bibitem[{Alibert(2019)}]{Alibert2019b}
Alibert, Y. 2019, A{\&}A, 624, A45

\bibitem[{Alibert \& Benz(2017)}]{Alibert2017}
Alibert, Y. \& Benz, W. 2017, A{\&}A, 598, L5

\bibitem[{Alibert {et~al.}(2013)Alibert, Carron, Fortier, Pfyffer, Benz,
  Mordasini, \& Swoboda}]{Alibert2013}
Alibert, Y., Carron, F., Fortier, A., {et~al.} 2013, A{\&}A, 558, A109

\bibitem[{Alibert {et~al.}(2004)Alibert, Mordasini, \& Benz}]{Alibert2004a}
Alibert, Y., Mordasini, C., \& Benz, W. 2004, A{\&}A, 417, L25

\bibitem[{Alibert {et~al.}(2011)Alibert, Mordasini, \& Benz}]{Alibert2011}
Alibert, Y., Mordasini, C., \& Benz, W. 2011, A{\&}A, 526, A63

\bibitem[{Alibert {et~al.}(2005)Alibert, Mordasini, Benz, \&
  Winisdoerffer}]{Alibert2005}
Alibert, Y., Mordasini, C., Benz, W., \& Winisdoerffer, C. 2005, A{\&}A, 434,
  343

\bibitem[{Andrews {et~al.}(2013)Andrews, Rosenfeld, Kraus, \&
  Wilner}]{Andrews2013}
Andrews, S.~M., Rosenfeld, K.~A., Kraus, A.~L., \& Wilner, D.~J. 2013, ApJ,
  771, 129

\bibitem[{Andrews {et~al.}(2018)Andrews, Terrell, Tripathi, Ansdell, Williams,
  \& Wilner}]{Andrews2018}
Andrews, S.~M., Terrell, M., Tripathi, A., {et~al.} 2018, ApJ, 865, 157

\bibitem[{Andrews {et~al.}(2010)Andrews, Wilner, Hughes, Qi, \&
  Dullemond}]{Andrews2010}
Andrews, S.~M., Wilner, D.~J., Hughes, A.~M., Qi, C., \& Dullemond, C.~P. 2010,
  ApJ, 723, 1241

\bibitem[{Anglada-Escud{\'{e}} {et~al.}(2016)Anglada-Escud{\'{e}}, Amado,
  Barnes, Berdi{\~{n}}as, Butler, Coleman, de~la Cueva, Dreizler, Endl,
  Giesers, Jeffers, Jenkins, Jones, Kiraga, K{\"{u}}rster,
  L{\'{o}}pez-Gonz{\'{a}}lez, Marvin, Morales, Morin, Nelson, Ortiz, Ofir,
  Paardekooper, Reiners, Rodr{\'{i}}guez, Rodrίguez-L{\'{o}}pez, Sarmiento,
  Strachan, Tsapras, Tuomi, \& Zechmeister}]{Anglada-Escude2016}
Anglada-Escud{\'{e}}, G., Amado, P.~J., Barnes, J., {et~al.} 2016, Nature, 536,
  437

\bibitem[{Anglada-Escud{\'{e}} {et~al.}(2012)Anglada-Escud{\'{e}}, Boss,
  Weinberger, Thompson, Butler, Vogt, \& Rivera}]{Anglada-Escude2012}
Anglada-Escud{\'{e}}, G., Boss, A.~P., Weinberger, A.~J., {et~al.} 2012, ApJ,
  746, 37

\bibitem[{Ansdell {et~al.}(2017)Ansdell, Williams, Manara, Miotello, Facchini,
  van~der Marel, Testi, \& van Dishoeck}]{Ansdell2017}
Ansdell, M., Williams, J.~P., Manara, C.~F., {et~al.} 2017, The Astronomical
  Journal, 153, 240

\bibitem[{Ansdell {et~al.}(2018)Ansdell, Williams, Trapman, van Terwisga,
  Facchini, Manara, van~der Marel, Miotello, Tazzari, Hogerheijde, Guidi,
  Testi, van Dishoeck, van Terwisga, Facchini, Manara, {Van Der Marel},
  Miotello, Tazzari, Hogerheijde, Guidi, Testi, \& van Dishoeck}]{Ansdell2018}
Ansdell, M., Williams, J.~P., Trapman, L., {et~al.} 2018, ApJ, 859, 21

\bibitem[{Arimatsu {et~al.}(2019)Arimatsu, Tsumura, Usui, Shinnaka, Ichikawa,
  Ootsubo, Kotani, Wada, Nagase, \& Watanabe}]{Arimatsu2019}
Arimatsu, K., Tsumura, K., Usui, F., {et~al.} 2019, Nature Astronomy, 3, 301

\bibitem[{Artigau {et~al.}(2015)Artigau, Gagn{\'{e}}, Faherty, Malo, Naud,
  Doyon, Lafreni{\`{e}}re, \& Beletsky}]{Artigau2015}
Artigau, {\'{E}}., Gagn{\'{e}}, J., Faherty, J., {et~al.} 2015, ApJ, 806, 254

\bibitem[{Baraffe {et~al.}(2015)Baraffe, Homeier, Allard, \&
  Chabrier}]{Baraffe2015}
Baraffe, I., Homeier, D., Allard, F., \& Chabrier, G. 2015, A{\&}A, 577, A42

\bibitem[{Barenfeld {et~al.}(2016)Barenfeld, Carpenter, Ricci, \&
  Isella}]{Barenfeld2016}
Barenfeld, S.~A., Carpenter, J.~M., Ricci, L., \& Isella, A. 2016, ApJ, 827,
  142

\bibitem[{Bate(2012)}]{Bate2012}
Bate, M.~R. 2012, MNRAS, 419, 3115

\bibitem[{Bell \& Lin(1994)}]{Bell1994}
Bell, K.~R. \& Lin, D. N.~C. 1994, ApJ, 427, 987

\bibitem[{Ben{\'{i}}tez-Llambay {et~al.}(2011)Ben{\'{i}}tez-Llambay, Masset, \&
  Beaug{\'{e}}}]{Benitez-Llambay2011}
Ben{\'{i}}tez-Llambay, P., Masset, F., \& Beaug{\'{e}}, C. 2011, A{\&}A, 528,
  A2

\bibitem[{Benz {et~al.}(2014)Benz, Ida, Alibert, Lin, \& Mordasini}]{Benz2014}
Benz, W., Ida, S., Alibert, Y., Lin, D. N.~C., \& Mordasini, C. 2014, in
  Protostars Planets VI, ed. H.~Beuther, R.~S. Klessen, C.~P. Dullemond, \&
  T.~Henning (Tucson: University of Arizona Press), 697--713

\bibitem[{Bodenheimer \& Pollack(1986)}]{Bodenheimer1986}
Bodenheimer, P. \& Pollack, J.~B. 1986, Icarus, 67, 391

\bibitem[{Bonfils {et~al.}(2018)Bonfils, Almenara, Cloutier, W{\"{u}}nsche,
  Astudillo-Defru, Berta-Thompson, Bouchy, Charbonneau, Delfosse, D{\'{i}}az,
  Dittmann, Doyon, Forveille, Irwin, Lovis, Mayor, Menou, Murgas, Newton, Pepe,
  Santos, \& Udry}]{Bonfils2018}
Bonfils, X., Almenara, J.-M., Cloutier, R., {et~al.} 2018, A{\&}A, 618, A142

\bibitem[{Bonfils {et~al.}(2005)Bonfils, Forveille, Delfosse, Udry, Mayor,
  Perrier, Bouchy, Pepe, Queloz, \& Bertaux}]{Bonfils2005}
Bonfils, X., Forveille, T., Delfosse, X., {et~al.} 2005, A{\&}A, 443, L15

\bibitem[{Boss(1997)}]{Boss1997}
Boss, A.~P. 1997, Science, 276, 1836

\bibitem[{Bouchet {et~al.}(2013)Bouchet, Mazevet, Morard, Guyot, \&
  Musella}]{Bouchet2013}
Bouchet, J., Mazevet, S., Morard, G., Guyot, F., \& Musella, R. 2013, Physical
  Review B, 87, 094102

\bibitem[{Bouchy {et~al.}(2017)Bouchy, Doyon, Artigau, Melo, Hernandez, Wildi,
  Delfosse, Lovis, Figueira, {Canto Martins}, {Gonz{\'{a}}lez Hern{\'{a}}ndez},
  Thibault, Reshetov, Pepe, Santos, de~Medeiros, Rebolo, Abreu, Adibekyan,
  Bandy, Benz, Blind, Bohlender, Boisse, Bovay, Broeg, Brousseau, Cabral,
  Chazelas, Cloutier, Coelho, Conod, Cumming, Delabre, Genolet, Hagelberg,
  Jayawardhana, K{\"{a}}ufl, Lafreni{\`{e}}re, {de Castro Le{\~{a}}o}, Malo,
  {de Medeiros Martins}, Matthews, Metchev, Oshagh, Ouellet, Parro, {Rasilla
  Pi{\~{n}}eiro}, Santos, Sarajlic, Segovia, Sordet, Udry, Valencia,
  Vall{\'{e}}e, Venn, Wade, \& Saddlemyer}]{Bouchy2017}
Bouchy, F., Doyon, R., Artigau, {\'{E}}., {et~al.} 2017, The Messenger, 169, 21

\bibitem[{Bouvier {et~al.}(2007)Bouvier, Alencar, Harries, Johns-Krull, \&
  Romanova}]{Bouvier2007}
Bouvier, J., Alencar, S. H.~P., Harries, T.~J., Johns-Krull, C.~M., \&
  Romanova, M.~M. 2007, in Protostars Planets V, ed. B.~Reipurth, D.~Jewitt, \&
  K.~Keil (Tucson: University of Arizona Press), 479--494

\bibitem[{Br{\"{u}}gger {et~al.}(2018)Br{\"{u}}gger, Alibert, Ataiee, \&
  Benz}]{Brugger2018}
Br{\"{u}}gger, N., Alibert, Y., Ataiee, S., \& Benz, W. 2018, A{\&}A, 619, A174

\bibitem[{Br{\"{u}}gger {et~al.}(2020)Br{\"{u}}gger, Burn, Coleman, Alibert, \&
  Benz}]{Brugger2020}
Br{\"{u}}gger, N., Burn, R., Coleman, G. A.~L., Alibert, Y., \& Benz, W. 2020,
  A{\&}A, 640, A21

\bibitem[{Bryson {et~al.}(2021)Bryson, Kunimoto, Kopparapu, Coughlin, Borucki,
  Koch, Aguirre, Allen, Barentsen, Batalha, Berger, Boss, Buchhave, Burke,
  Caldwell, Campbell, Catanzarite, Chandrasekaran, Chaplin, Christiansen,
  Christensen-Dalsgaard, Ciardi, Clarke, Cochran, Dotson, Doyle, Duarte,
  Dunham, Dupree, Endl, Fanson, Ford, Fujieh, {Gautier III}, Geary, Gilliland,
  Girouard, Gould, Haas, Henze, Holman, Howard, Howell, Huber, Hunter, Jenkins,
  Kjeldsen, Kolodziejczak, Larson, Latham, Li, Mathur, Meibom, Middour, Morris,
  Morton, Mullally, Mullally, Pletcher, Prsa, Quinn, Quintana, Ragozzine,
  Ramirez, Sanderfer, Sasselov, Seader, Shabram, Shporer, Smith, Steffen,
  Still, Torres, Troeltzsch, Twicken, Uddin, {Van Cleve}, Voss, Weiss, Welsh,
  Wohler, \& Zamudio}]{Bryson2020}
Bryson, S., Kunimoto, M., Kopparapu, R.~K., {et~al.} 2021, The Astronomical
  Journal, 161, 36

\bibitem[{Burdanov {et~al.}(2018)Burdanov, Delrez, Gillon, \&
  Jehin}]{Burdanov2018}
Burdanov, A., Delrez, L., Gillon, M., \& Jehin, E. 2018, in Handbook of
  Exoplanets, ed. H.~J. Deeg \& J.~A. Belmonte (Cham: Springer International
  Publishing), 1--17

\bibitem[{Butler {et~al.}(2006)Butler, Johnson, Marcy, Wright, Vogt, \&
  Fischer}]{Butler2006}
Butler, R.~P., Johnson, J.~A., Marcy, G.~W., {et~al.} 2006, Publications of the
  Astronomical Society of the Pacific, 118, 1685

\bibitem[{Cameron(1978)}]{Cameron1978}
Cameron, A.~G. 1978, Moon Planets, 18, 5

\bibitem[{Chambers {et~al.}(1996)Chambers, Wetherill, \& Boss}]{Chambers1996}
Chambers, J., Wetherill, G., \& Boss, A. 1996, Icarus, 119, 261

\bibitem[{Chambers(1999)}]{Chambers1999}
Chambers, J.~E. 1999, MNRAS, 304, 793

\bibitem[{Chambers(2006)}]{Chambers2006}
Chambers, J.~E. 2006, Icarus, 180, 496

\bibitem[{Choi {et~al.}(2016)Choi, Dotter, Conroy, Cantiello, Paxton, \&
  Johnson}]{Choi2016}
Choi, J., Dotter, A., Conroy, C., {et~al.} 2016, ApJ, 823, 102

\bibitem[{Clarke {et~al.}(2001)Clarke, Gendrin, \& Sotomayor}]{Clarke2001}
Clarke, C.~J., Gendrin, A., \& Sotomayor, M. 2001, MNRAS, 328, 485

\bibitem[{Coleman {et~al.}(2019)Coleman, Leleu, Alibert, \& Benz}]{Coleman2019}
Coleman, G. A.~L., Leleu, A., Alibert, Y., \& Benz, W. 2019, A{\&}A, 631, A7

\bibitem[{Coleman \& Nelson(2014)}]{Coleman2014}
Coleman, G. A.~L. \& Nelson, R.~P. 2014, MNRAS, 445, 479

\bibitem[{Coleman \& Nelson(2016)}]{Coleman2016a}
Coleman, G. A.~L. \& Nelson, R.~P. 2016, MNRAS, 460, 2779

\bibitem[{Crida {et~al.}(2006)Crida, Morbidelli, \& Masset}]{Crida2006}
Crida, A., Morbidelli, A., \& Masset, F. 2006, Icarus, 181, 587

\bibitem[{Delfosse {et~al.}(1998)Delfosse, Forveille, Mayor, Perrier, Naef, \&
  Queloz}]{Delfosse1998}
Delfosse, X., Forveille, T., Mayor, M., {et~al.} 1998, A{\&}A, 338, L67

\bibitem[{Delrez {et~al.}(2018)Delrez, Gillon, Queloz, Demory, Almleaky,
  de~Wit, Jehin, Triaud, Barkaoui, Burdanov, Burgasser, Ducrot, McCormac,
  Murray, {Silva Fernandes}, Sohy, Thompson, {Van Grootel}, Alonso,
  Benkhaldoun, \& Rebolo}]{Delrez2018a}
Delrez, L., Gillon, M., Queloz, D., {et~al.} 2018, in Society of Photo-Optical
  Instrumentation Engineers (SPIE) Conference Series, Vol. 10700, Ground-based
  Airborne Telescopes VII, ed. H.~K. Marshall \& J.~Spyromilio, 107001I

\bibitem[{Demory {et~al.}(2020)Demory, Pozuelos, {G{\'{o}}mez Maqueo Chew},
  Sabin, Petrucci, Schroffenegger, Grimm, Sestovic, Gillon, McCormac, Barkaoui,
  Benz, Bieryla, Bouchy, Burdanov, Collins, de~Wit, Dressing, Garcia,
  Giacalone, Guerra, Haldemann, Heng, Jehin, Jofr{\'{e}}, Kane, Lillo-Box,
  Maign{\'{e}}, Mordasini, Morris, Niraula, Queloz, Rackham, Savel, Soubkiou,
  Srdoc, Stassun, Triaud, Zambelli, Ricker, Latham, Seager, Winn, Jenkins,
  Calvario-Vel{\'{a}}squez, {Franco Herrera}, Colorado, {Cadena Zepeda},
  Figueroa, Watson, Lugo-Ibarra, Carigi, Guisa, Herrera, {Sierra D{\'{i}}az},
  Su{\'{a}}rez, Barrado, Batalha, Benkhaldoun, Chontos, Dai, Essack, Ghachoui,
  Huang, Huber, Isaacson, Lissauer, Morales-Calder{\'{o}}n, Robertson, Roy,
  Twicken, Vanderburg, \& Weiss}]{Demory2020}
Demory, B.-O., Pozuelos, F.~J., {G{\'{o}}mez Maqueo Chew}, Y., {et~al.} 2020,
  A{\&}A, 642, A49

\bibitem[{Dittkrist {et~al.}(2014)Dittkrist, Mordasini, Klahr, Alibert, \&
  Henning}]{Dittkrist2014}
Dittkrist, K.-M. K.-M., Mordasini, C., Klahr, H., Alibert, Y., \& Henning, T.
  2014, A{\&}A, 567, A121

\bibitem[{Dorn {et~al.}(2015)Dorn, Khan, Heng, Connolly, Alibert, Benz, \&
  Tackley}]{Dorn2015}
Dorn, C., Khan, A., Heng, K., {et~al.} 2015, A{\&}A, 577, A83

\bibitem[{Dorn {et~al.}(2018)Dorn, Mosegaard, Grimm, \& Alibert}]{Dorn2018}
Dorn, C., Mosegaard, K., Grimm, S.~L., \& Alibert, Y. 2018, ApJ, 865, 20

\bibitem[{Dr{\c{a}}{\.{z}}kowska \& Alibert(2017)}]{Drazkowska2017}
Dr{\c{a}}{\.{z}}kowska, J. \& Alibert, Y. 2017, A{\&}A, 608, A92

\bibitem[{Dr{\c{a}}{\.{z}}kowska {et~al.}(2016)Dr{\c{a}}{\.{z}}kowska, Alibert,
  \& Moore}]{Drazkowska2016}
Dr{\c{a}}{\.{z}}kowska, J., Alibert, Y., \& Moore, B. 2016, A{\&}A, 594, A105

\bibitem[{Dressing \& Charbonneau(2013)}]{Dressing2013}
Dressing, C.~D. \& Charbonneau, D. 2013, ApJ, 767, 95

\bibitem[{Dressing \& Charbonneau(2015)}]{Dressing2015}
Dressing, C.~D. \& Charbonneau, D. 2015, ApJ, 807, 45

\bibitem[{Emsenhuber {et~al.}(2020)Emsenhuber, Cambioni, Asphaug, Gabriel,
  Schwartz, \& Furfaro}]{Emsenhuber2020}
Emsenhuber, A., Cambioni, S., Asphaug, E., {et~al.} 2020, ApJ, 891, 6

\bibitem[{Emsenhuber {et~al.}(2021{\natexlab{a}})Emsenhuber, Mordasini, Burn,
  Alibert, Benz, \& Asphaug}]{Emsenhuber2020a}
Emsenhuber, A., Mordasini, C., Burn, R., {et~al.} 2021{\natexlab{a}}, A{\&}A,
  656, A69

\bibitem[{Emsenhuber {et~al.}(2021{\natexlab{b}})Emsenhuber, Mordasini, Burn,
  Alibert, Benz, \& Asphaug}]{Emsenhuber2020b}
Emsenhuber, A., Mordasini, C., Burn, R., {et~al.} 2021{\natexlab{b}}, A{\&}A,
  656, A70

\bibitem[{Endl {et~al.}(2006)Endl, Cochran, Kuerster, Paulson, Wittenmyer,
  MacQueen, \& Tull}]{Endl2006}
Endl, M., Cochran, W.~D., Kuerster, M., {et~al.} 2006, ApJ, 649, 436

\bibitem[{Fedele {et~al.}(2010)Fedele, van~den Ancker, Henning, Jayawardhana,
  \& Oliveira}]{Fedele2010}
Fedele, D., van~den Ancker, M.~E., Henning, T., Jayawardhana, R., \& Oliveira,
  J.~M. 2010, A{\&}A, 510, A72

\bibitem[{Feiden(2016)}]{Feiden2016}
Feiden, G.~A. 2016, A{\&}A, 593, A99

\bibitem[{Fortier {et~al.}(2013)Fortier, Alibert, Carron, Benz, \&
  Dittkrist}]{Fortier2013}
Fortier, A., Alibert, Y., Carron, F., Benz, W., \& Dittkrist, K.-M. 2013,
  A{\&}A, 549, A44

\bibitem[{Fulton \& Petigura(2018)}]{Fulton2018}
Fulton, B.~J. \& Petigura, E.~A. 2018, The Astronomical Journal, 156, 264

\bibitem[{Gaidos {et~al.}(2013)Gaidos, Fischer, Mann, \& Howard}]{Gaidos2013}
Gaidos, E., Fischer, D.~A., Mann, A.~W., \& Howard, A.~W. 2013, ApJ, 771, 18

\bibitem[{Gaidos {et~al.}(2016)Gaidos, Mann, Kraus, \& Ireland}]{Gaidos2016}
Gaidos, E., Mann, A.~W., Kraus, A.~L., \& Ireland, M. 2016, MNRAS, 457, 2877

\bibitem[{Gibbs {et~al.}(2020)Gibbs, Bixel, Rackham, Apai, Schlecker, Espinoza,
  Mancini, Chen, Henning, Gabor, Boyle, {Perez Chavez}, Mousseau, Dietrich,
  {Jay Socia}, Ip, Ngeow, Tsai, Bhandare, Marian, Baehr, Brown, H{\"{a}}berle,
  Keppler, Molaverdikhani, \& Sarkis}]{Gibbs2020}
Gibbs, A., Bixel, A., Rackham, B.~V., {et~al.} 2020, The Astronomical Journal,
  159, 169

\bibitem[{Gillon {et~al.}(2016)Gillon, Jehin, Lederer, Delrez, de~Wit,
  Burdanov, {Van Grootel}, Burgasser, Triaud, Opitom, Demory, Sahu, {Bardalez
  Gagliuffi}, Magain, \& Queloz}]{Gillon2016}
Gillon, M., Jehin, E., Lederer, S.~M., {et~al.} 2016, Nature, 533, 221

\bibitem[{Gillon {et~al.}(2017)Gillon, Triaud, Demory, Jehin, Agol, Deck,
  Lederer, de~Wit, Burdanov, Ingalls, Bolmont, Leconte, Raymond, Selsis,
  Turbet, Barkaoui, Burgasser, Burleigh, Carey, Chaushev, Copperwheat, Delrez,
  Fernandes, Holdsworth, Kotze, {Van Grootel}, Almleaky, Benkhaldoun, Magain,
  Queloz, Grootel, Bolmont, Leconte, Raymond, Selsis, Turbet, Barkaoui,
  Burgasser, Burleigh, Carey, Chaushev, Copperwheat, Delrez, Fernandes,
  Holdsworth, Kotze, {Van Grootel}, Almleaky, Benkhaldoun, Magain, \&
  Queloz}]{Gillon2017}
Gillon, M., Triaud, A. H. M.~J., Demory, B.-O., {et~al.} 2017, Nature, 542, 456

\bibitem[{Ginzburg {et~al.}(2016)Ginzburg, Schlichting, \& Sari}]{Ginzburg2016}
Ginzburg, S., Schlichting, H.~E., \& Sari, R. 2016, ApJ, 825, 29

\bibitem[{Grimm {et~al.}(2018)Grimm, Demory, Gillon, Dorn, Agol, Burdanov,
  Delrez, Sestovic, Triaud, Turbet, Bolmont, Caldas, {De Wit}, Jehin, Leconte,
  Raymond, Grootel, Burgasser, Carey, Fabrycky, Heng, Hernandez, Ingalls,
  Lederer, Selsis, \& Queloz}]{Grimm2018}
Grimm, S.~L., Demory, B.-O., Gillon, M., {et~al.} 2018, A{\&}A, 613, A68

\bibitem[{Grootel {et~al.}(2018)Grootel, Fernandes, Gillon, Jehin, Manfroid,
  Scuflaire, Burgasser, Barkaoui, Benkhaldoun, Burdanov, Delrez, Demory,
  de~Wit, Queloz, \& Triaud}]{Grootel2018}
Grootel, V.~V., Fernandes, C.~S., Gillon, M., {et~al.} 2018, ApJ, 853, 30

\bibitem[{Guillot(2010)}]{Guillot2010}
Guillot, T. 2010, A{\&}A, 520, A27

\bibitem[{G{\"{u}}nther(2013)}]{Gunther2013}
G{\"{u}}nther, H. 2013, Astronomische Nachrichten, 334, 67

\bibitem[{Haisch {et~al.}(2001)Haisch, Lada, \& Lada}]{Haisch2001}
Haisch, K.~E., Lada, E.~A., \& Lada, C.~J. 2001, ApJ, 553, L153

\bibitem[{Hakim {et~al.}(2018)Hakim, Rivoldini, {Van Hoolst}, Cottenier,
  Jaeken, Chust, \& Steinle-Neumann}]{Hakim2018}
Hakim, K., Rivoldini, A., {Van Hoolst}, T., {et~al.} 2018, Icarus, 313, 61

\bibitem[{Haldemann {et~al.}(2020)Haldemann, Alibert, Mordasini, \&
  Benz}]{Haldemann2020}
Haldemann, J., Alibert, Y., Mordasini, C., \& Benz, W. 2020, A{\&}A, 643, A105

\bibitem[{Hamann {et~al.}(2019)Hamann, Montet, Fabrycky, Agol, \&
  Kruse}]{Hamann2019}
Hamann, A., Montet, B.~T., Fabrycky, D.~C., Agol, E., \& Kruse, E. 2019, The
  Astronomical Journal, 158, 133

\bibitem[{Hansen(2008)}]{Hansen2008}
Hansen, B. M.~S. 2008, The Astrophysical Journal Supplement Series, 179, 484

\bibitem[{Hansen(2015)}]{Hansen2015}
Hansen, B. M.~S. 2015, International Journal of Astrobiology, 14, 267

\bibitem[{Harps{\o}e {et~al.}(2013)Harps{\o}e, Hardis, Hinse, J{\o}rgensen,
  Mancini, Southworth, Alsubai, Bozza, Browne, Burgdorf, {Calchi Novati},
  Dodds, Dominik, Fang, Finet, Gerner, Gu, Hundertmark, Jessen-Hansen, Kains,
  Kerins, Kjeldsen, Liebig, Lund, Lundkvist, Mathiasen, Nesvorn{\'{y}},
  Nikolov, Penny, Proft, Rahvar, Ricci, Sahu, Scarpetta, Sch{\"{a}}fer,
  Sch{\"{o}}nebeck, Snodgrass, Skottfelt, Surdej, Tregloan-Reed, \&
  Wertz}]{Harpsoe2012}
Harps{\o}e, K. B.~W., Hardis, S., Hinse, T.~C., {et~al.} 2013, A{\&}A, 549, A10

\bibitem[{Hasegawa \& Pudritz(2011)}]{Hasegawa2011}
Hasegawa, Y. \& Pudritz, R.~E. 2011, MNRAS, 417, 1236

\bibitem[{Haworth {et~al.}(2018)Haworth, Clarke, Rahman, Winter, \&
  Facchini}]{Haworth2018}
Haworth, T.~J., Clarke, C.~J., Rahman, W., Winter, A.~J., \& Facchini, S. 2018,
  MNRAS, 481, 452

\bibitem[{Henderson \& Stassun(2011)}]{Henderson2011}
Henderson, C.~B. \& Stassun, K.~G. 2011, ApJ, 747, 51

\bibitem[{Herbst {et~al.}(2002)Herbst, Bailer-Jones, Mundt, Meisenheimer, \&
  Wackermann}]{Herbst2002}
Herbst, W., Bailer-Jones, C. A.~L., Mundt, R., Meisenheimer, K., \& Wackermann,
  R. 2002, A{\&}A, 396, 513

\bibitem[{Hirose {et~al.}(2013)Hirose, Labrosse, \& Hernlund}]{Hirose2013}
Hirose, K., Labrosse, S., \& Hernlund, J. 2013, Annual Review of Earth and
  Planetary Sciences, 41, 657

\bibitem[{Howard {et~al.}(2010)Howard, Johnson, Marcy, Fischer, Wright, Bernat,
  Henry, Peek, Isaacson, Apps, Endl, Cochran, Valenti, Anderson, \&
  Piskunov}]{Howard2010}
Howard, A.~W., Johnson, J.~A., Marcy, G.~W., {et~al.} 2010, ApJ, 721, 1467

\bibitem[{Hueso \& Guillot(2005)}]{Hueso2005}
Hueso, R. \& Guillot, T. 2005, A{\&}A, 442, 703

\bibitem[{Hunter(2007)}]{Hunter2007}
Hunter, J.~D. 2007, Computing in Science {\&} Engineering, 9, 90

\bibitem[{Ida \& Lin(2004{\natexlab{a}})}]{Ida2004}
Ida, S. \& Lin, D. N.~C. 2004{\natexlab{a}}, ApJ, 604, 388

\bibitem[{Ida \& Lin(2004{\natexlab{b}})}]{Ida2004a}
Ida, S. \& Lin, D. N.~C. 2004{\natexlab{b}}, ApJ, 616, 567

\bibitem[{Ida \& Lin(2005)}]{Ida2005}
Ida, S. \& Lin, D. N.~C. 2005, ApJ, 626, 1045

\bibitem[{Ida \& Makino(1992{\natexlab{a}})}]{Ida1992a}
Ida, S. \& Makino, J. 1992{\natexlab{a}}, Icarus, 96, 107

\bibitem[{Ida \& Makino(1992{\natexlab{b}})}]{Ida1992b}
Ida, S. \& Makino, J. 1992{\natexlab{b}}, Icarus, 98, 28

\bibitem[{Inaba {et~al.}(2001)Inaba, Tanaka, Nakazawa, Wetherill, \&
  Kokubo}]{Inaba2001}
Inaba, S., Tanaka, H., Nakazawa, K., Wetherill, G.~W., \& Kokubo, E. 2001,
  Icarus, 149, 235

\bibitem[{Irwin {et~al.}(2008)Irwin, Hodgkin, Aigrain, Bouvier, Hebb, Irwin, \&
  Moraux}]{Irwin2008}
Irwin, J., Hodgkin, S., Aigrain, S., {et~al.} 2008, MNRAS, 384, 675

\bibitem[{Irwin {et~al.}(2007)Irwin, Hodgkin, Aigrain, Bouvier, Hebb, \&
  Moraux}]{Irwin2007}
Irwin, J., Hodgkin, S., Aigrain, S., {et~al.} 2007, MNRAS, 383, 1588

\bibitem[{Izidoro {et~al.}(2017)Izidoro, Ogihara, Raymond, Morbidelli, Pierens,
  Bitsch, Cossou, \& Hersant}]{Izidoro2017}
Izidoro, A., Ogihara, M., Raymond, S.~N., {et~al.} 2017, MNRAS, 470, 1750

\bibitem[{Jin \& Mordasini(2018)}]{Jin2018}
Jin, S. \& Mordasini, C. 2018, ApJ, 853, 163

\bibitem[{Jin {et~al.}(2014)Jin, Mordasini, Parmentier, van Boekel, Henning, \&
  Ji}]{Jin2014}
Jin, S., Mordasini, C., Parmentier, V., {et~al.} 2014, ApJ, 795, 65

\bibitem[{Johansen {et~al.}(2012)Johansen, Youdin, \& Lithwick}]{Johansen2012}
Johansen, A., Youdin, A.~N., \& Lithwick, Y. 2012, A{\&}A, 537, A125

\bibitem[{Johnson {et~al.}(2010)Johnson, Aller, Howard, \& Crepp}]{Johnson2010}
Johnson, J.~A., Aller, K.~M., Howard, A.~W., \& Crepp, J.~R. 2010, Publications
  of the Astronomical Society of the Pacific, 122, 905

\bibitem[{Johnson {et~al.}(2007)Johnson, Butler, Marcy, Fischer, Vogt, Wright,
  \& Peek}]{Johnson2007}
Johnson, J.~A., Butler, R.~P., Marcy, G.~W., {et~al.} 2007, ApJ, 670, 833

\bibitem[{Kaltenegger(2017)}]{Kaltenegger2017}
Kaltenegger, L. 2017, Annual Review of Astronomy and Astrophysics, 55, 433

\bibitem[{Kasting {et~al.}(1993)Kasting, Whitmire, \& Reynolds}]{Kasting1993}
Kasting, J.~F., Whitmire, D.~P., \& Reynolds, R.~T. 1993, Icarus, 101, 108

\bibitem[{Kenyon \& Bromley(2020)}]{Kenyon2020}
Kenyon, S.~J. \& Bromley, B.~C. 2020, The Planetary Science Journal, 1, 40

\bibitem[{Kimura {et~al.}(2016)Kimura, Kunitomo, \& Takahashi}]{Kimura2016}
Kimura, S.~S., Kunitomo, M., \& Takahashi, S.~Z. 2016, MNRAS, 461, 2257

\bibitem[{Klahr \& Schreiber(2021)}]{Klahr2020}
Klahr, H. \& Schreiber, A. 2021, ApJ, 911, 9

\bibitem[{Kokubo \& Ida(1998)}]{Kokubo1998}
Kokubo, E. \& Ida, S. 1998, Icarus, 131, 171

\bibitem[{Kokubo \& Ida(2000)}]{Kokubo2000}
Kokubo, E. \& Ida, S. 2000, Icarus, 143, 15

\bibitem[{Kopparapu {et~al.}(2013{\natexlab{a}})Kopparapu, Ramirez, Kasting,
  Eymet, Robinson, Mahadevan, Terrien, Domagal-Goldman, Meadows, \&
  Deshpande}]{Kopparapu2013a}
Kopparapu, R.~K., Ramirez, R., Kasting, J.~F., {et~al.} 2013{\natexlab{a}},
  ApJ, 770, 82

\bibitem[{Kopparapu {et~al.}(2013{\natexlab{b}})Kopparapu, Ramirez, Kasting,
  Eymet, Robinson, Mahadevan, Terrien, Domagal-Goldman, Meadows, \&
  Deshpande}]{Kopparapu2013}
Kopparapu, R.~K., Ramirez, R., Kasting, J.~F., {et~al.} 2013{\natexlab{b}},
  ApJ, 765, 131

\bibitem[{Kopparapu {et~al.}(2014)Kopparapu, Ramirez, SchottelKotte, Kasting,
  Domagal-Goldman, \& Eymet}]{Kopparapu2014}
Kopparapu, R.~K., Ramirez, R.~M., SchottelKotte, J., {et~al.} 2014, ApJ, 787,
  L29

\bibitem[{Lamm {et~al.}(2005)Lamm, Mundt, Bailer-Jones, \& Herbst}]{Lamm2005}
Lamm, M.~H., Mundt, R., Bailer-Jones, C. A.~L., \& Herbst, W. 2005, A{\&}A,
  430, 1005

\bibitem[{Laughlin {et~al.}(2004)Laughlin, Bodenheimer, \&
  Adams}]{Laughlin2004}
Laughlin, G., Bodenheimer, P., \& Adams, F.~C. 2004, ApJ, 612, L73

\bibitem[{Lenz {et~al.}(2019)Lenz, Klahr, \& Birnstiel}]{Lenz2019}
Lenz, C.~T., Klahr, H., \& Birnstiel, T. 2019, ApJ, 874, 36

\bibitem[{Li {et~al.}(2019)Li, Youdin, \& Simon}]{Li2019}
Li, R., Youdin, A.~N., \& Simon, J.~B. 2019, ApJ, 885, 69

\bibitem[{Lichtenberg {et~al.}(2019)Lichtenberg, Golabek, Burn, Meyer, Alibert,
  Gerya, \& Mordasini}]{Lichtenberg2019}
Lichtenberg, T., Golabek, G.~J., Burn, R., {et~al.} 2019, Nature Astronomy, 3,
  307

\bibitem[{Lissauer \& Stewart(1993)}]{Lissauer1993}
Lissauer, J.~J. \& Stewart, G.~R. 1993, in Protostars planets III, ed. E.~H.
  Levy \& J.~I. Lunine (University of Arizona Press), 1061--1088

\bibitem[{Liu {et~al.}(2019)Liu, Lambrechts, Johansen, \& Liu}]{Liu2019}
Liu, B., Lambrechts, M., Johansen, A., \& Liu, F. 2019, A{\&}A, 632, A7

\bibitem[{Liu {et~al.}(2020)Liu, Lambrechts, Johansen, Pascucci, \&
  Henning}]{Liu2020a}
Liu, B., Lambrechts, M., Johansen, A., Pascucci, I., \& Henning, T. 2020,
  A{\&}A, 638, A88

\bibitem[{Lodders(2003)}]{Lodders2003}
Lodders, K. 2003, ApJ, 591, 1220

\bibitem[{Lopez \& Fortney(2013)}]{Lopez2013}
Lopez, E.~D. \& Fortney, J.~J. 2013, ApJ, 776, 2

\bibitem[{Machida {et~al.}(2010)Machida, Inutsuka, \& Matsumoto}]{Machida2010}
Machida, M.~N., Inutsuka, S.-i., \& Matsumoto, T. 2010, ApJ, 724, 1006

\bibitem[{Mamajek {et~al.}(2009)Mamajek, Usuda, Tamura, \& Ishii}]{Mamajek2009}
Mamajek, E.~E., Usuda, T., Tamura, M., \& Ishii, M. 2009, in AIP Conference
  Proceedings, Vol. 1158 (AIP), 3--10

\bibitem[{Manara {et~al.}(2019)Manara, Mordasini, Testi, Williams, Miotello,
  Lodato, \& Emsenhuber}]{Manara2019}
Manara, C.~F., Mordasini, C., Testi, L., {et~al.} 2019, A{\&}A, 631, L2

\bibitem[{Manara {et~al.}(2020)Manara, Natta, Rosotti, Alcal{\'{a}}, Nisini,
  Lodato, Testi, Pascucci, Hillenbrand, Carpenter, Scholz, Fedele, Frasca,
  Mulders, Rigliaco, Scardoni, \& Zari}]{Manara2020}
Manara, C.~F., Natta, A., Rosotti, G.~P., {et~al.} 2020, A{\&}A, 639, A58

\bibitem[{Manara {et~al.}(2012)Manara, Robberto, {Da Rio}, Lodato, Hillenbrand,
  Stassun, \& Soderblom}]{Manara2012}
Manara, C.~F., Robberto, M., {Da Rio}, N., {et~al.} 2012, ApJ, 755, 154

\bibitem[{Marboeuf {et~al.}(2014)Marboeuf, Thiabaud, Alibert, Cabral, \&
  Benz}]{Marboeuf2014b}
Marboeuf, U., Thiabaud, A., Alibert, Y., Cabral, N., \& Benz, W. 2014, A{\&}A,
  570, A36

\bibitem[{Marcy {et~al.}(2001)Marcy, Butler, Fischer, Vogt, Lissauer, \&
  Rivera}]{Marcy2001}
Marcy, G.~W., Butler, R.~P., Fischer, D., {et~al.} 2001, ApJ, 556, 296

\bibitem[{Marcy {et~al.}(1998)Marcy, Butler, Vogt, Fischer, \&
  Lissauer}]{Marcy1998}
Marcy, G.~W., Butler, R.~P., Vogt, S.~S., Fischer, D., \& Lissauer, J.~J. 1998,
  ApJ, 505, L147

\bibitem[{Matsuyama {et~al.}(2003)Matsuyama, Johnstone, \&
  Hartmann}]{Matsuyama2003}
Matsuyama, I., Johnstone, D., \& Hartmann, L. 2003, ApJ, 582, 893

\bibitem[{Mayor {et~al.}(2011)Mayor, Marmier, Lovis, Udry, S{\'{e}}gransan,
  Pepe, Benz, Bertaux, Bouchy, Dumusque, Curto, Mordasini, Queloz, \&
  Santos}]{Mayor2011}
Mayor, M., Marmier, M., Lovis, C., {et~al.} 2011, ArXiv e-prints
  [\eprint[arXiv]{1109.2497}]

\bibitem[{Mazevet {et~al.}(2019)Mazevet, Licari, Chabrier, \&
  Potekhin}]{Mazevet2019}
Mazevet, S., Licari, A., Chabrier, G., \& Potekhin, A.~Y. 2019, A{\&}A, 621,
  A128

\bibitem[{Ment {et~al.}(2019)Ment, Dittmann, Astudillo-Defru, Charbonneau,
  Irwin, Bonfils, Murgas, Almenara, Forveille, Agol, Ballard, Berta-Thompson,
  Bouchy, Cloutier, Delfosse, Doyon, Dressing, Esquerdo, Haywood, Kipping,
  Latham, Lovis, Newton, Pepe, Rodriguez, Santos, Tan, Udry, Winters, \&
  W{\"{u}}nsche}]{Ment2019}
Ment, K., Dittmann, J.~A., Astudillo-Defru, N., {et~al.} 2019, The Astronomical
  Journal, 157, 32

\bibitem[{Miguel {et~al.}(2020)Miguel, Cridland, Ormel, Fortney, \&
  Ida}]{Miguel2020}
Miguel, Y., Cridland, A., Ormel, C., Fortney, J., \& Ida, S. 2020, MNRAS, 491,
  1998

\bibitem[{Millholland {et~al.}(2018)Millholland, Laughlin, Teske, Butler, Burt,
  Holden, Vogt, Crane, Shectman, \& Thompson}]{Millholland2018}
Millholland, S., Laughlin, G., Teske, J., {et~al.} 2018, The Astronomical
  Journal, 155, 106

\bibitem[{Mishra {et~al.}(2021)Mishra, Alibert, Leleu, Emsenhuber, Mordasini,
  Burn, Udry, \& Benz}]{Mishra2021}
Mishra, L., Alibert, Y., Leleu, A., {et~al.} 2021, A{\&}A
  [\eprint[arXiv]{2105.12745}]

\bibitem[{Mizuno(1980)}]{Mizuno1980}
Mizuno, H. 1980, Progress of Theoretical Physics, 64, 544

\bibitem[{Mizuno {et~al.}(1978)Mizuno, Nakazawa, \& Hayashi}]{Mizuno1978}
Mizuno, H., Nakazawa, K., \& Hayashi, C. 1978, Progress of Theoretical Physics,
  60, 699

\bibitem[{Montet {et~al.}(2014)Montet, Crepp, Johnson, Howard, \&
  Marcy}]{Montet2014}
Montet, B.~T., Crepp, J.~R., Johnson, J.~A., Howard, A.~W., \& Marcy, G.~W.
  2014, ApJ, 781, 28

\bibitem[{Morales {et~al.}(2019)Morales, Mustill, Ribas, Davies, Reiners,
  Bauer, Kossakowski, Herrero, Rodr{\'{i}}guez, L{\'{o}}pez-Gonz{\'{a}}lez,
  Rodr{\'{i}}guez-L{\'{o}}pez, B{\'{e}}jar, Gonz{\'{a}}lez-Cuesta, Luque,
  Pall{\'{e}}, Perger, Baroch, Johansen, Klahr, Mordasini,
  Anglada-Escud{\'{e}}, Caballero, Cort{\'{e}}s-Contreras, Dreizler, Lafarga,
  Nagel, Passegger, Reffert, Rosich, Schweitzer, Tal-Or, Trifonov, Zechmeister,
  Quirrenbach, Amado, Guenther, Hagen, Henning, Jeffers, Kaminski,
  K{\"{u}}rster, Montes, Seifert, Abell{\'{a}}n, Abril, Aceituno, Aceituno,
  Alonso-Floriano, {Ammler-von Eiff}, Antona, Arroyo-Torres, Azzaro, Barrado,
  Becerril-Jarque, Ben{\'{i}}tez, Berdi{\~{n}}as, Bergond, Brinkm{\"{o}}ller,
  del Burgo, Burn, Calvo-Ortega, Cano, C{\'{a}}rdenas, Guill{\'{e}}n, Carro,
  Casal, Casanova, Casasayas-Barris, Chaturvedi, Cifuentes, Claret,
  Colom{\'{e}}, Czesla, D{\'{i}}ez-Alonso, Dorda, Emsenhuber, Fern{\'{a}}ndez,
  Fern{\'{a}}ndez-Mart{\'{i}}n, Ferro, Fuhrmeister,
  Galad{\'{i}}-Enr{\'{i}}quez, Cava, Vargas, Garcia-Piquer, Gesa,
  Gonz{\'{a}}lez-{\'{A}}lvarez, Hern{\'{a}}ndez, Gonz{\'{a}}lez-Peinado,
  Gu{\`{a}}rdia, Guijarro, de~Guindos, Hatzes, Hauschildt, Hedrosa, Hermelo,
  Arabi, Otero, Hintz, Holgado, Huber, Huke, Johnson, de~Juan, Kehr, Kemmer,
  Kim, Kl{\"{u}}ter, Klutsch, Labarga, Labiche, Lalitha, Lamp{\'{o}}n, Lara,
  Launhardt, L{\'{a}}zaro, Lizon, Llamas, Lodieu, {L{\'{o}}pez del Fresno},
  Salas, L{\'{o}}pez-Santiago, Madinabeitia, Mall, Mancini, Mandel, Marfil,
  Molina, Mart{\'{i}}n, Mart{\'{i}}n-Fern{\'{a}}ndez, Mart{\'{i}}n-Ruiz,
  Mart{\'{i}}nez-Rodr{\'{i}}guez, Marvin, Mirabet, Moya, Naranjo, Nelson,
  Nortmann, Nowak, Ofir, Pascual, Pavlov, Pedraz, Medialdea,
  P{\'{e}}rez-Calpena, Perryman, Rabaza, Ballesta, Rebolo, Redondo, Rix,
  Rodler, Trinidad, Sabotta, Sadegi, Salz, S{\'{a}}nchez-Blanco, Carrasco,
  S{\'{a}}nchez-L{\'{o}}pez, Sanz-Forcada, Sarkis, Sarmiento, Sch{\"{a}}fer,
  Schlecker, Schmitt, Sch{\"{o}}fer, Solano, Sota, Stahl, Stock, Stuber,
  St{\"{u}}rmer, Su{\'{a}}rez, Tabernero, Tulloch, Veredas, Vico-Linares,
  Vilardell, Wagner, Winkler, Wolthoff, Yan, \& Osorio}]{Morales2019}
Morales, J.~C., Mustill, A.~J., Ribas, I., {et~al.} 2019, Science, 365, 1441

\bibitem[{Morbidelli {et~al.}(2021)Morbidelli, Nesvorny, Bottke, \&
  Marchi}]{Morbidelli2021}
Morbidelli, A., Nesvorny, D., Bottke, W., \& Marchi, S. 2021, Icarus, 356,
  114256

\bibitem[{Mordasini(2014)}]{Mordasini2014}
Mordasini, C. 2014, A{\&}A, 572, A118

\bibitem[{Mordasini(2018)}]{Mordasini2018}
Mordasini, C. 2018, in Handbook of Exoplanets (Springer International
  Publishing), 143

\bibitem[{Mordasini {et~al.}(2009{\natexlab{a}})Mordasini, Alibert, \&
  Benz}]{Mordasini2009}
Mordasini, C., Alibert, Y., \& Benz, W. 2009{\natexlab{a}}, A{\&}A, 501, 1139

\bibitem[{Mordasini {et~al.}(2009{\natexlab{b}})Mordasini, Alibert, Benz, \&
  Naef}]{Mordasini2009b}
Mordasini, C., Alibert, Y., Benz, W., \& Naef, D. 2009{\natexlab{b}}, A{\&}A,
  501, 1161

\bibitem[{Mordasini {et~al.}(2012{\natexlab{a}})Mordasini, Alibert, Georgy,
  Dittkrist, Klahr, \& Henning}]{Mordasini2012b}
Mordasini, C., Alibert, Y., Georgy, C., {et~al.} 2012{\natexlab{a}}, A{\&}A,
  547, A112

\bibitem[{Mordasini {et~al.}(2012{\natexlab{b}})Mordasini, Alibert, Klahr, \&
  Henning}]{Mordasini2012c}
Mordasini, C., Alibert, Y., Klahr, H., \& Henning, T. 2012{\natexlab{b}},
  A{\&}A, 547, A111

\bibitem[{Mordasini {et~al.}(2010)Mordasini, Klahr, Alibert, Benz, \&
  Dittkrist}]{Mordasini}
Mordasini, C., Klahr, H., Alibert, Y., Benz, W., \& Dittkrist, K.-M. 2010, in
  Circumstellar disks and planets, Kiel

\bibitem[{Mordasini {et~al.}(2015)Mordasini, Molli{\`{e}}re, Dittkrist, Jin, \&
  Alibert}]{Mordasini2015}
Mordasini, C., Molli{\`{e}}re, P., Dittkrist, K.-M. K.-M., Jin, S., \& Alibert,
  Y. 2015, International Journal of Astrobiology, 14, 201

\bibitem[{Mulders {et~al.}(2019)Mulders, Mordasini, Pascucci, Ciesla,
  Emsenhuber, \& Apai}]{Mulders2019}
Mulders, G.~D., Mordasini, C., Pascucci, I., {et~al.} 2019, ApJ, 887, 157

\bibitem[{Mulders {et~al.}(2015)Mulders, Pascucci, \& Apai}]{Mulders2015a}
Mulders, G.~D., Pascucci, I., \& Apai, D. 2015, ApJ, 814, 130

\bibitem[{Nakamoto \& Nakagawa(1994)}]{Nakamoto1994}
Nakamoto, T. \& Nakagawa, Y. 1994, ApJ, 421, 640

\bibitem[{Nutzman \& Charbonneau(2008)}]{Nutzman2008}
Nutzman, P. \& Charbonneau, D. 2008, Publications of the Astronomical Society
  of the Pacific, 120, 317

\bibitem[{Ogihara {et~al.}(2018)Ogihara, Kokubo, Suzuki, \&
  Morbidelli}]{Ogihara2018}
Ogihara, M., Kokubo, E., Suzuki, T.~K., \& Morbidelli, A. 2018, A{\&}A, 615,
  A63

\bibitem[{Ogilvie(2014)}]{Ogilvie2014}
Ogilvie, G.~I. 2014, Annual Review of Astronomy and Astrophysics, 52, 171

\bibitem[{Ohtsuki {et~al.}(2002)Ohtsuki, Stewart, \& Ida}]{Ohtsuki2002}
Ohtsuki, K., Stewart, G.~R., \& Ida, S. 2002, Icarus, 155, 436

\bibitem[{Ormel {et~al.}(2010)Ormel, Dullemond, \& Spaans}]{Ormel2010b}
Ormel, C.~W., Dullemond, C.~P., \& Spaans, M. 2010, ApJ, 714, L103

\bibitem[{Ormel {et~al.}(2017)Ormel, Liu, \& Schoonenberg}]{Ormel2017}
Ormel, C.~W., Liu, B., \& Schoonenberg, D. 2017, A{\&}A, 604, A1

\bibitem[{Owen \& Wu(2013)}]{Owen2013}
Owen, J.~E. \& Wu, Y. 2013, ApJ, 775, 105

\bibitem[{Paardekooper {et~al.}(2011)Paardekooper, Baruteau, \&
  Kley}]{Paardekooper2011}
Paardekooper, S.-J., Baruteau, C., \& Kley, W. 2011, MNRAS, 410, 293

\bibitem[{Pascucci {et~al.}(2018)Pascucci, Mulders, Gould, \&
  Fernandes}]{Pascucci2018}
Pascucci, I., Mulders, G.~D., Gould, A., \& Fernandes, R. 2018, ApJ, 856, L28

\bibitem[{Pascucci {et~al.}(2016)Pascucci, Testi, Herczeg, Long, Manara,
  Hendler, Mulders, Krijt, Ciesla, Henning, Mohanty, Drabek-Maunder, Apai,
  Szűcs, Sacco, \& Olofsson}]{Pascucci2016}
Pascucci, I., Testi, L., Herczeg, G.~J., {et~al.} 2016, ApJ, 831, 125

\bibitem[{Payne \& Lodato(2007)}]{Payne2007}
Payne, M.~J. \& Lodato, G. 2007, MNRAS, 381, 1597

\bibitem[{Pinilla {et~al.}(2018)Pinilla, Natta, Manara, Ricci, Scholz, \&
  Testi}]{Pinilla2018}
Pinilla, P., Natta, A., Manara, C.~F., {et~al.} 2018, A{\&}A, 615, A95

\bibitem[{Pollack {et~al.}(1996)Pollack, Hubickyj, Bodenheimer, Lissauer,
  Podolak, \& Greenzweig}]{Pollack1996}
Pollack, J.~B., Hubickyj, O., Bodenheimer, P.~H., {et~al.} 1996, Icarus, 124,
  62

\bibitem[{Pringle(1981)}]{Pringle1981}
Pringle, J.~E. 1981, Annual Review of Astronomy and Astrophysics, 19

\bibitem[{Quirrenbach \& {CARMENES Consortium}(2020)}]{Quirrenbach2020}
Quirrenbach, A. \& {CARMENES Consortium}. 2020, in Ground-based Airborne
  Instrumentation Astronomy VIII, ed. C.~J. Evans, J.~J. Bryant, \& K.~Motohara
  (SPIE), 262

\bibitem[{Rafikov(2003)}]{Rafikov2003}
Rafikov, R.~R. 2003, The Astronomical Journal, 125, 942

\bibitem[{Rafikov(2004)}]{Rafikov2004}
Rafikov, R.~R. 2004, The Astronomical Journal, 128, 1348

\bibitem[{Rafikov(2017)}]{Rafikov2017}
Rafikov, R.~R. 2017, ApJ, 837, 163

\bibitem[{Raymond {et~al.}(2007)Raymond, Scalo, \& Meadows}]{Raymond2007}
Raymond, S.~N., Scalo, J., \& Meadows, V.~S. 2007, ApJ, 669, 606

\bibitem[{Reffert {et~al.}(2015)Reffert, Bergmann, Quirrenbach, Trifonov, \&
  K{\"{u}}nstler}]{Reffert2015}
Reffert, S., Bergmann, C., Quirrenbach, A., Trifonov, T., \& K{\"{u}}nstler, A.
  2015, A{\&}A, 574, A116

\bibitem[{Ribas {et~al.}(2015)Ribas, Bouy, \& Mer{\'{i}}n}]{Ribas2015}
Ribas, {\'{A}}., Bouy, H., \& Mer{\'{i}}n, B. 2015, A{\&}A, 576, A52

\bibitem[{Ribas {et~al.}(2014)Ribas, Mer{\'{i}}n, Bouy, \& Maud}]{Ribas2014}
Ribas, {\'{A}}., Mer{\'{i}}n, B., Bouy, H., \& Maud, L.~T. 2014, A{\&}A, 561,
  A54

\bibitem[{Richert {et~al.}(2018)Richert, Getman, Feigelson, Kuhn, Broos,
  Povich, Bate, \& Garmire}]{Richert2018}
Richert, A. J.~W., Getman, K.~V., Feigelson, E.~D., {et~al.} 2018, MNRAS, 477,
  5191

\bibitem[{Ricker {et~al.}(2014)Ricker, Winn, Vanderspek, Latham, Bakos, Bean,
  Berta-Thompson, Brown, Buchhave, Butler, Butler, Chaplin, Charbonneau,
  Christensen-Dalsgaard, Clampin, Deming, Doty, {De Lee}, Dressing, Dunham,
  Endl, Fressin, Ge, Henning, Holman, Howard, Ida, Jenkins, Jernigan, Johnson,
  Kaltenegger, Kawai, Kjeldsen, Laughlin, Levine, Lin, Lissauer, MacQueen,
  Marcy, McCullough, Morton, Narita, Paegert, Palle, Pepe, Pepper, Quirrenbach,
  Rinehart, Sasselov, Sato, Seager, Sozzetti, Stassun, Sullivan, Szentgyorgyi,
  Torres, Udry, \& Villasenor}]{Ricker2014}
Ricker, G.~R., Winn, J.~N., Vanderspek, R., {et~al.} 2014, Journal of
  Astronomical Telescopes, Instruments, and Systems, 1, 014003

\bibitem[{Rivera {et~al.}(2010)Rivera, Laughlin, Butler, Vogt, Haghighipour, \&
  Meschiari}]{Rivera2010}
Rivera, E.~J., Laughlin, G., Butler, R.~P., {et~al.} 2010, ApJ, 719, 890

\bibitem[{Rivera {et~al.}(2005)Rivera, Lissauer, Butler, Marcy, Vogt, Fischer,
  Brown, Laughlin, \& Henry}]{Rivera2005}
Rivera, E.~J., Lissauer, J.~J., Butler, R.~P., {et~al.} 2005, ApJ, 634, 625

\bibitem[{Sanchis {et~al.}(2021)Sanchis, Testi, Natta, Facchini, Manara,
  Miotello, Ercolano, Henning, Preibisch, Carpenter, de~Gregorio-Monsalvo,
  Jayawardhana, Lopez, Mu{\v{z}}i{\'{c}}, Pascucci, Santamar{\'{i}}a-Miranda,
  van Terwisga, \& Williams}]{Sanchis2021}
Sanchis, E., Testi, L., Natta, A., {et~al.} 2021, A{\&}A, 649, A19

\bibitem[{Sanchis {et~al.}(2020)Sanchis, Testi, Natta, Manara, Ercolano,
  Preibisch, Henning, Facchini, Miotello, de~Gregorio-Monsalvo, Lopez,
  Mu{\v{z}}i{\'{c}}, Pascucci, Santamar{\'{i}}a-Miranda, Scholz, Tazzari, van
  Terwisga, \& Williams}]{Sanchis2020}
Sanchis, E., Testi, L., Natta, A., {et~al.} 2020, A{\&}A, 633, A114

\bibitem[{Santos {et~al.}(2003)Santos, Israelian, Mayor, Rebolo, \&
  Udry}]{Santos2003}
Santos, N.~C., Israelian, G., Mayor, M., Rebolo, R., \& Udry, S. 2003, A{\&}A,
  398, 363

\bibitem[{Schlaufman {et~al.}(2010)Schlaufman, Lin, \& Ida}]{Schlaufman2010}
Schlaufman, K.~C., Lin, D.~N., \& Ida, S. 2010, ApJL, 724, L53

\bibitem[{Schlecker {et~al.}(2021)Schlecker, Mordasini, Emsenhuber, Klahr,
  Henning, Burn, Alibert, \& Benz}]{Schlecker2020}
Schlecker, M., Mordasini, C., Emsenhuber, A., {et~al.} 2021, A{\&}A, 656, A71

\bibitem[{Schlichting {et~al.}(2013)Schlichting, Fuentes, \&
  Trilling}]{Schlichting2013}
Schlichting, H.~E., Fuentes, C.~I., \& Trilling, D.~E. 2013, The Astronomical
  Journal, 146, 36

\bibitem[{Schoonenberg {et~al.}(2019)Schoonenberg, Liu, Ormel, \&
  Dorn}]{Schoonenberg2019}
Schoonenberg, D., Liu, B., Ormel, C.~W., \& Dorn, C. 2019, A{\&}A, 627, A149

\bibitem[{Schoonenberg \& Ormel(2017)}]{Schoonenberg2017}
Schoonenberg, D. \& Ormel, C.~W. 2017, A{\&}A, 602, A21

\bibitem[{Scott(1992)}]{Scott1992}
Scott, D.~W. 1992, {Multivariate Density Estimation: Theory, Practice, and
  Visualization}, Wiley Series in Probability and Statistics (New York: John
  Wiley {\&} Sons, Inc.), 317

\bibitem[{Seager {et~al.}(2007)Seager, Kuchner, Hier-Majumder, \&
  Militzer}]{Seager2007}
Seager, S., Kuchner, M., Hier-Majumder, C.~A., \& Militzer, B. 2007, ApJ, 669,
  1279

\bibitem[{Seifahrt {et~al.}(2016)Seifahrt, Bean, St{\"{u}}rmer, Gers, Grobler,
  Reed, \& Jones}]{Seifahrt2016}
Seifahrt, A., Bean, J.~L., St{\"{u}}rmer, J., {et~al.} 2016, in Ground-based
  Airborne Instrumentation Astronomy VI, ed. C.~J. Evans, L.~Simard, \&
  H.~Takami, Vol. 9908, 990818

\bibitem[{Sestovic \& Demory(2020)}]{Sestovic2020}
Sestovic, M. \& Demory, B.-O. 2020, A{\&}A, 641, A170

\bibitem[{Shah {et~al.}(2021)Shah, Alibert, Helled, \& Mezger}]{Shah2021}
Shah, O., Alibert, Y., Helled, R., \& Mezger, K. 2021, A{\&}A, 646, A162

\bibitem[{Shakura \& Sunyaev(1973)}]{Shakura1973}
Shakura, N.~I. \& Sunyaev, R.~A. 1973, A{\&}A, 24, 337

\bibitem[{Shields {et~al.}(2016)Shields, Ballard, \& Johnson}]{Shields2016}
Shields, A.~L., Ballard, S., \& Johnson, J.~A. 2016, Physics Reports, 663, 1

\bibitem[{Siess {et~al.}(2000)Siess, Dufour, \& Forestini}]{Siess2000}
Siess, L., Dufour, E., \& Forestini, M. 2000, A{\&}A, 358, 593

\bibitem[{Stassun {et~al.}(2016)Stassun, Collins, \& Gaudi}]{Stassun2016}
Stassun, K.~G., Collins, K.~A., \& Gaudi, B.~S. 2016, The Astronomical Journal,
  153, 136

\bibitem[{Stock {et~al.}(2020)Stock, Cai, Spurzem, Kouwenhoven, \& {Portegies
  Zwart}}]{Stock2020}
Stock, K., Cai, M.~X., Spurzem, R., Kouwenhoven, M. B.~N., \& {Portegies
  Zwart}, S. 2020, MNRAS, 497, 1807

\bibitem[{Strom {et~al.}(1989)Strom, Strom, Edwards, Cabrit, \&
  Skrutskie}]{Strom1989}
Strom, K.~M., Strom, S.~E., Edwards, S., Cabrit, S., \& Skrutskie, M.~F. 1989,
  The Astronomical Journal, 97, 1451

\bibitem[{Sumi {et~al.}(2011)Sumi, Kamiya, Bennett, Bond, Abe, Botzler, Fukui,
  Furusawa, Hearnshaw, Itow, Kilmartin, Korpela, Lin, Ling, Masuda, Matsubara,
  Miyake, Motomura, Muraki, Nagaya, Nakamura, Ohnishi, Okumura, Perrott,
  Rattenbury, Saito, Sako, Sullivan, Sweatman, Tristram, Yock, Udalski,
  Szyma{\'{n}}ski, Kubiak, Pietrzy{\'{n}}ski, Poleski, SoszyA{\'{n}}ski,
  Wyrzykowski, \& Ulaczyk}]{Sumi2011}
Sumi, T., Kamiya, K., Bennett, D.~P., {et~al.} 2011, Nature, 473, 349

\bibitem[{Suzuki {et~al.}(2016)Suzuki, Ogihara, Morbidelli, Crida, \&
  Guillot}]{Suzuki2016Winds}
Suzuki, T.~K., Ogihara, M., Morbidelli, A., Crida, A., \& Guillot, T. 2016,
  A{\&}A, 596, A74

\bibitem[{Tasker {et~al.}(2017)Tasker, Tan, Heng, Kane, Spiegel, Brasser,
  Casey, Desch, Dorn, Hernlund, Houser, Laneuville, Lasbleis, Libert, Noack,
  Unterborn, \& Wicks}]{Tasker2017}
Tasker, E., Tan, J., Heng, K., {et~al.} 2017, Nature Astronomy, 1, 0042

\bibitem[{Taubner {et~al.}(2020)Taubner, Olsson-Francis, Vance, Ramkissoon,
  Postberg, de~Vera, Antunes, {Camprubi Casas}, Sekine, Noack, Barge, Goodman,
  Jebbar, Journaux, Karatekin, Klenner, Rabbow, Rettberg, R{\"{u}}ckriemen-Bez,
  Saur, Shibuya, \& Soderlund}]{Taubner2020}
Taubner, R.-S., Olsson-Francis, K., Vance, S.~D., {et~al.} 2020, Space Science
  Reviews, 216, 9

\bibitem[{Testi {et~al.}(2016)Testi, Natta, Scholz, Tazzari, Ricci, \& {de
  Gregorio Monsalvo}}]{Testi2016}
Testi, L., Natta, A., Scholz, A., {et~al.} 2016, A{\&}A, 593, A111

\bibitem[{Thiabaud {et~al.}(2014)Thiabaud, Marboeuf, Alibert, Cabral, Leya, \&
  Mezger}]{Thiabaud2014}
Thiabaud, A., Marboeuf, U., Alibert, Y., {et~al.} 2014, A{\&}A, 562, A27

\bibitem[{Turbet {et~al.}(2020)Turbet, Bolmont, Ehrenreich, Gratier, Leconte,
  Selsis, Hara, \& Lovis}]{Turbet2020}
Turbet, M., Bolmont, E., Ehrenreich, D., {et~al.} 2020, A{\&}A, 638, A41

\bibitem[{Tychoniec {et~al.}(2018)Tychoniec, Tobin, Karska, Chandler, Dunham,
  Harris, Kratter, Li, Looney, Melis, P{\'{e}}rez, Sadavoy, Segura-Cox, \& van
  Dishoeck}]{Tychoniec2018}
Tychoniec, {\L}., Tobin, J.~J., Karska, A., {et~al.} 2018, The Astrophysical
  Journal Supplement Series, 238, 19

\bibitem[{Udry {et~al.}(2007)Udry, Bonfils, Delfosse, Forveille, Mayor,
  Perrier, Bouchy, Lovis, Pepe, Queloz, \& Bertaux}]{Udry2007}
Udry, S., Bonfils, X., Delfosse, X., {et~al.} 2007, A{\&}A, 469, L43

\bibitem[{{Van Hoolst} {et~al.}(2019){Van Hoolst}, Noack, \&
  Rivoldini}]{VanHoolst2019}
{Van Hoolst}, T., Noack, L., \& Rivoldini, A. 2019, Advances Physics X, 4,
  1630316

\bibitem[{Venturini {et~al.}(2020)Venturini, Guilera, Haldemann, Ronco, \&
  Mordasini}]{Venturini2020}
Venturini, J., Guilera, O.~M., Haldemann, J., Ronco, M.~P., \& Mordasini, C.
  2020, A{\&}A, 643, L1

\bibitem[{Venuti {et~al.}(2017)Venuti, Bouvier, Cody, Stauffer, Micela, Rebull,
  Alencar, Sousa, Hillenbrand, \& Flaccomio}]{Venuti2017}
Venuti, L., Bouvier, J., Cody, A.~M., {et~al.} 2017, A{\&}A, 599, A23

\bibitem[{Veras \& Armitage(2004)}]{Veras2004}
Veras, D. \& Armitage, P.~J. 2004, MNRAS, 347, 613

\bibitem[{Veras \& Raymond(2012)}]{Veras2012}
Veras, D.~M. \& Raymond, S.~N. 2012, Monthly Notices of the Royal Astronomical
  Society: Letters, 421, 117

\bibitem[{Voelkel {et~al.}(2021)Voelkel, Deienno, Kretke, \&
  Klahr}]{Voelkel2021}
Voelkel, O., Deienno, R., Kretke, K., \& Klahr, H. 2021, A{\&}A, 645, A131

\bibitem[{Voelkel {et~al.}(2020)Voelkel, Klahr, Mordasini, Emsenhuber, \&
  Lenz}]{Voelkel2020}
Voelkel, O., Klahr, H., Mordasini, C., Emsenhuber, A., \& Lenz, C. 2020,
  A{\&}A, 642, A75

\bibitem[{von Braun {et~al.}(2013)von Braun, Boyajian, van Belle, Kane, Jones,
  Farrington, Schaefer, Vargas, Scott, ten Brummelaar, Kephart, Gies, Ciardi,
  Lopez-Morales, Mazingue, McAlister, Ridgway, Goldfinger, Turner, \&
  Sturmann}]{vonBraun2013}
von Braun, K., Boyajian, T.~S., van Belle, G.~T., {et~al.} 2013, MNRAS, 438,
  2413

\bibitem[{Vorobyov \& Basu(2009)}]{Vorobyov2009}
Vorobyov, E.~I. \& Basu, S. 2009, ApJ, 703, 922

\bibitem[{Wheatley {et~al.}(2018)Wheatley, West, Goad, Jenkins, Pollacco,
  Queloz, Rauer, Udry, Watson, Chazelas, Eigm{\"{u}}ller, Lambert, Genolet,
  McCormac, Walker, Armstrong, Bayliss, Bento, Bouchy, Burleigh, Cabrera,
  Casewell, Chaushev, Chote, Csizmadia, Erikson, Faedi, Foxell, G{\"{a}}nsicke,
  Gillen, Grange, G{\"{u}}nther, Hodgkin, Jackman, Jord{\'{a}}n, Louden,
  Metrailler, Moyano, Nielsen, Osborn, Poppenhaeger, Raddi, Raynard, Smith,
  Soto, \& Titz-Weider}]{Wheatley2018}
Wheatley, P.~J., West, R.~G., Goad, M.~R., {et~al.} 2018, MNRAS, 475, 4476

\bibitem[{Williams {et~al.}(2019)Williams, Cieza, Hales, Ansdell,
  Ruiz-Rodriguez, Casassus, Perez, \& Zurlo}]{Williams2019}
Williams, J.~P., Cieza, L., Hales, A., {et~al.} 2019, ApJ, 875, L9

\bibitem[{Winters {et~al.}(2014)Winters, Henry, Lurie, Hambly, Jao, Bartlett,
  Boyd, Dieterich, Finch, Hosey, Ianna, Riedel, Slatten, \&
  Subasavage}]{Winters2014}
Winters, J.~G., Henry, T.~J., Lurie, J.~C., {et~al.} 2014, The Astronomical
  Journal, 149, 5

\bibitem[{Wu(2019)}]{Wu2019}
Wu, Y. 2019, ApJ, 874, 91

\bibitem[{Wyatt {et~al.}(2020)Wyatt, Kral, \& Sinclair}]{Wyatt2020}
Wyatt, M.~C., Kral, Q., \& Sinclair, C.~A. 2020, MNRAS, 491, 782

\bibitem[{Zeng {et~al.}(2019)Zeng, Jacobsen, Sasselov, Petaev, Vanderburg,
  Lopez-Morales, Perez-Mercader, Mattsson, Li, Heising, Bonomo, Damasso,
  Berger, Cao, Levi, \& Wordsworth}]{Zeng2019}
Zeng, L., Jacobsen, S.~B., Sasselov, D.~D., {et~al.} 2019, Proceedings of the
  National Academy of Sciences, 116, 9723

\end{thebibliography}

\end{document}